\documentclass[12pt,a4paper]{article}

\usepackage[margin=1in]{geometry}
\usepackage{graphicx, float, subfigure, subcaption, xparse, pgffor}
\usepackage{amsmath, amssymb}
\usepackage[
  colorlinks=true,
  linkcolor=blue,
  citecolor=blue,
  filecolor=blue,
  urlcolor=blue,
  pdfborder={0 0 0}
]{hyperref}

\graphicspath{{./figures/}}

\title{Orbits in the Integrable Hénon-Heiles Systems}

\author{
A.C. Tzemos\thanks{Corresponding author: \texttt{atzemos@academyofathens.gr}}$^1$, 
G. Contopoulos$^1$, 
F. Zanias$^{1,2}$\thanks{\texttt{foivos.zanias@student.uva.nl}}
}

\date{
$^1$Research Center for Astronomy and Applied Mathematics of the Academy of Athens,\\
Soranou Efessiou 4, GR-11527 Athens, Greece\\[1mm]
$^2$University of Amsterdam, Science Park 904, 1098 XH Amsterdam, The Netherlands
}

\begin{document}

\maketitle
\begin{abstract}
We study in detail the form of the orbits in integrable generalized Hénon-Heiles systems with Hamiltonians of the form 
$
H = \frac{1}{2}(\dot{x}^2 + Ax^2 + \dot{y}^2 + By^2) + \epsilon(xy^2 + \alpha x^3).
$
In particular, we focus on the invariant curves on Poincaré surfaces of section ($ y = 0$) and the corresponding orbits on the $x-y$ plane. We provide a detailed analysis of the transition from bounded to escaping orbits in each integrable system case, highlighting the mechanism behind the escape to infinity. Then, we investigate the form of the non-escaping orbits, conducting a comparative analysis across various integrable cases and physical parameters.

\end{abstract}

\section{Introduction}\label{sec1}

The phase space of a generic nonlinear Hamiltonian system is mixed: it contains regions with ordered orbits and regions of chaotic orbits (that show high sensitivity on the initial conditions). 

A $2N$-dimensional Hamiltonian system is called integrable when there exist $N$ independent integrals of motion, i.e. functions of the phase-space variables and of the physical parameters of the system. Integrability plays a key role in Hamiltonian Dynamics since all the orbits of an integrable system are ordered. Namely, there is no chaos in integrable systems. Various cases of integrable systems have been studied in the field of Dynamical Astronomy \cite{stackel1890,stackel1893,lynden1962stellar}. For a comprehensive bibliography, see \cite{contopoulos2002order}.

The orbits of a nonlinear integrable Hamiltonian system are divided into  bounded and escaping and they are, in general, computed numerically. However there are also several cases where they can be computed analytically. A major part of the results in Hamiltonian Dynamics has been acquired by use of perturbation theory. By introducing various forms of perturbations in non integrable systems,  we  can gradually increase the strength of the perturbation parameter and study the departure from integrability and the  emergence of chaos.

Moreover, it has been shown that for a nearly integrable system we can construct formal (approximate) integrals of motion in the form of power series with respect to the perturbation parameter that are non-convergent  \cite{contopoulos2002order,giorgilli2022notes}. In such systems there are exact integral surfaces as shown by Kolmogorov \cite{kolmogorov1954conservation}, Arnold \cite{arnold1961} and Moser \cite{Möser:430015} (KAM theorem) which appear near stable equilibria or stable periodic orbits. However, between any two integral surfaces there are in general regions of chaos.

It is remarkable that near unstable equilibria or unstable periodic orbits (which are responsible for the generation of chaos) there are convergent integrals that extend along chaotic orbits as shown by Moser \cite{Moser1956,Moser1958} and Giorgilli  \cite{Giorgilli2001}. 

One of the most well studied systems in Hamiltonian Dynamics is that introduced by H\'{e}non and Heiles \cite{Henon1964}, which is represented by the Hamiltonian
\begin{equation}\label{H2}
    H=\frac{1}{2}\left(\dot{x}^2+Ax^2+\dot{y}^2+By^2\right)+\epsilon(xy^2+\alpha x^3)=E
\end{equation}
{and evolves according to the Newtonian equations of motion}
\begin{align}
\ddot{x} &= -A x - \epsilon (y^2 + 3\alpha x^2) \\
\ddot{y} &= -B y - 2\epsilon x y.
\end{align}
The original system has  $A=B=E=1$ and  $\alpha=-1/3$ and was used as a simple model of a disc galaxy. Since then, it has been extensively studied in the field of chaotic dynamics, {both} classical \cite{conte2005explicit, zhao2007threshold, blesa2012escape} and quantum \cite{Waite1981,sengupta1996quantum,contopoulos2024}.

However, through extensive use of Painlevé analysis (see \cite{ablowitz1980connection,ramani1989painleve, lakshmanan1993painleve} for a review of this method and \cite{wang2016integrability} for a modern application), it was shown that the general system \eqref{H2} is integrable in three cases
 \cite{chang1982analytic, grammaticos1982painleve, bountis1982integrable}:
 \begin{enumerate}
    \item $A=B$, $\alpha = 1/3$
    \item Arbitrary $A, B$ and $\alpha=2$
    \item $A=16B$, $\alpha = 16/3$
\end{enumerate}
In \cite{fordy1991henon} it was shown that these are the only integrable cases, while in \cite{sarlet1991new}  a first step was made toward their generalization in a extended class of H\'{e}non-Heiles Hamiltonians. Finally,  alternative methods for the detection integrability were given in \cite{hietarinta1987direct,smirnov1998integrability}.

{In our previous study \cite{contarx}, we used the theory of the ``third integral of motion'' \cite{contopoulos2002order,Contopoulos1960} in the Hénon-Heiles system for the integrable case (i), examining in detail the form of the invariant curves, both as approximated by the third integral and as derived from the exact equations of motion. We also  investigated the onset of chaos through the resonance overlap mechanism \cite{Rosenbluth1966,Contopoulos1966c} in the case of the non integrable H\'{e}non-Heiles case with $\alpha=-1/3$.}

{In the present work, we extend our analysis to the cases (ii) and (iii), comparing their invariant curves and orbits with those of case (i). In integrable cases, where chaos is absent, the orbits are strictly ordered and can be either escaping or bounded. A key focus of this paper is the distinction between escaping and bounded orbits \cite{blesa2012escape}. We conduct a thorough investigation of the transition to escape, analyzing the role of key physical parameters, namely the frequencies and the magnitude of the perturbation.}

{In particular, we make a detailed study of the form  of the invariant curves and of  the curves of zero velocity (CZV). The invariant curves are determined by incorporating the second integral of motion into the Hamiltonian and imposing the condition $(y = 0$), while the CZV is derived directly from the Hamiltonian when $\dot{x}=\dot{y}=0$. By systematically examining the resulting analytical equations,  we identify the regions of escaping and bounded orbits and determine the escape routes of the trajectories. Our results are accompanied with representative orbits on the $x-y$ plane found by direct numerical integration of the Hamiltonian equations (Runge-Kutta 4-5 Dormand-Prince method with  absolute and relative error tolerances equal to $10^{-10}$).}

The structure of the paper is the following: In section 2 we present the integrable system (ii) for $A=B=1$ and then for $B\neq A\, (E=A=1)$ and compare the corresponding orbits and invariant curves. In section 3, we move on to the system (iii) and present again the form of the corresponding orbits and invariant cures. In section 4, we briefly review the main properties of the system (i). Finally, in section 5 we summarize our results by comparing the three cases and draw our conclusions.

\section{Case (ii) $\alpha=2$ }

\subsection{E=A=1, B=1}

For $\alpha=2$ and arbitrary $A,B$ there is a second integral of motion of the form:
\begin{equation}
     Q=4\, \left( y\dot{x}-x\dot{y} \right) \dot{y}+ \left(4Bx + \epsilon\,y^2+4\,\epsilon x^2\right) {y}^{2}+{\frac { \left( 4\,B-A
  \right)  \left( B{y}^{2}+{\dot{y}}^{2} \right) }{\epsilon}}.
\end{equation}
In the present subsection we  work in a resonant case where $A=B=1$ and $E=1$. Then
\begin{equation}\label{H3}
    H=\frac{1}{2}\left(\dot{x}^2+x^2+\dot{y}^2+y^2\right)+\epsilon(xy^2+2x^3)=1\
\end{equation}
and 
\begin{equation}
    Q= \left( 4{x}^{2}{y}^{2}+{y}^{4} \right) {\epsilon}^{2}+ \left( 4\dot{x}\dot{y}y-4{\dot{y}}^{2}x+4x{y}^{2} \right) \epsilon+3{y}^{2}+3{\dot{y}}^{2}=K',
\end{equation}
where $K'=K\epsilon$.
First we find the invariant curves $(x,\dot{x})$ on the surface of section $y=0$. 

The invariant curves are given by the equation
\begin{equation}
    (3-4\epsilon x)\dot{y}^2=K'=K\epsilon,\label{eqKE}
\end{equation}
where from Eq.~\eqref{H3}:
\begin{equation}
    \dot{y}^2=2-\dot{x}^2-x^2-4\epsilon x^3. \label{yd}
\end{equation}

\begin{figure}[H]
    \centering
    \includegraphics[width=0.4\linewidth]{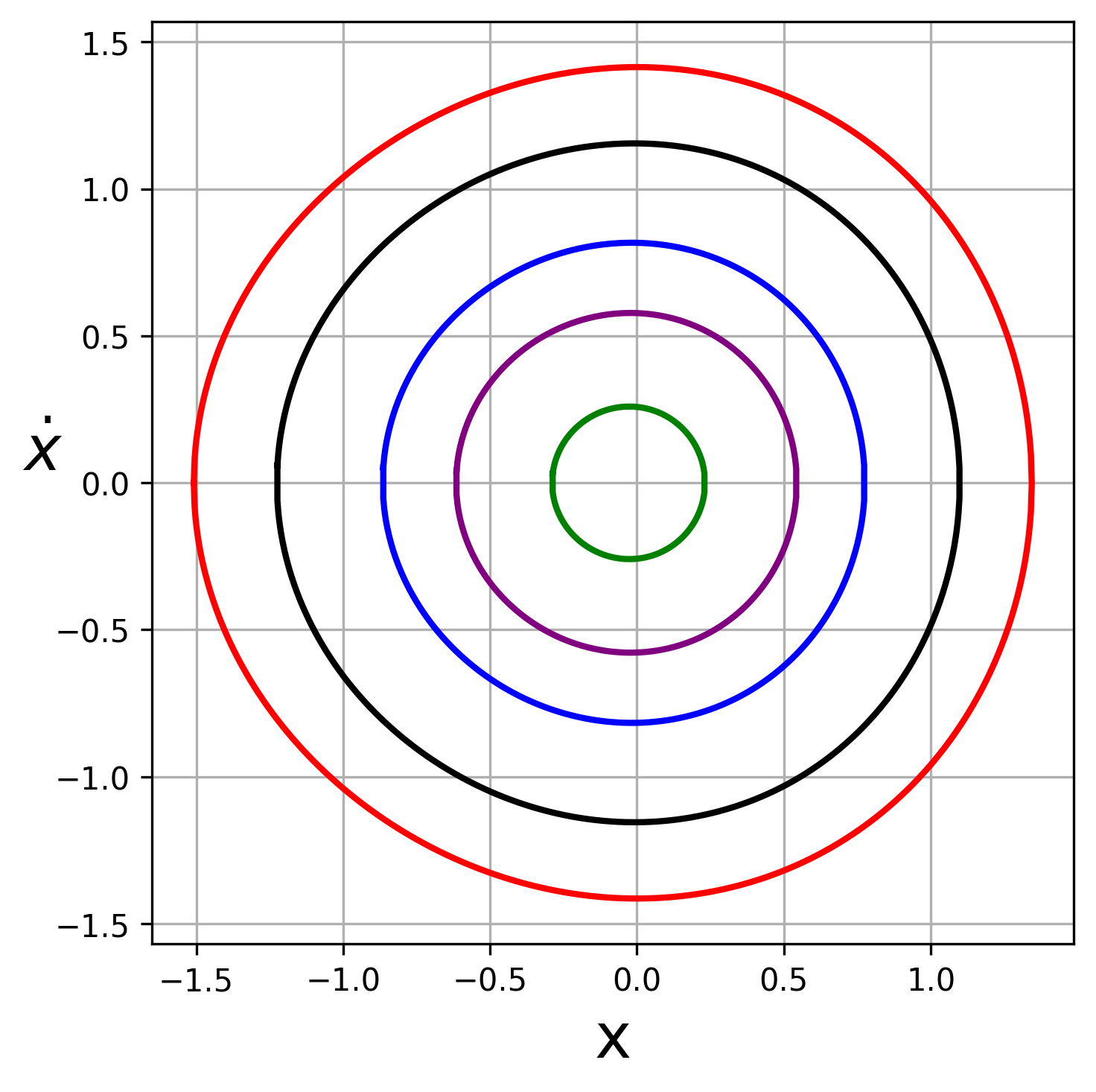}
    \caption{Invariant curves on the $x-\dot{x}$ plane for $\epsilon=0.02$ and $A=B=1$ and for various values of $K$: $K=0$ (red), $K=100$ (black), $K=200$ (blue), $K=250$ (purple) and $K=290$ (green). }
    \label{fig:0_002}
\end{figure}

\begin{figure}[H]
    \centering
    \includegraphics[width=0.4\linewidth]{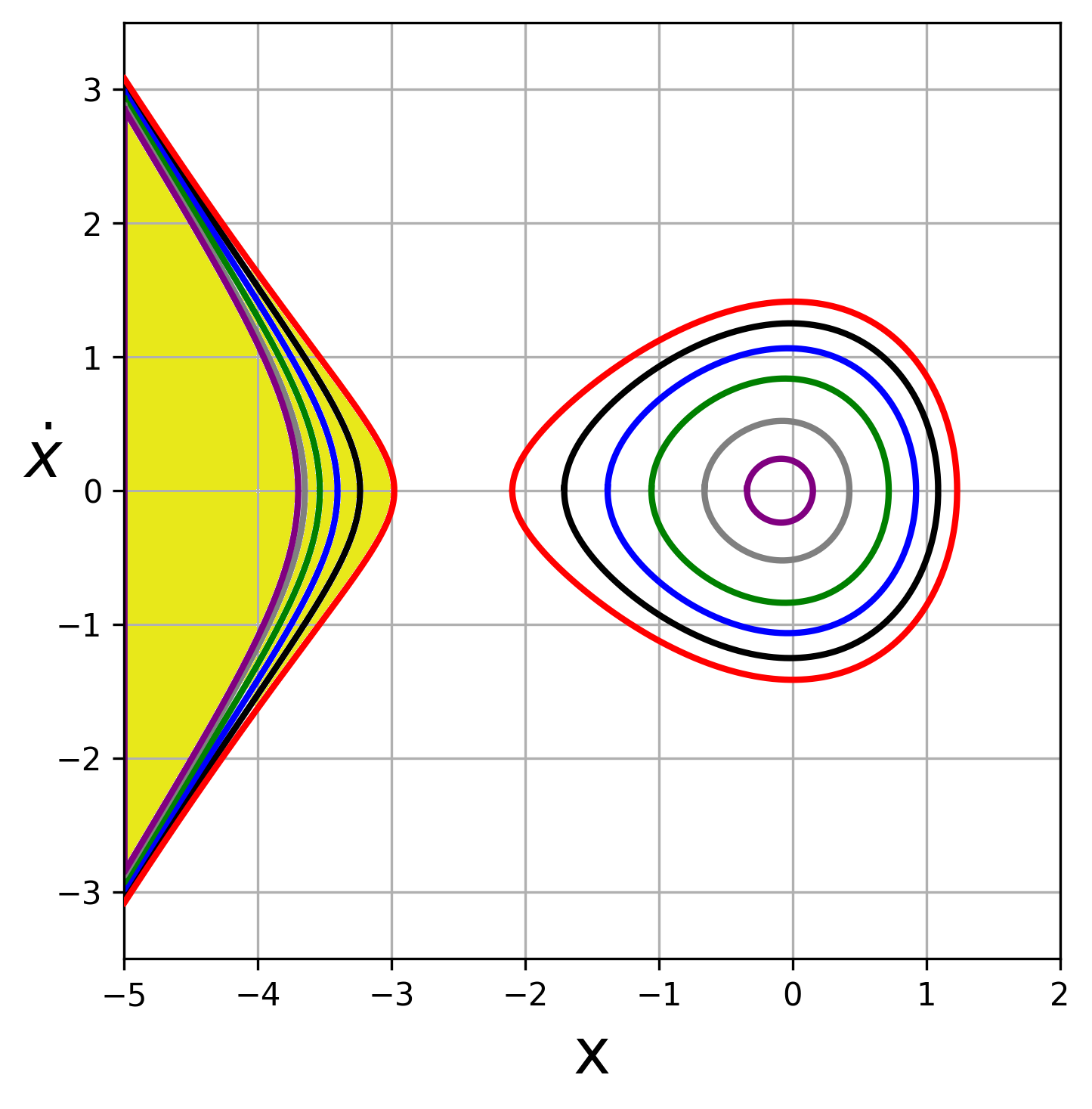}[a]
    \includegraphics[width=0.4\linewidth]{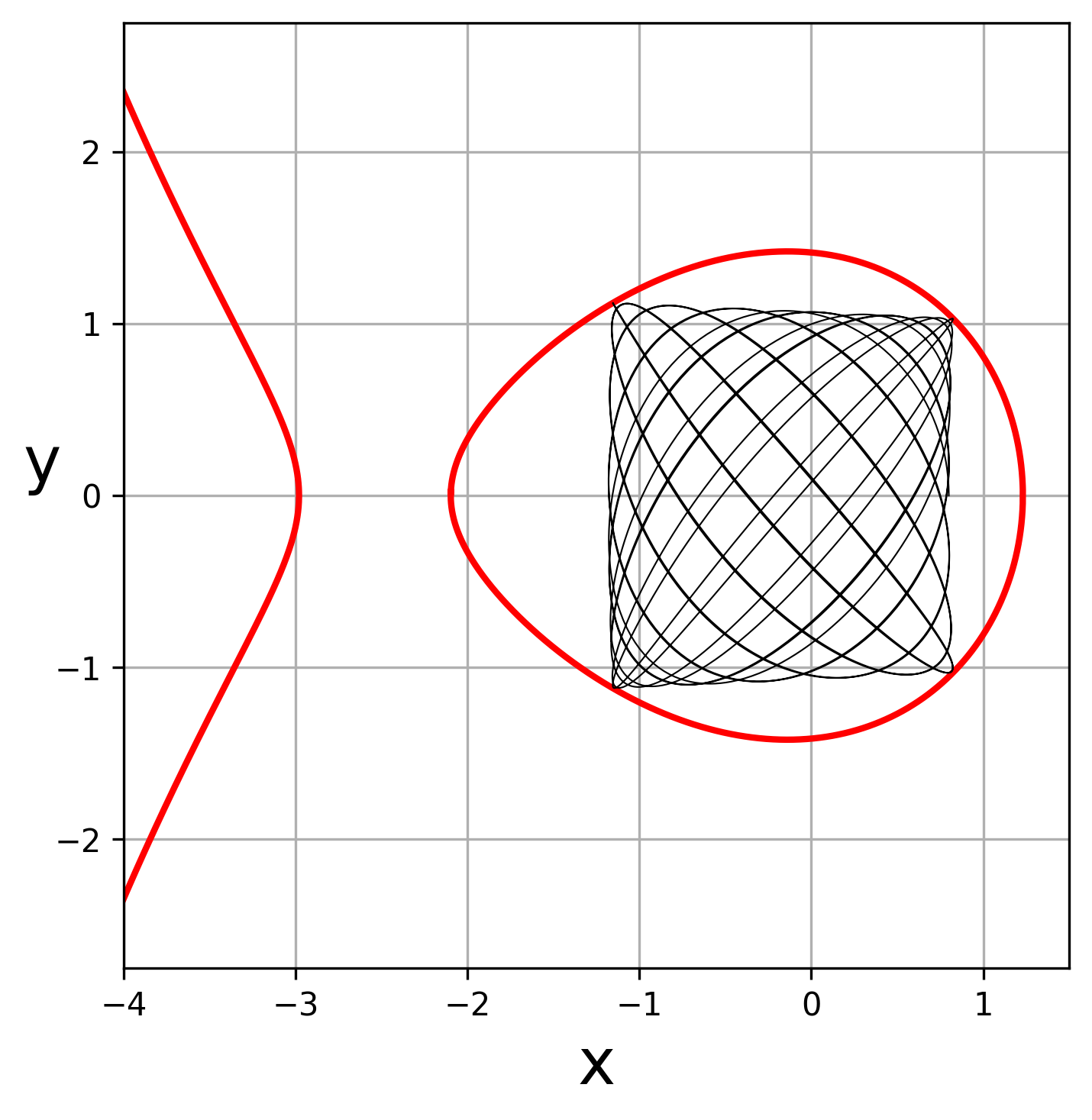}[b]
    \includegraphics[width=0.4\linewidth]{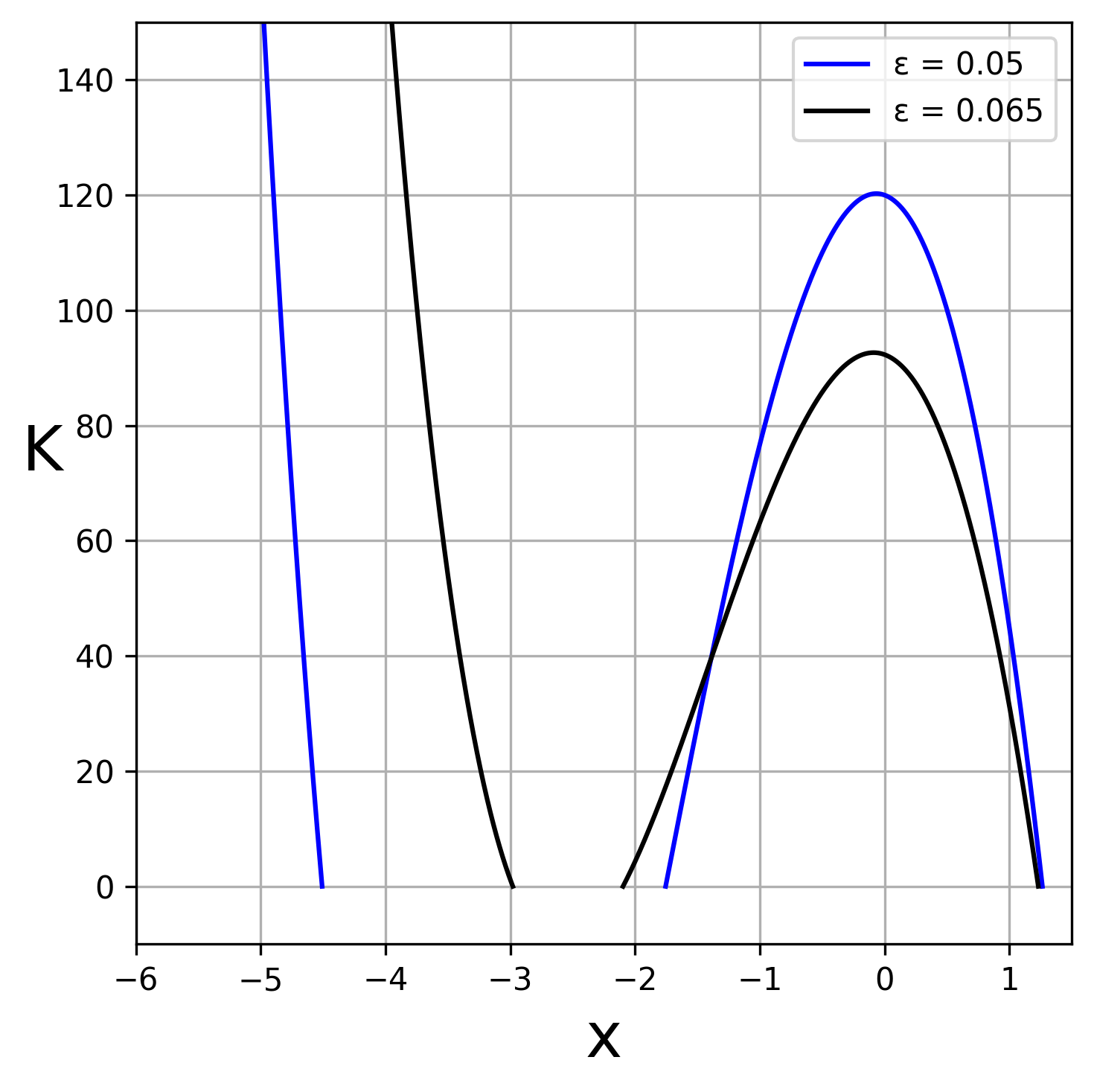}[c]
    
    \caption{a) Invariant curves on the $x-\dot{x}$ plane  for $\epsilon=0.065$ and for various values of $K$: $K=0$ (red), $K=10$ (black), $K=20$ (blue), $K=30$ (green), $K=40$ (gray) and $K=50$ (burgundy). The yellow region contains escaping orbits. b)  A generalized Lissajous curve for $\epsilon=0.065$, $x_0=0.8$, $y_0=0$, $\dot{x}_0=0$. c) The value of $K$ as a function of $x$ for $\epsilon=0.05$ and $\epsilon=0.065$.}
    \label{fig:0_065}
\end{figure}

\begin{figure}[H]
    \centering
    \includegraphics[width=0.4\linewidth]{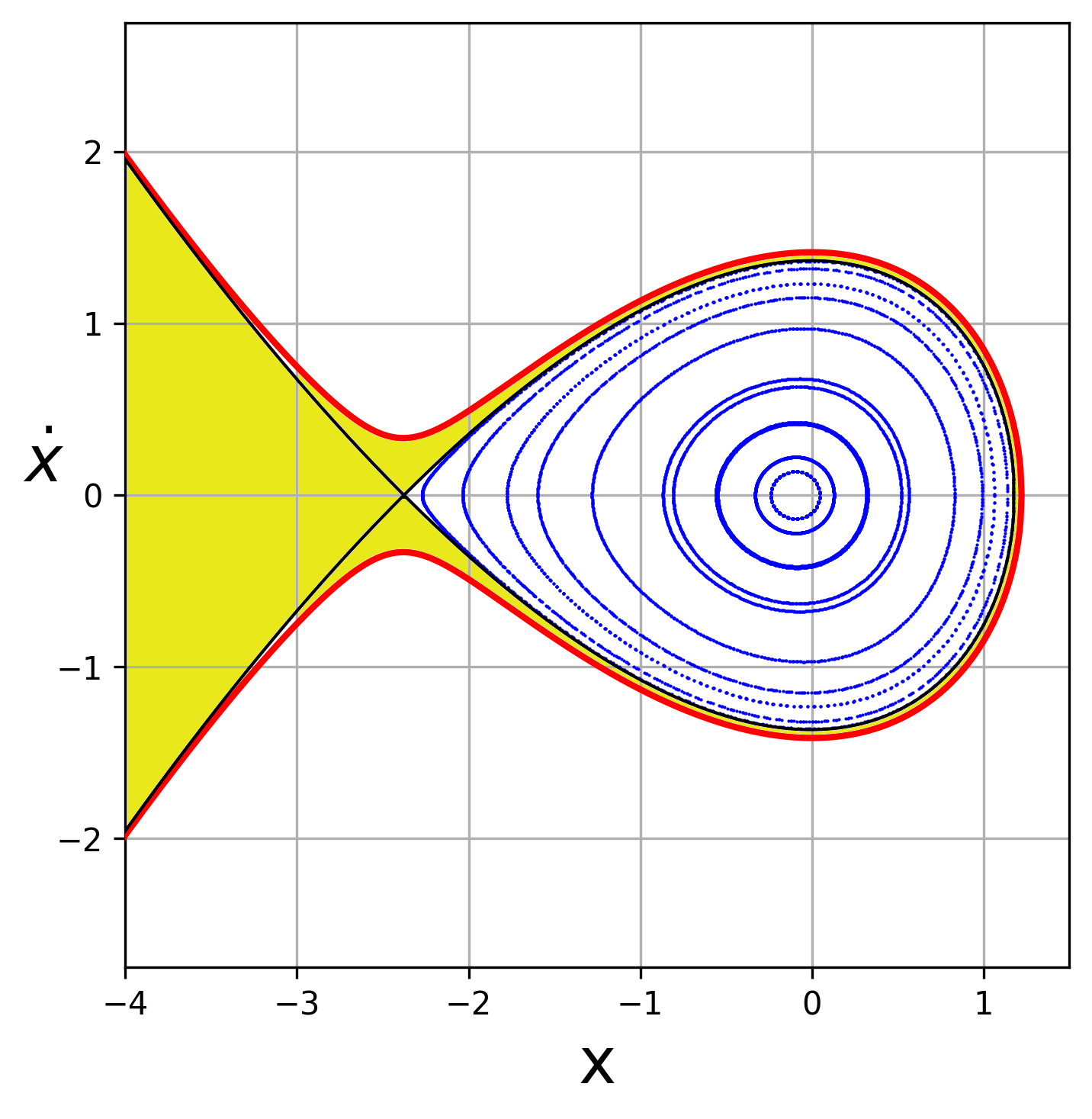}
    \includegraphics[width=0.4\linewidth]{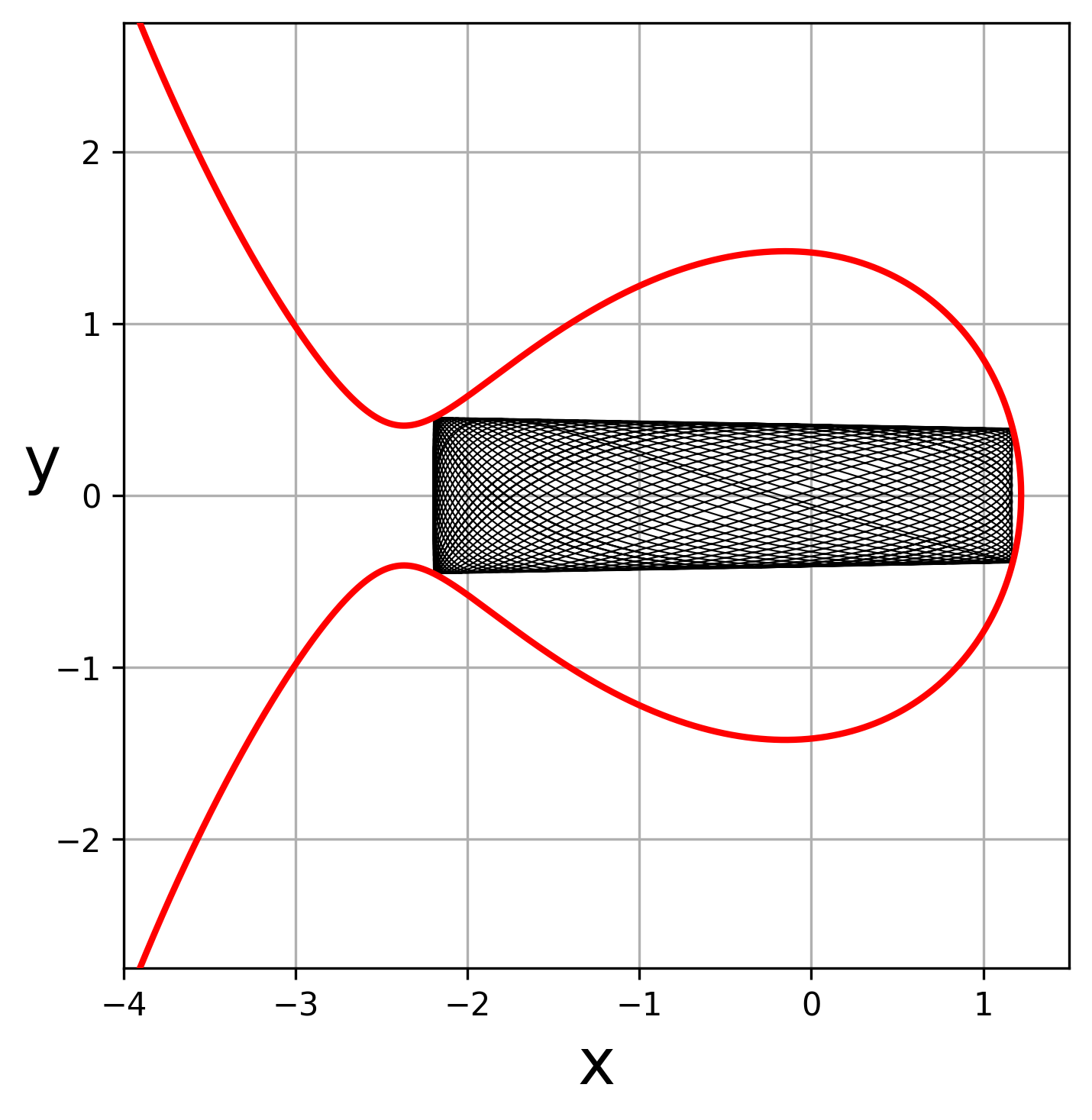}
    \includegraphics[width=0.4\linewidth]{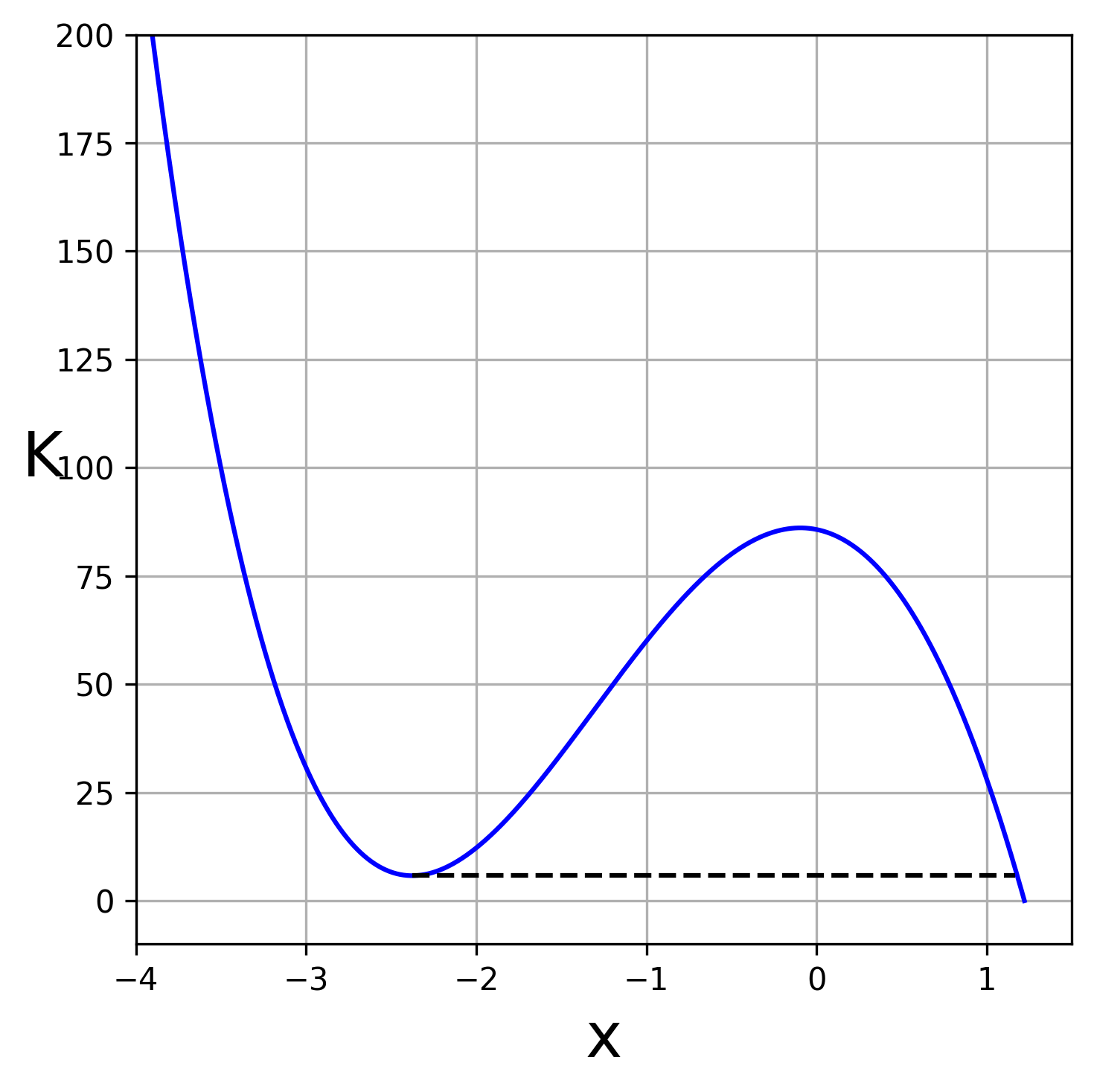}
    \caption{a) The invariant curves for $\epsilon=0.07$ ($>\epsilon_\text{esc}$) for various values of $K$. The outermost red curve  corresponds to $K=0$. The curve crossing itself ($K=5.78$) forms a loop on the right whose center represents a periodic orbit. The yellow regions contain escaping orbits. b) A Lissajous orbit in this case. c) The values of $K$ of the invariant curves as a function of $x$. The dashed line from the minimum of $K$ defines on the right the maximum $x$ of the loop. }
    \label{fig:007}
\end{figure}

 For $\epsilon=0$, Eq.~\eqref{eqKE} gives $\dot{y}^2=0$ and from Eq.~\eqref{yd} we find the circle
\begin{equation}
\dot{x}^2+x^2=2
\end{equation}
for all values of $K$.
For small values of $\epsilon$ we have a set of closed invariant curves for various values of $K$ (Figs.~\ref{fig:0_002}, \ref{fig:0_065}a) and an open curve on the left which is outside the limits of the Fig.~\ref{fig:0_002}. The invariant curves are derived from equations \eqref{eqKE} and \eqref{yd} and are given by
\begin{equation}
    \dot{x}^2=2-x^2-4\epsilon x^3-\frac{K'}{3-4\epsilon x}.\label{xdK}
\end{equation}
For a fixed $\epsilon$ and various values of $K'$ we plot some invariant curves given by various colors in Figs.~\ref{fig:0_002}, \ref{fig:0_065}a and \ref{fig:007}a. 
The outermost red curve of Figs.~\ref{fig:0_002}, \ref{fig:0_065}a is an oval corresponding to $K=0$. Then Eq.~\eqref{xdK} becomes
\begin{equation} 
    \dot{x}^2=2-x^2-4\epsilon x^3\label{outerm}
\end{equation}
and for $\dot{x}^2=0$ (on the $x$-axis) it acquires  three roots $x_1>0>x_2>x_3$. In particular, Eq.~\eqref{outerm} corresponds to an oval between $x_2$ and $x_1$ and an open curve on the left of $x_3$ (this curve is outside the limits of Fig.~\ref{fig:0_002}).

We note that it is not possible for $K$ to take negative values  ($K<0$) because then $\dot{y}^2$ would be negative. In fact, the factor $(3-4\epsilon x)$ does not become negative unless $x>\frac{3}{4\epsilon}$ and this is well beyond the larger root $x_1$ of Eq.~\eqref{outerm}  for $\dot{x}=\dot{y}=0$. $K$ increases as the size of the closed invariant curve decreases and tends to a maximum at the center of the oval. For example, for $\epsilon=0.02$ (Fig.~\ref{fig:0_002}) $K$ increases  from $K=0$ up to approximately $K=300$, and $\epsilon=0.05$. As $\epsilon$ increases the maximum value $K_{max}$ decreases. 

In Fig.~\ref{fig:0_065}c we show $K$ as a function of $x$ for $\epsilon=0.065$. The central point represents a periodic orbit at $x=-0.089$ for $\epsilon=0.05$ and has $K_\text{max}=92.65$. We also have open curves on the left outside $x_3$ (Fig.~\ref{fig:0_065}a). Similar curves exist for $\epsilon$ up to the escape value (see below) $\epsilon=\epsilon_\text{esc}=0.068$.

All the orbits inside the oval are generalized Lissajous curves (Fig.~\ref{fig:0_065}b), i.e.  they fill a region close to a parallelogram, but with curved boundaries, almost parallel to the axes $y=0$ and $x=0$, symmetric with respect to the $x$-axis. Their corners are located on the curve of zero velocity (CZV) on  the $(x-y)$ plane,  which is given by setting $\dot{x}=\dot{y}=0$ in Eq.~\eqref{H3} and has the form
\begin{equation}
    y^2=\frac{2-x^2-4\epsilon x^3}{1+2\epsilon x}.
\end{equation}
The roots of $y^2=0$ are $x_1>0>x_2>x_3$. We have $y^2>0$ when $x_2<x<x_3$ and when $x<x_3$. These roots are the same as those of Eq.~\ref{outerm} for $\dot{x}=\dot{y}=0$.

The limits of the red curves ($K=0$) in Figs.~\ref{fig:0_002} and \ref{fig:0_065}a at $\dot{x}=0$ are the roots of Eq.~\eqref{yd} for $\dot{y}^2=0$. 
E.g. if $\epsilon=0.065$, we have $x_3=-2.980, x_2=-2.097, x_1=1.231$. The roots $x_2$ and $x_3$ join at $\dot{y}=0$ when
\begin{equation}
    \frac{\partial \dot{y}^2}{\partial \dot{x}}=-2\dot{x}=0,\quad \frac{\partial \dot{y}^2}{\partial x}=-2x-12\epsilon x^2=0.\label{part}
\end{equation}
Hence $\dot{x}=0$ and 
\begin{equation}
    x=-\frac{1}{6\epsilon}.
\end{equation}
Then $\dot{y}^2=0$ for $\epsilon=\epsilon_\text{esc}=1/(6\sqrt{6})\simeq 0.06804$. This is the escape perturbation.

For $\epsilon > \epsilon_\text{esc}$, Eq.~\eqref{outerm} for $\dot{x}^2=0$ has only one root $x_1$ (e.g. for $\epsilon=0.07$, $x_1=1.22$) (Fig.~\ref{fig:007}a). Then, there are escapes to the left infinity ($x=-\infty$) of the orbits starting outside the loop formed by the curve crossing itself at the point $(x=-1.376$, $\dot{x}=0)$ (corresponding to $K'=0.406$) as well as those in a thin layer around the loop on the right of the crossing point. On the other hand, all the orbits starting inside the loop formed by this curve are Lissajous figures and do not escape (Fig.~\ref{fig:007}b). 

The values of $K$ in this case ($\epsilon=0.07$) are given as a function of $x$ in Fig.~\ref{fig:007}c. $K$ is positive on the left of $x_1=1.22$ but it takes a minimum and a maximum value. $K_\text{max}$ corresponds to a periodic orbit while $K_\text{min}$ corresponds to the crossing invariant curve of Fig.~\ref{fig:007}a. $K_\text{max}$ of the loop is at $x=-0.096$. The dashed line from $K_{min}$ reaches a point $x=x_{max}=1.18$ on the right, which gives the maximum of the loop for $\dot{x}=0$. If $x$ is between this maximum $x$ and $x_1$ then the Lissajous orbits are thinner than in Fig.~\ref{fig:007}b and escape to the left. This happens for $0<K<K_\text{min}$ (Fig.~\ref{fig:007}c).

\begin{figure}[H]
    \centering
    \includegraphics[width=0.4\linewidth]{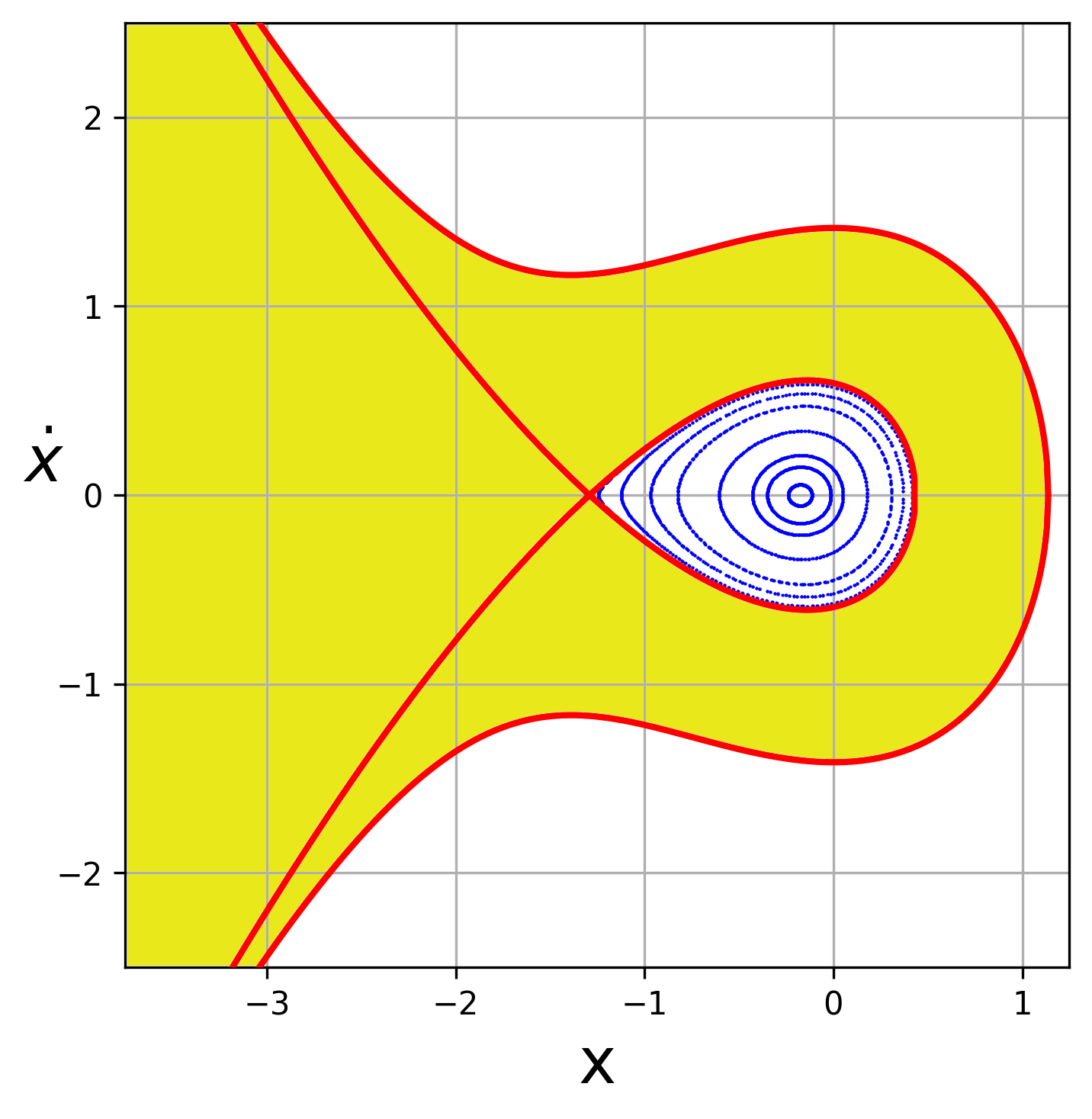}
    \includegraphics[width=0.4\linewidth]{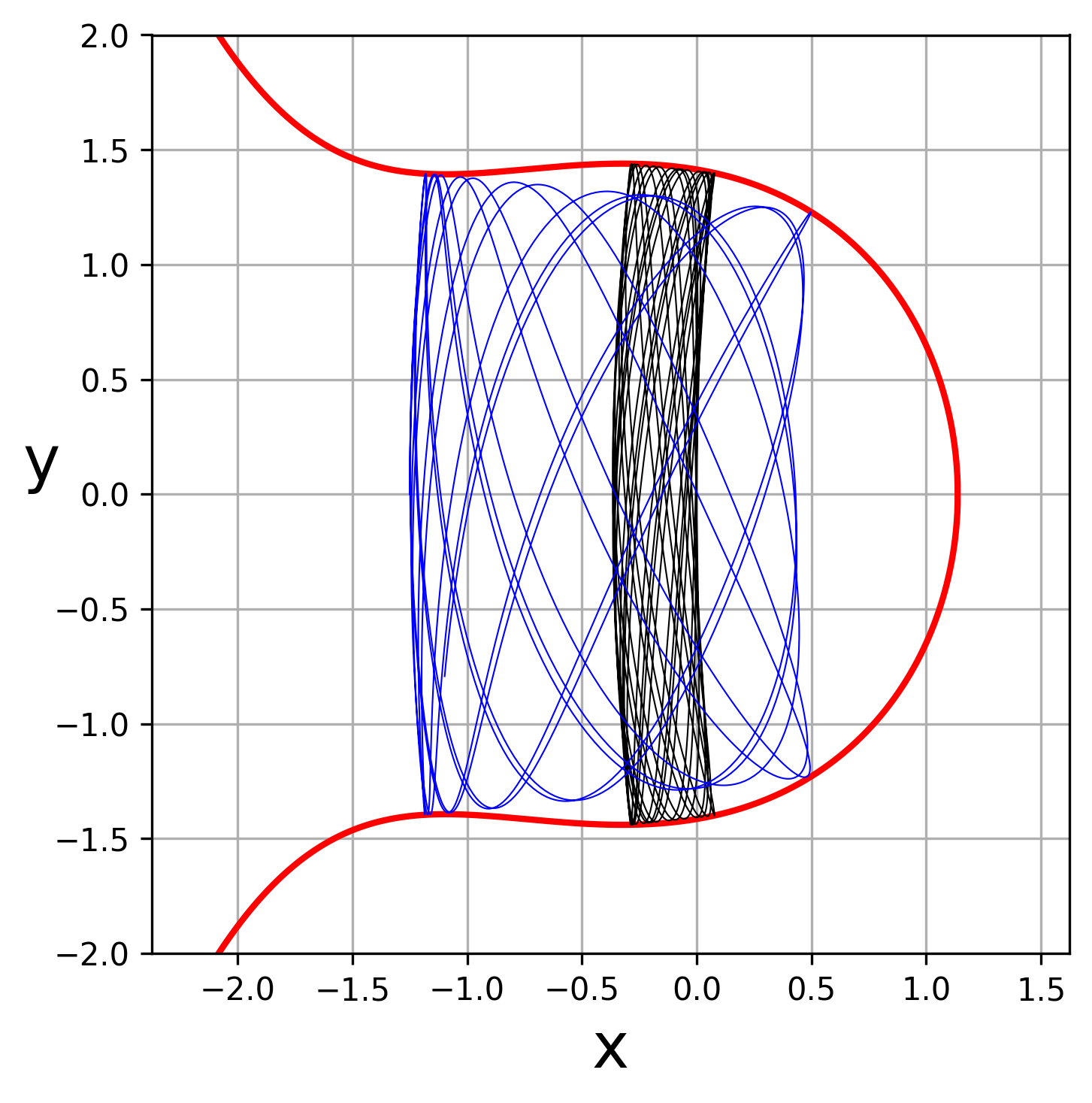}
    \caption{a) Invariant curves for $\epsilon=0.12$. The orbits starting outside the loop escape (yellow). b) Two Lissajous orbits in this case. The larger curve is a limiting case with its upper and lower boundaries tangent to the CZV.}
    \label{fig:a1}
\end{figure}

For larger $\epsilon$, the boundary of the invariant curves ($\dot{y}^2=0$) is wider (Fig.~\ref{fig:a1}a for $\epsilon=0.12$). In this case, the value of $K$ for the invariant curve that crosses itself and forms a loop is $K = 41.21$. The loop on the right reaches $x_\text{esc}=0.43$ for $\dot{x}=0$. The invariant curves inside the loop (Fig.~\ref{fig:a1}a) become smaller in size as $K'=K\epsilon$ increases, and the corresponding Lissajous orbits become thinner in $x$ and larger in $y$. The limit is a periodic orbit at $x_\text{per}=-0.176$ ($\dot{x}=0$). The corners of the Lissajous figures are on the CZV on the $x-y$ plane and the limiting Lissajous curve has its upper and lower boundaries tangent to the CZV (Fig.~\ref{fig:a1}b). If we move outside the closed loop, then the boundaries of the orbits do not cross the CZV on the left anymore, and extend to $-\infty$. Thus, the corresponding orbits escape (yellow area in Fig.~\ref{fig:a1}a).

\begin{figure}[H]
    \centering
    \includegraphics[width=0.4\linewidth]{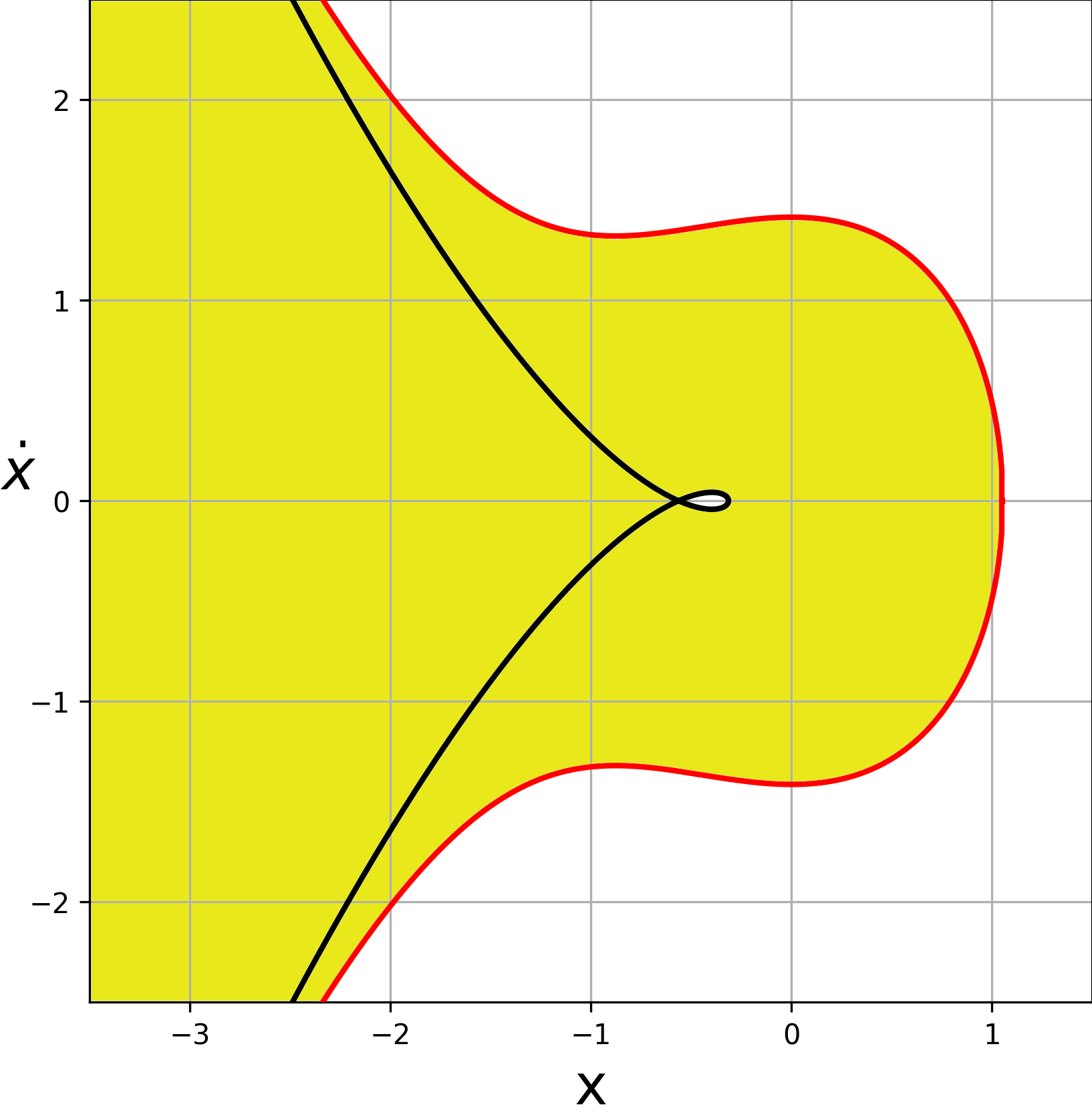}[a]
    \includegraphics[width=0.4\linewidth]{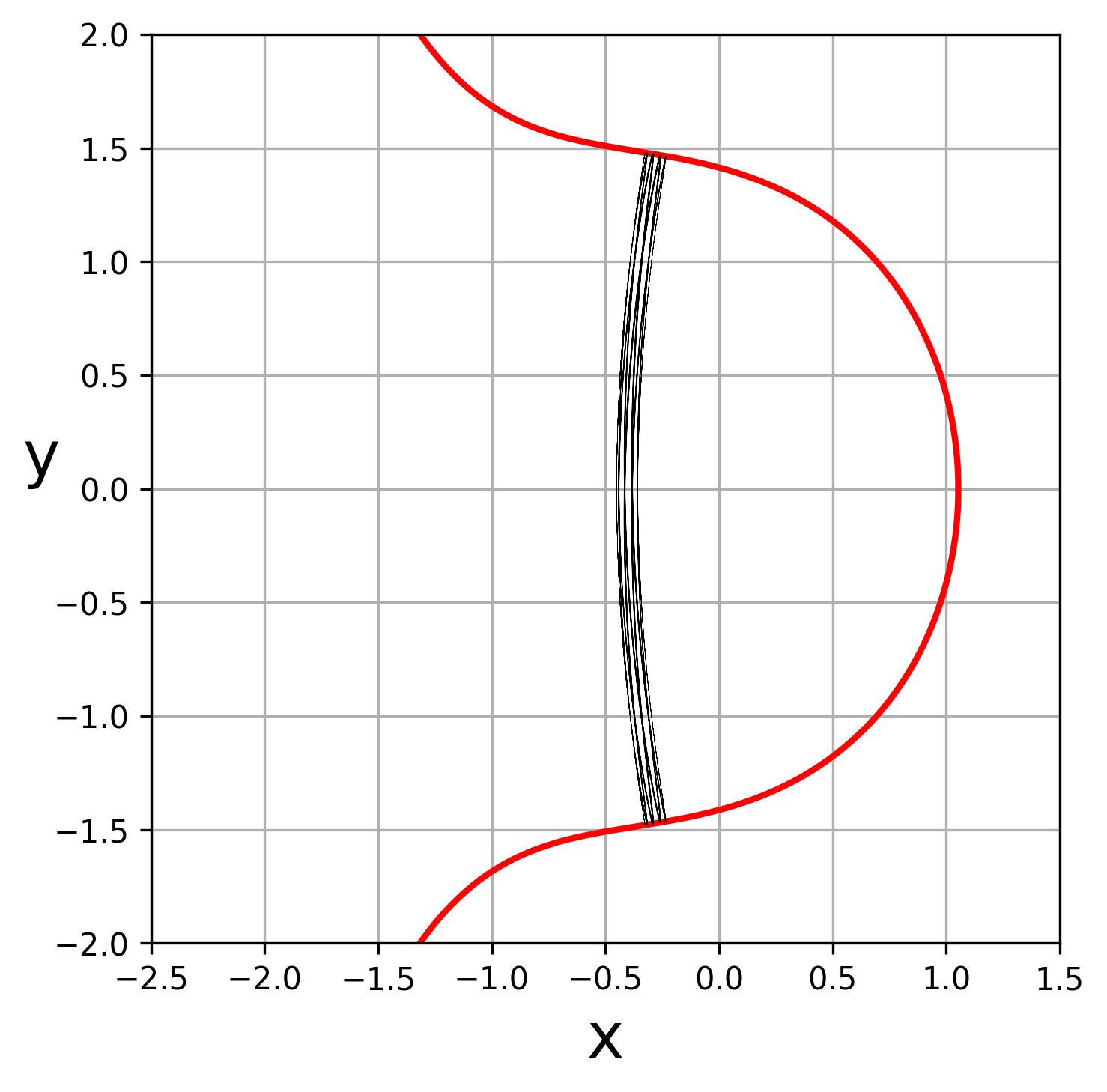}[b]
    \caption{a) Invariant curves in the case $\epsilon=0.19$. Most orbits escape (yellow) except for those in the small loop. b) The corresponding Lissajous curves exist only in a small interval of $x$.}
    \label{fig:a3_eps019}
\end{figure}

As $\epsilon$ increases, the size of the loop decreases and tends to zero. E.g. for $\epsilon=0.19$ we have a very small island (Fig.~\ref{fig:a3_eps019}a) and very thin Lissajous figures (Fig.~\ref{fig:a3_eps019}b). For $\epsilon$ beyond $0.193$ there is no invariant curve crossing itself and no island. All the invariant curves are open on the left and all the orbits are escaping.

\begin{figure}[H]
    \centering
    \includegraphics[width=0.4\linewidth]{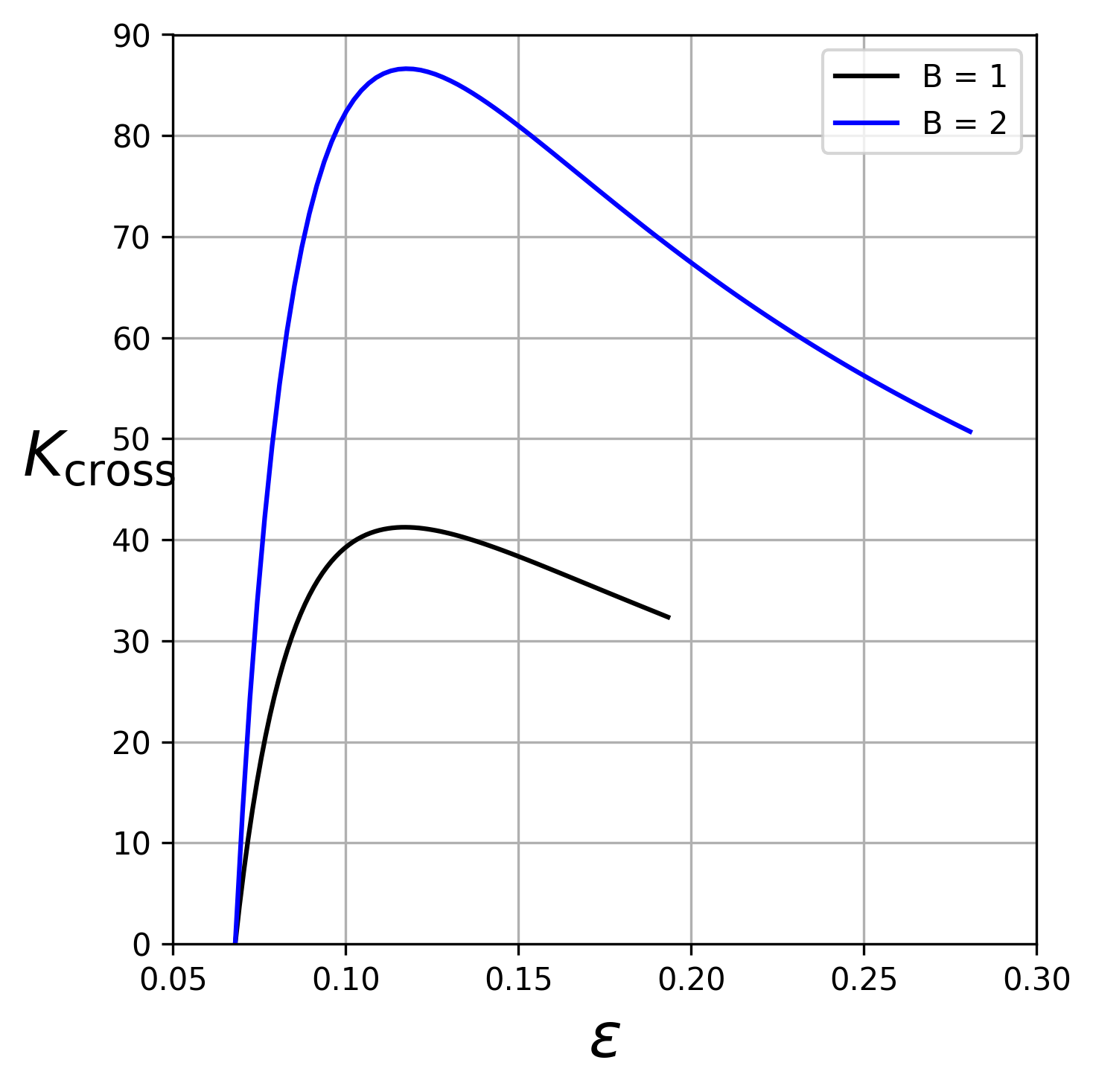}[a]
    \includegraphics[width=0.4\linewidth]{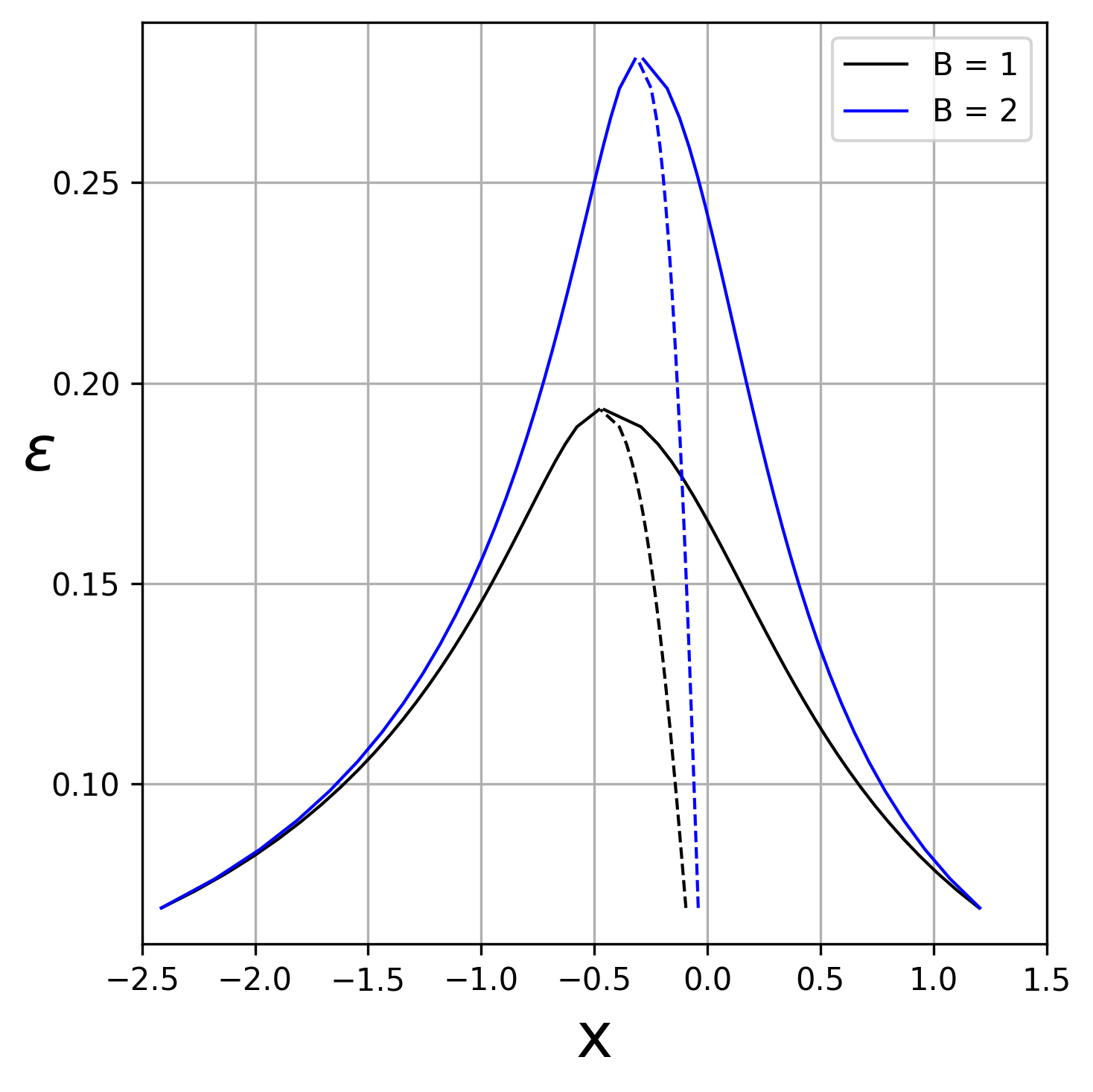}[b]
    \caption{a) The values of $K_{cross}$ of the crossing point as a function of $x$ for $B=1$ and $B=2$. b) The corresponding values of $x=x_\text{cross}$ at the crossing point as a function of $\epsilon$ (left curves). We also give the value of $x_\text{max}$ at the end of the loop (right curve) and $x=x_\text{per}$ at the periodic orbit in the middle (we plot $\epsilon$ as a function of $x$ in order to emphasize the limits of the loop).}
    \label{fig:k_x_epsilon}
\end{figure}

It is of interest to plot $K_\text{esc}$ and $x_\text{esc}$ at the crossing point of the invariant curve that crosses itself at $\dot{x}=0$ as well as the $x_{max}$   of the loop in $\dot{x}=0$ as functions of $\epsilon$. (Figs.~\ref{fig:k_x_epsilon}a,b). There, we see that the curves of $K$ for $B=1$ (Fig.~\ref{fig:k_x_epsilon}a) start at $\epsilon=0.068$ with $K=0$ and terminate at $\epsilon=0.193$ and $K = 32.35$. This curve goes through a maximum $K$. However, if we plot $K'$ as a function of $\epsilon$, this has no intermediate maximum, but $K'$ always increases with increasing $\epsilon$.

\subsection{$E=A=1, B\neq 1$}
In this case  the integral $Q$ for $y=0$ is
\begin{equation}
    (4B-A-4\epsilon x)\dot{y}^2=K'=K\epsilon.
\end{equation}
If we set $B=2$ while keeping $A=1$, we find similar results. In fact, Eq.~\eqref{yd} expressing $\dot{y}^2$ as a function of $x,\epsilon, K'$ is the same. Eq.~\eqref{xdK} giving the CZV gives now $2y^2$ instead of $y^2$, i.e. the CZV is altered only by a scale factor. Therefore, the results are qualitatively quite similar.  E.g. the maximum $\epsilon$ for which the invariant curve crosses itself and the loop on its right shrinks to a point  at $\epsilon_\text{max}=0.281$ instead of $\epsilon_\text{max}=0.193$ of the case $B=1$  (Fig.~\ref{fig:k_x_epsilon}b). The limiting $K$ for $B=2$ is $K=50.7$. Similar results are found for larger values of $B$.

If we change the values of $A,B$ and $E$ and keep $\alpha=2$, the system remains integrable. In fact, if the ratio $A/B$ remains constant then only the scales change. E.g. the escape perturbation is now 
\begin{equation}
    \epsilon_{esc}=\frac{1}{6\sqrt{6}}\left(\frac{A}{E}\right)^{1/2}.
\end{equation}
Therefore $\epsilon_{esc}$ is the same for all values of $B$.
However, when $A/B$ changes, the forms of the figures may change qualitatively. E.g.  if we keep $A=E=1$ and change $B$, the value of $y^2$ is 
\begin{equation}
    y^2=\frac{2-\dot{x}^2-\dot{y}^2-x^2-4\epsilon x^3}{B+2\epsilon x}.\label{eq15}
\end{equation}
Then the roots of the numerator for $\dot{x}=\dot{y}=0$ remain $x_1,x_2,x_3$, the same as above, but the value of $x$ for which the CZV goes to infinity is 
\begin{equation}
   x_{\infty}=-\frac{B}{2\epsilon}.\label{xinf}
\end{equation}
We note that in the previous case ($B=1$) the value of $x_{\infty}$ was to the left of $x_2$ for all the values of $\epsilon$ up to $\epsilon_\text{esc}$ (Fig.~\ref{fig:0_065}). But if $B$ is sufficiently small then $x_{\infty}$ may be on the right of $x_2$, as in the case with $B=0.08$ and $\epsilon=0.03$, where $x_1=1.31$, $x_2=-1.57$, $x_3=-8.08$ and $x_{\infty}=-1.33$. {As regards the invariant curves, they are either closed (on the right) or open (on the left)}.

The transition to the case $x_{\infty}>x_2$ happens when $x_{esc}=-B/2\epsilon$ is a root of the numerator of Eq.~\eqref{eq15} for $\dot{x}=\dot{y}=0$. This happens when 
\begin{equation}
    \epsilon=\epsilon_{esc}'=B\sqrt{\frac{1-2B}{8}}.
\end{equation}
E.g. if $B=0.08$ we find $\epsilon_{esc}'=0.026$. This value is smaller than $\epsilon_{esc}=0.068$. Thus if $\epsilon>\epsilon_{esc}'$, we have escapes to $y^2=\infty$ for $\epsilon<\epsilon_{esc}$.

\begin{figure}[H]
    \centering
    \includegraphics[width=0.4\linewidth]{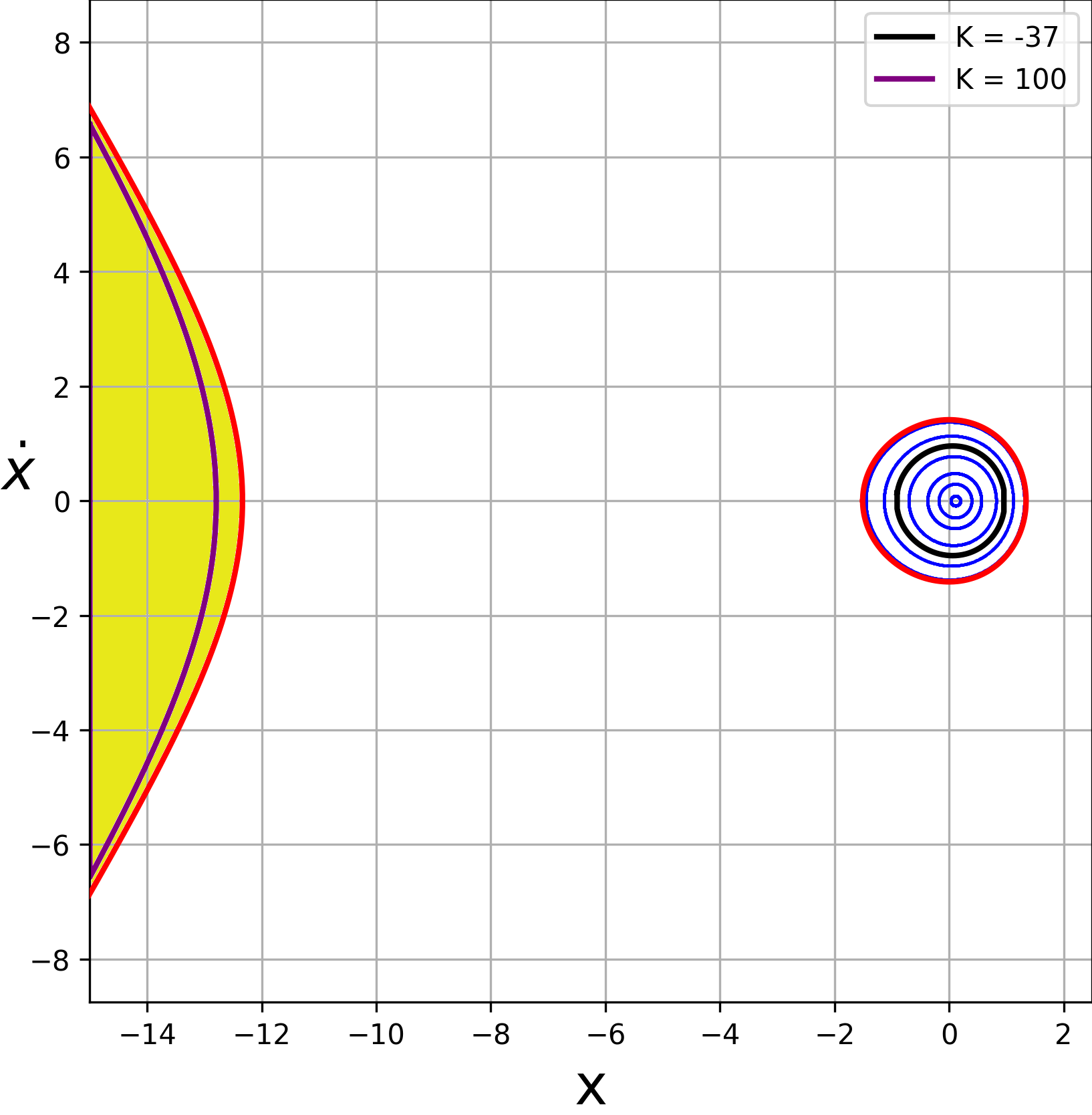}[a]
    \includegraphics[width=0.4\linewidth]{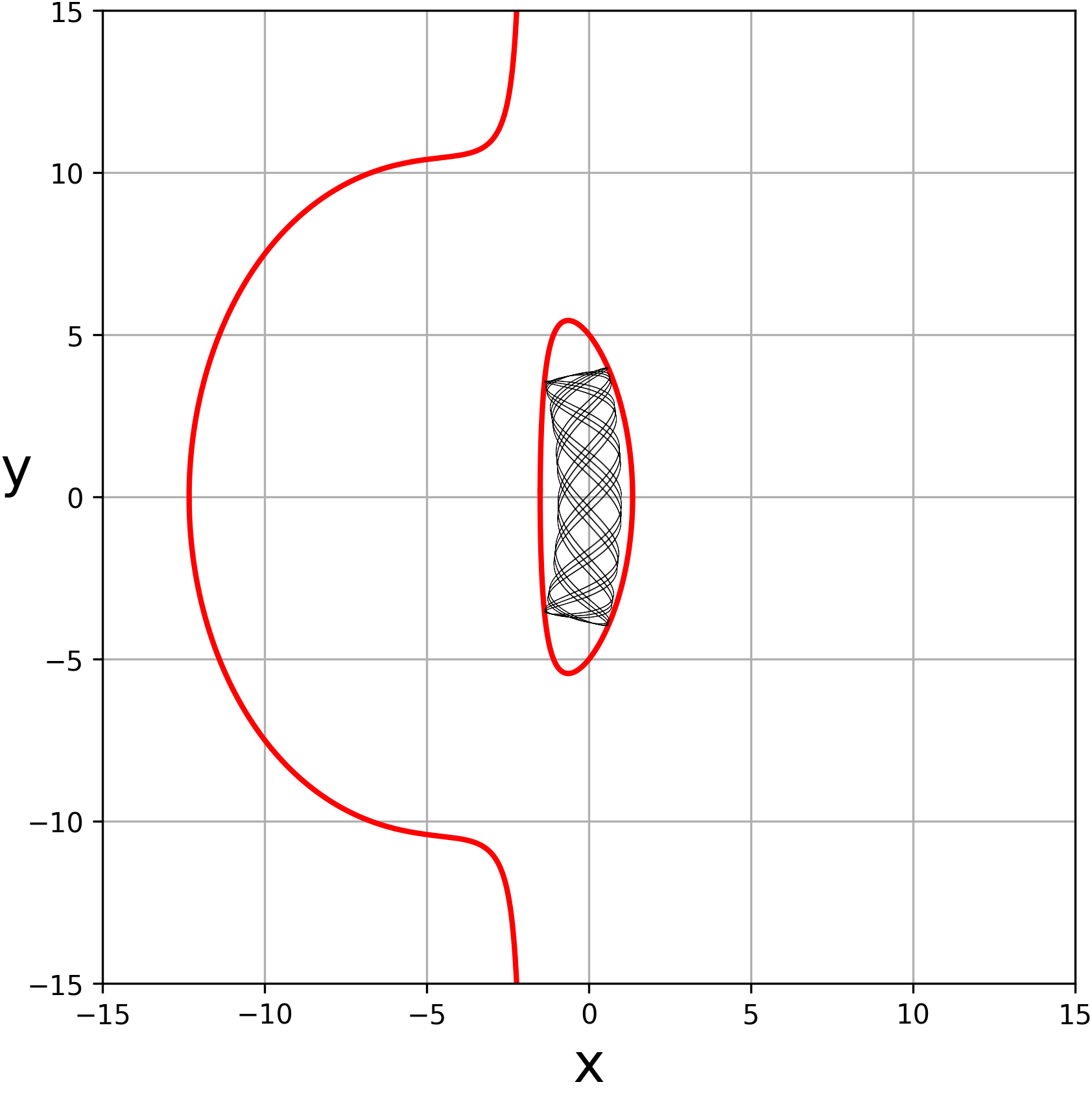}[b]
    \includegraphics[width=0.4\linewidth]{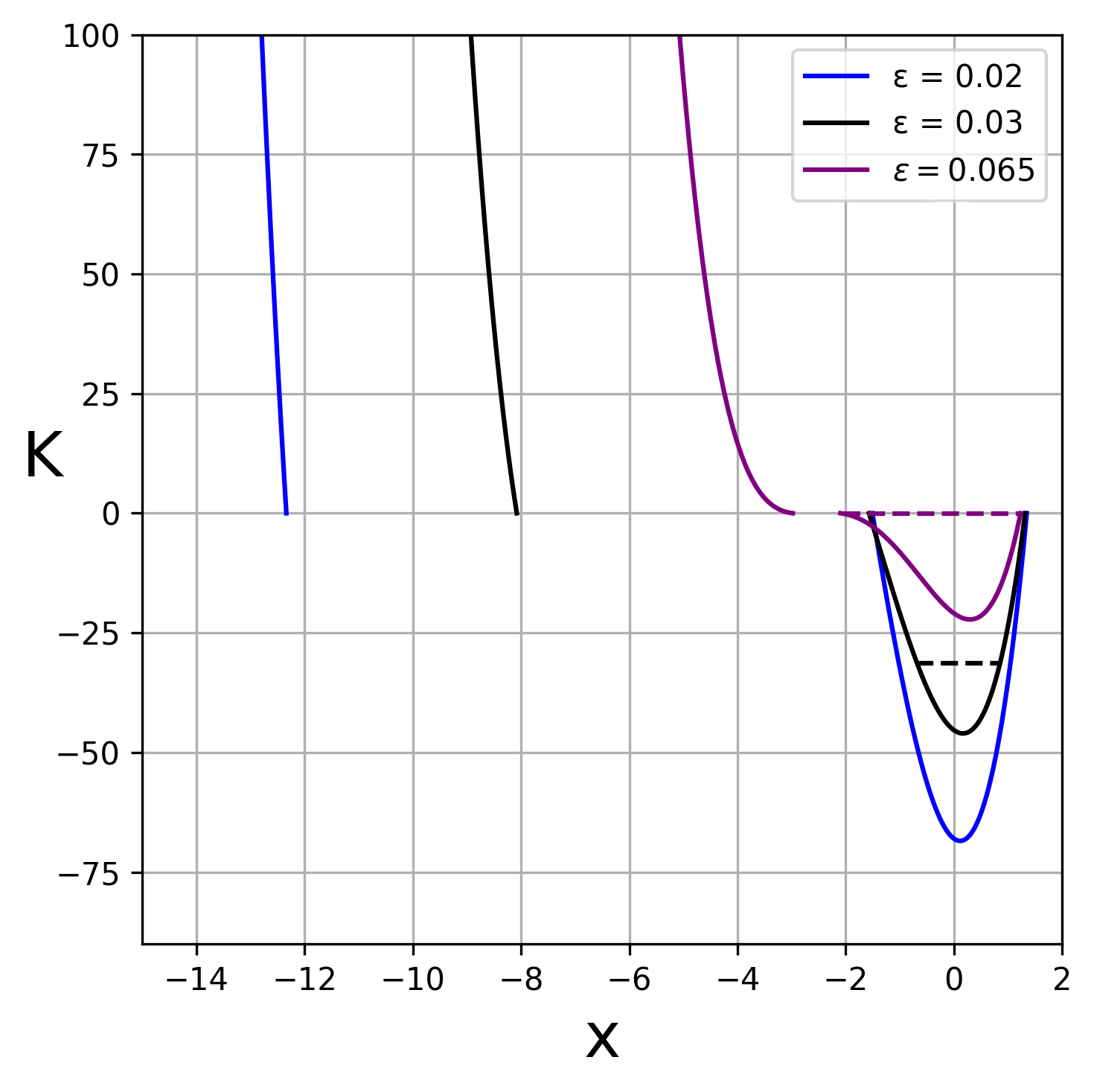}[c]
    \caption{a) Invariant curves on the $x-\dot{x}$ plane in the case with $B=0.08\,(\ll 1)$ and $\epsilon=0.02$. The red colored limiting curve  is an oval on the right and an open curve on the left. Between these red curves there are no orbits.  b) The open and closed part of the CZV (red curves).  All the orbits inside the oval are of Lissajous form, while those on the left of the open curve are escaping. c) The values of $K$ as functions of $x$ for $\epsilon=0.02$, $\epsilon=0.03$ and $\epsilon=0.065$. In the second and the third case the orbits with $K$ below the dashed lines are escaping.}
    \label{fig:002_a2}
\end{figure}

When $B<1$ and $\epsilon=0.02$ the invariant curves on the plane $x-\dot{x}$ are given in (Fig.~\ref{fig:002_a2}a). We find that $x_{\infty}<x_2$ and the corresponding CZV consists of an oval between $x_2$ and $x_1$ and an open curve starting at $x_3$ and reaching $y=\pm \infty$ at $x_{\infty}$ (Fig.~\ref{fig:002_a2}b). In this case all the orbits starting inside the oval are of Lissajous form  and there are no escapes. Of course, the orbits starting on the left of $x_3$ are all escaping. The values of $K$ as functions of $x$ are in Fig.~\ref{fig:002_a2}c.

\begin{figure}[H]
    \centering
    \includegraphics[width=0.4\linewidth]{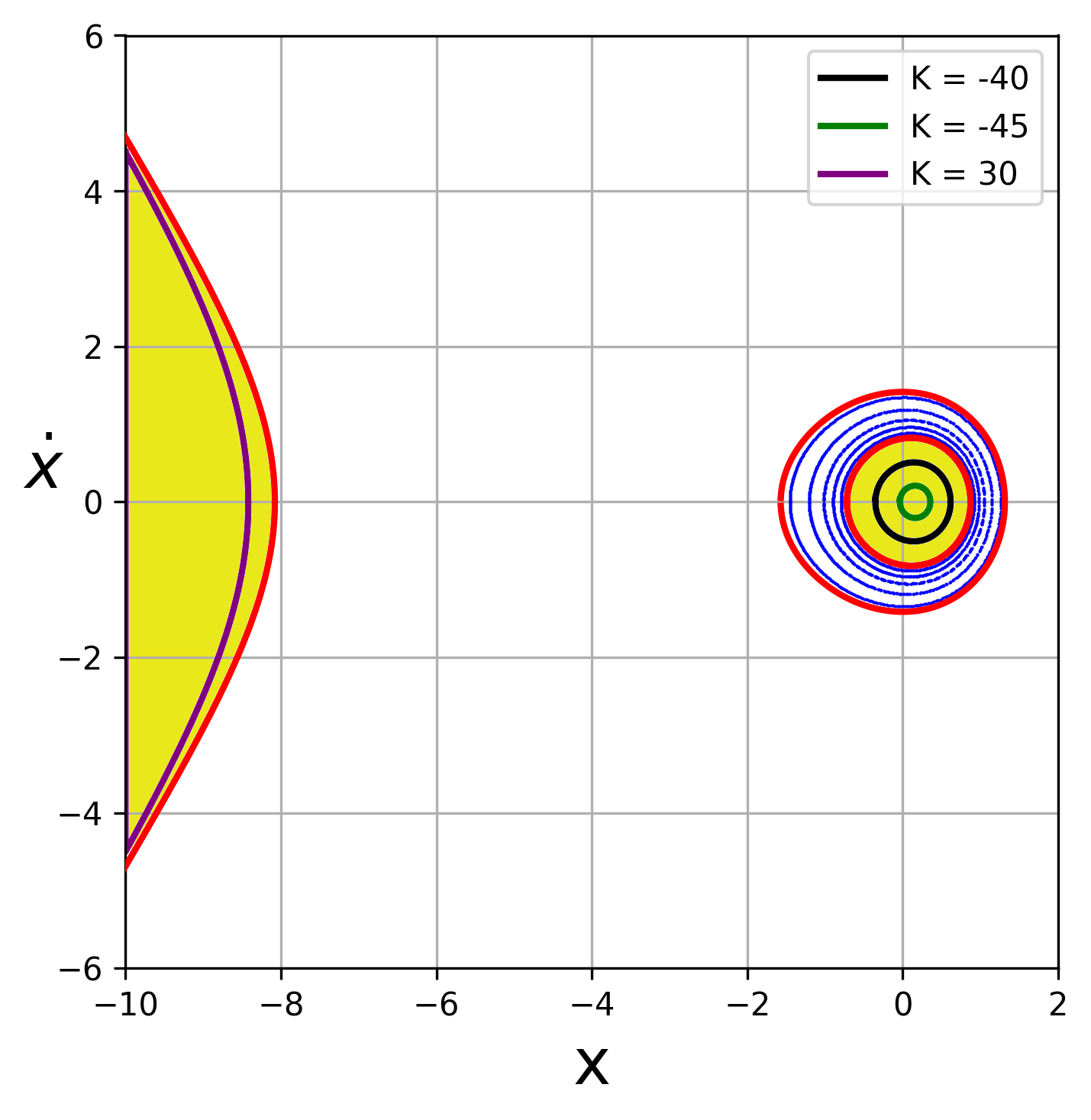}[a]
    \includegraphics[width=0.4\linewidth]{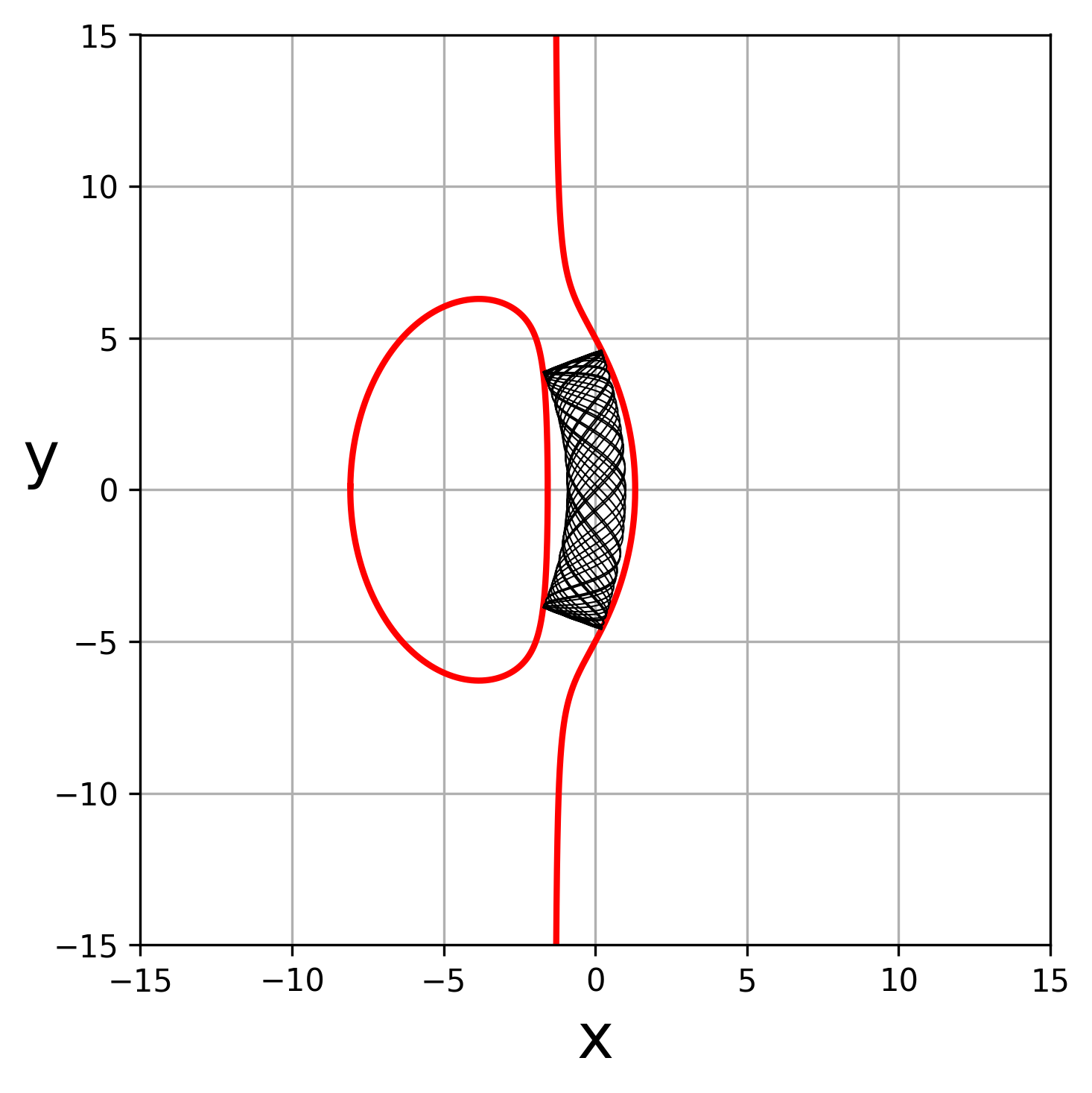}[b]
        \caption{a) Invariant curves on the $x-\dot{x}$ plane in the case with $B=0.08$ and $\epsilon=0.03$. b) The open and closed parts of the CZV (red curves) and a Lissajous orbit. The closed part contains no orbits because there $\dot{x}^2+\dot{y}^2<0$. The middle  red invariant curve in a) corresponds to the transition between Lissajous and escaping orbits. The invariant curves inside this red curve (yellow region)  correspond to escaping orbits while those outside it correspond to Lissajous orbits. }
    \label{fig:003_a2}
\end{figure}

If we increase the perturbation to $\epsilon=0.03$ then we find an island of closed invariant curves between $x_2$ and $x_3$ on the right (with $x=y=0$) or open on the left starting at $x<x_3$ (Fig.~\ref{fig:003_a2}a). The open ones correspond to escaping orbits while the  closed ones correspond to escaping orbits if they are close to the center and Lissajous orbits if they are close to the boundary of the island. The transition between escaping and Lissajous orbits is represented by a red curve. We note that every Lissajous orbit has its intersection with the $y=0$ axis on the same invariant curve, but any invariant curve with the same $K$ inside the red curve represents the initial conditions of different escaping orbits.

Furthermore, the corresponding  CZV (Fig. ~\ref{fig:003_a2}b) consists of an open part starting now at $x=x_1$ and going to $y=\pm \infty$ at $x_{\infty}$, and of an oval between $x_2$ and $x_3$. Inside the oval we have $\dot{x}^2+\dot{y}^2<0$, i.e. it is a forbidden region (Fig.~\ref{fig:003_a2}a,b). Between the two pats of the CZV there are two openings pointing upwards and downwards. Orbits starting at $y=0$ are  generalized Lissajous curves symmetric with respect to the $x$-axis, with their corners on the two parts of the CZV (Fig. ~\ref{fig:003_a2}b). The Lissajous figures become thinner if the initial  $x$ (for $\dot{x}=0$) becomes larger and the orbits starting beyond $x_{\infty}$ escape to infinity through the upper or the lower opening. The transition happens when the left and right boundaries of the orbits become tangent to the CZV. The transition invariant curve (through $\dot{x}=0, x=x_{\infty}$) is marked red in Fig.~\ref{fig:003_a2}a and all the invariant curves inside it correspond to different initial conditions of escaping orbits.

The corresponding values of $K$ as a function of $x$  for $\epsilon=0.03$ are shown in  Fig.~\ref{fig:002_a2}c. In this case the orbits with $K$  below the horizontal dashed lines  are escaping. In the case $\epsilon=0.065$ (larger $\epsilon$) almost all the orbits are escaping and only a few orbits with $K$ slightly negative are of Lissajous form.

We note that the forms of the boundaries of the Lissajous figures  can be found theoretically  by eliminating $\dot{x}, \dot{y}$ from the Hamiltonian (Eq.~\eqref{H2}), the second integral of motion, and the equation
\begin{equation}
    \frac{\partial Q}{\partial \dot{x}}\dot{y}-\frac{\partial Q}{\partial \dot{y}}\dot{x}=0,
\end{equation}
as shown in \cite{Contopoulos1960,ContopMouts1965}. This procedure gives, in general, a complicated relation between $x$ and $y$ that depends on $\epsilon, K'$ and $B$.

If $B\neq 1$, then the invariant curves for $y=0$ are given by the equation
\begin{equation}\label{eqp}
    \left(4B-A-4\epsilon x\right)\dot{y}^2=K',
\end{equation}
which is reduced to Eq.~\eqref{eqKE} if $A=B=1$.

However, if $A=1$ but $B<1$, the coefficient $F\equiv(4B-A-4\epsilon x)$ of $\dot{y}^2$ is positive only if 
\begin{equation}
        x<x_\text{trans}=\frac{(4B-A)}{4\epsilon}
\end{equation}
for any positive $K'=K\epsilon$. If $\epsilon$ is larger than $(4B-A)/4\epsilon$, then this coefficient is negative.

On the other hand, we have real orbits starting at $y=0$ if $x$ is between $x_2$ and $x_3$ (because $\dot{y}^2=2-x^2-4\epsilon x^3$ is positive). Therefore, if for a given $B$ and $\epsilon$, the transition value $x_\text{trans}=(4B-A)/4\epsilon$ is to the left of the corresponding $x_2$, then for all values of interest $x_2<x<x_1$ we have $x>x_\text{trans}$, hence $F$ and $K'$ are negative. Such is the case $B=0.08$ where $x_\text{trans}=-0.17/\epsilon$. This value is $x_\text{trans}=-8.5$ for $\epsilon=0.02$ and $x_\text{trans}=-2.5$ for $\epsilon=0.068$. All values of $x_\text{trans}$ are then to the left of $x_2$.

{On the contrary}, in the case $B=1$ we have seen that $x_\text{trans}$ is larger than $x_1$ for all values of $\epsilon$ up to $\epsilon=0.068$, and therefore both $F$ and $K$ are positive.

\begin{figure}[H]
    \centering
    \includegraphics[width=0.4\linewidth]{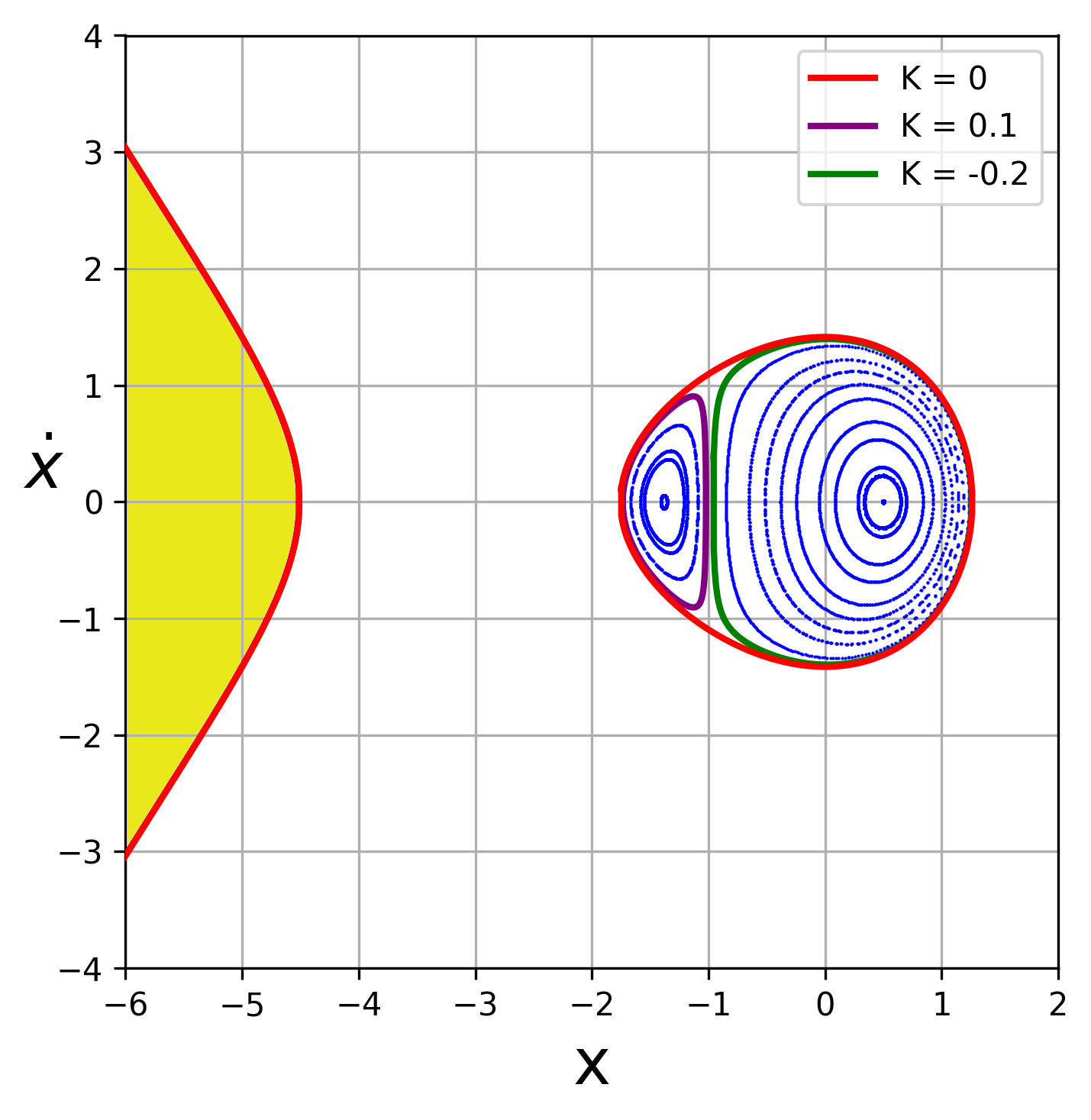}[a]
    \includegraphics[width=0.4\linewidth]{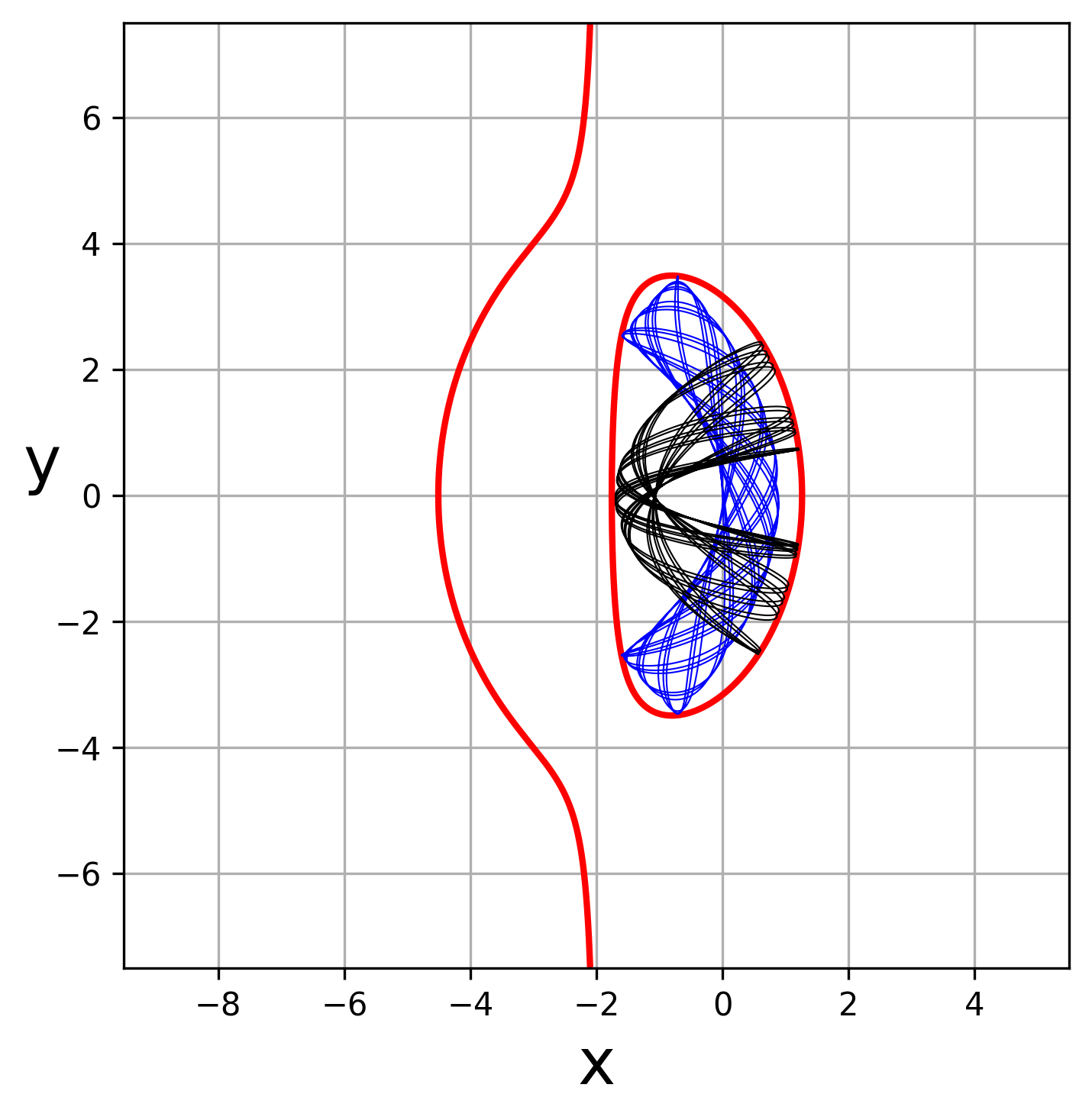}[b]
    \includegraphics[width=0.4\linewidth]{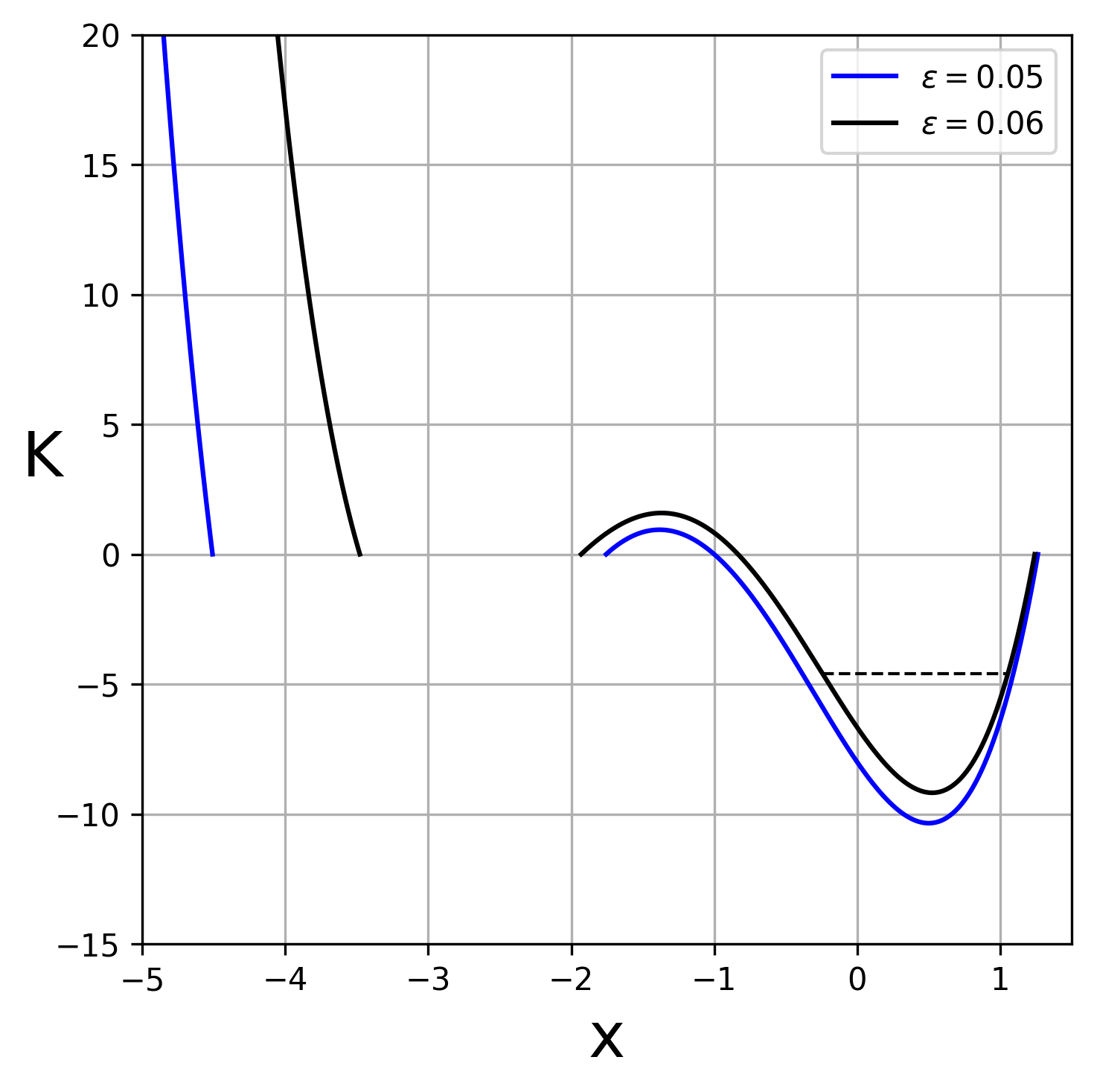}[c]
    \caption{a) Invariant curves inside a limiting curve $\dot{y}^2=2-\dot{x}^2-x^2-4\epsilon x^3$ with $B=0.2$ and $\epsilon=0.05$ around two centers, one on the left ($K>0$) and one on the right ($K<0$), separated by a vertical line. b) Two Lissajous curves, one on the left turning to the right above and below the $x$-axis, and one on the right turning to the left above and below the $x$-axis. c) The values of $K$ as functions of $x$ for $\epsilon=0.05$ and $\epsilon=0.06$. The values below the dashed line of the case $\epsilon=0.06$ correspond to escaping orbits. The orbits on the left of the open red curve in a) are escaping (yellow).}
    \label{fig:11}
\end{figure}

But for some intermediate values of $B$ (between $B=0.08$ and $B=1$) we have a transition value of the coefficient of $\dot{y}^2$ between $x_2$ and $x_3$. Such is the case $B=0.2$, where the transition value is $x_\text{trans}=-0.05/\epsilon$ and for $\epsilon=0.05$ we have $x_\text{trans}=-1$. Then we have two types of invariant curves (Fig.~\ref{fig:11}a), those on the left of $x=x_\text{trans}=-1$ and those on the right of $x_\text{trans}$. $F$ is positive when $x<x_\text{trans}$ $F$ and negative for $x>x_\text{trans}$ (Fig.~\ref{fig:11}b). The curve which gives $K$ as a function of $x$ for $\dot{x}=0$ has a positive part for $x$ between $x_2$ and $x_\text{trans}$ and then becomes zero for $x=x_\text{trans}$. Between $x_\text{trans}$ and $x_1$, $K$ is negative. The corresponding Lissajous figures are given in Fig.~\ref{fig:11}b. On the left we have Lissajous orbits turning to the right upwards and downwards and on the right we have Lissajous orbits turning to the left upwards and downwards. The corresponding CZV forms a closed oval and all the orbits inside it are of Lissajous form (Fig.~\ref{fig:11}b).

\begin{figure}[H]
    \centering
    \includegraphics[width=0.4\linewidth]{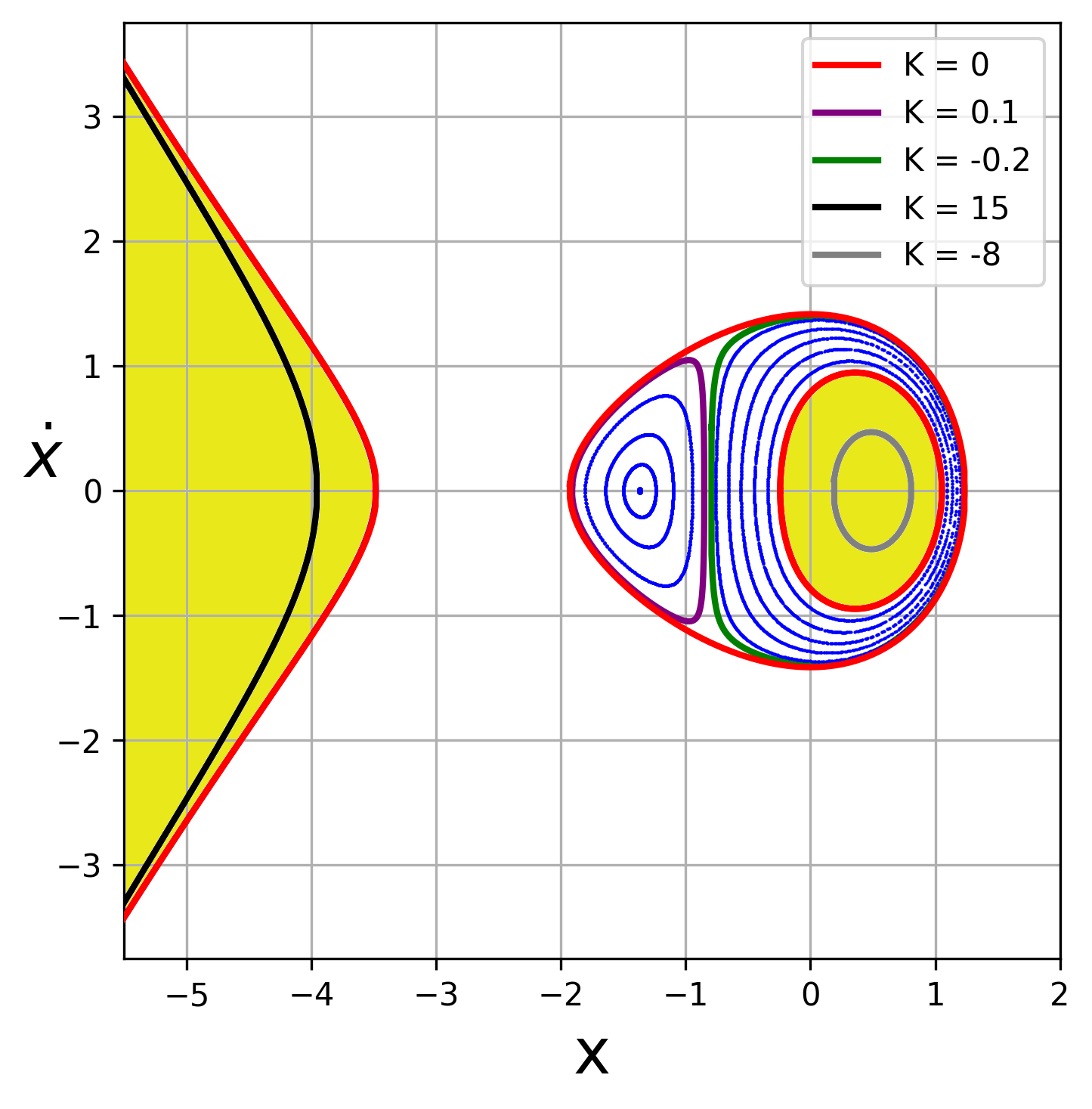}[a]
    \includegraphics[width=0.4\linewidth]{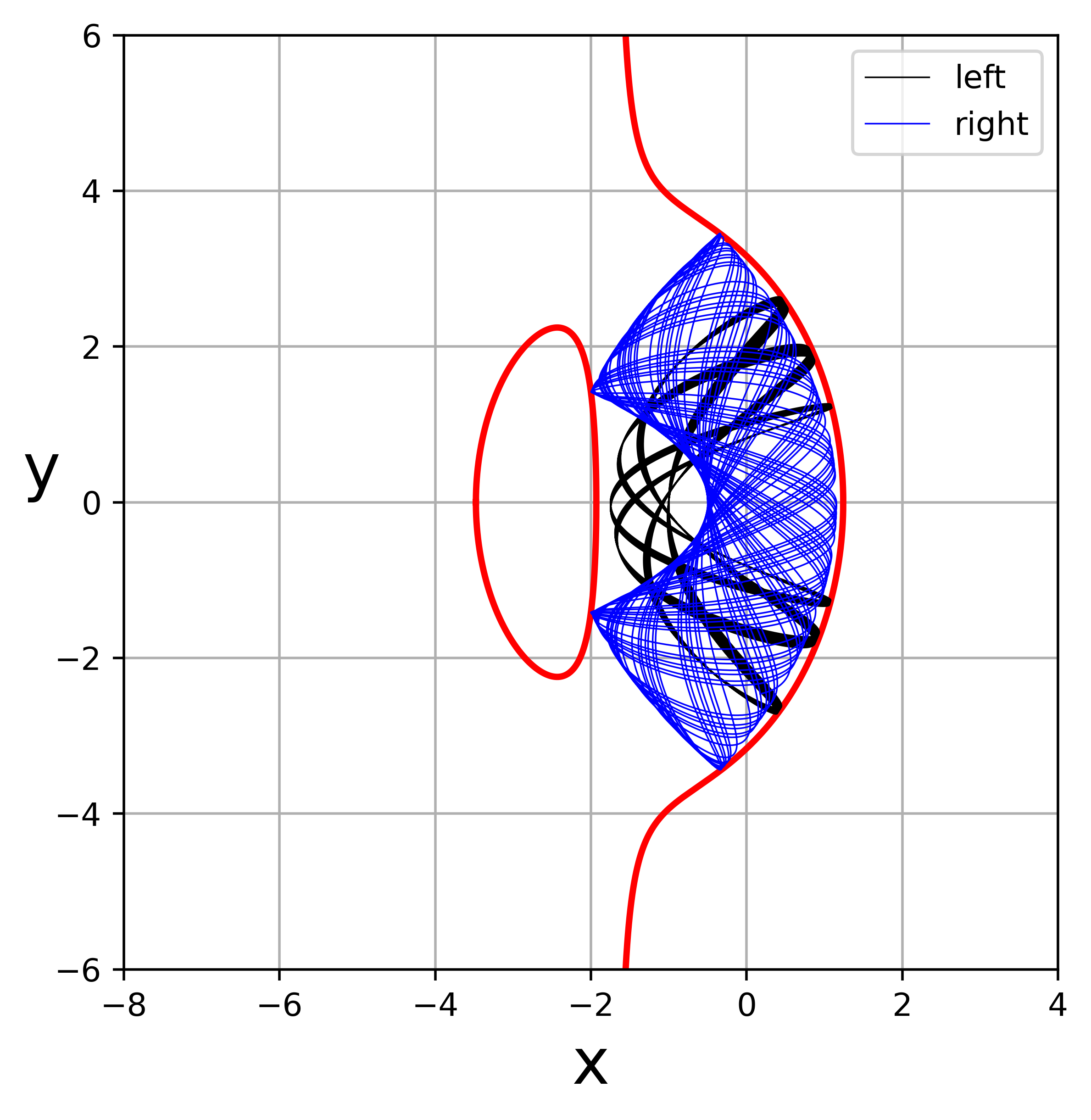}[b]
    \caption{a) Invariant curves as in Fig.~\ref{fig:11}a for $B=0.2$ and $\epsilon=0.06$. b) Lissajous figures on the left and on the right. The invariant curves inside the red curve of a) correspond to escaping orbits. The orbits on the left of the open red curve in a) are escaping (yellow).}
    \label{fig:006}
\end{figure}

However, in the case $B=0.2$ $\epsilon=0.06$ the CZV goes to infinity between $x_2$ and $x_1$ and there are both Lissajous orbits and orbits escaping. The separation is a red curve in Fig.~\ref{fig:006}a.

The orbits starting inside the inner red curve of Fig.~\ref{fig:006}a are escaping (yellow), while the orbits starting outside it, but inside the oval surrounded by the red curve, are of Lissajous form, as the one starting on the right of $x_\text{trans}$  which is given in Fig.~\ref{fig:006}b (blue). The  limiting invariant curve (red) of Fig.~\ref{fig:006}a corresponds to a limiting Lissajous orbit on the $x-y$ plane similar to the blue orbit of b) that has its left and right boundaries tangent to the CZV. All the black orbits of Fig.~\ref{fig:006}b are bounded.

\begin{figure}[H]
    \centering    \includegraphics[width=0.4\linewidth]{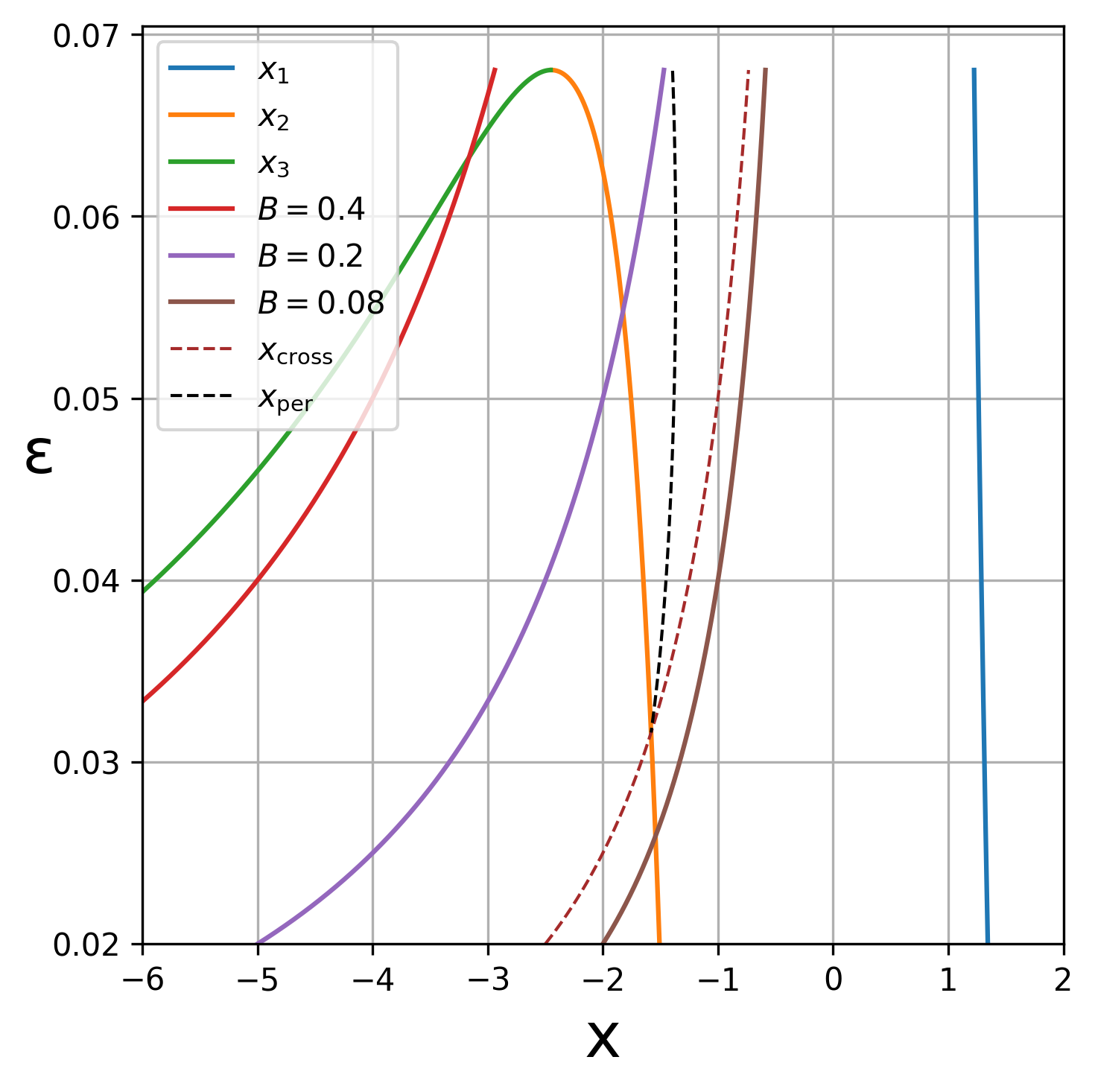}
    \caption{The roots $x_1$ (blue), $x_2$ (orange), $x_3$ (green) of the equation $2-x^2 - 4\epsilon x^3=0$ as  functions of $\epsilon$ (for any value of $B$), and the values of $x_\infty$ (where $y^2 \rightarrow \infty$) for $B=0.08$ (brown), $B=0.2$ (purple) and $B=0.4$ (red). We also provide for $B=0.2$ the values of $x_\text{cross}$ (dashed red) where the coefficient of $\dot{y}^2$ in eq.~\eqref{eqp} is zero, as well as the values of $x_\text{per}$ (dashed black) of the corresponding periodic orbits.}\label{fig:9}
\end{figure}

In Fig.~\ref{fig:9} we give the transition value of $x=x_\text{trans}$ as a function of $\epsilon$. For $B=0.2$ this curve  intersects the curve of  $x_2$ at $\epsilon\simeq 0.33.$ For smaller values of $\epsilon$ the transition value $x_\text{trans}$ is on the left of $x_2$. Therefore for $\epsilon<0.33$ there is no separation of the invariant curves on the surface of section (as in Figs.~\ref{fig:11} and \ref{fig:006}) and all the Lissajous orbits are turning to the left above and below the $x$-axis as in Fig.~\ref{fig:002_a2}b.

In Figs.~\ref{fig:11}b and \ref{fig:006}b there is a periodic orbit in the center of the left Lissajous orbits corresponding to the maximum $K$ of Fig.~\ref{fig:11}c.
Its position $x_\text{per}$ is given as a function of $\epsilon$ in Fig.~\ref{fig:9}. The curve $x_\text{per}$ terminates when the curve with $x=(4B-A)/4\epsilon$ for $B=0.2$ reaches the curve $x_2$ (for $\epsilon=0.33$). This is reasonable because for smaller $\epsilon$ there are no Lissajous orbits of the left type of Figs.~\ref{fig:11}b and \ref{fig:006}b.

\begin{figure}[H]
    \centering
    \includegraphics[width=0.32\linewidth]{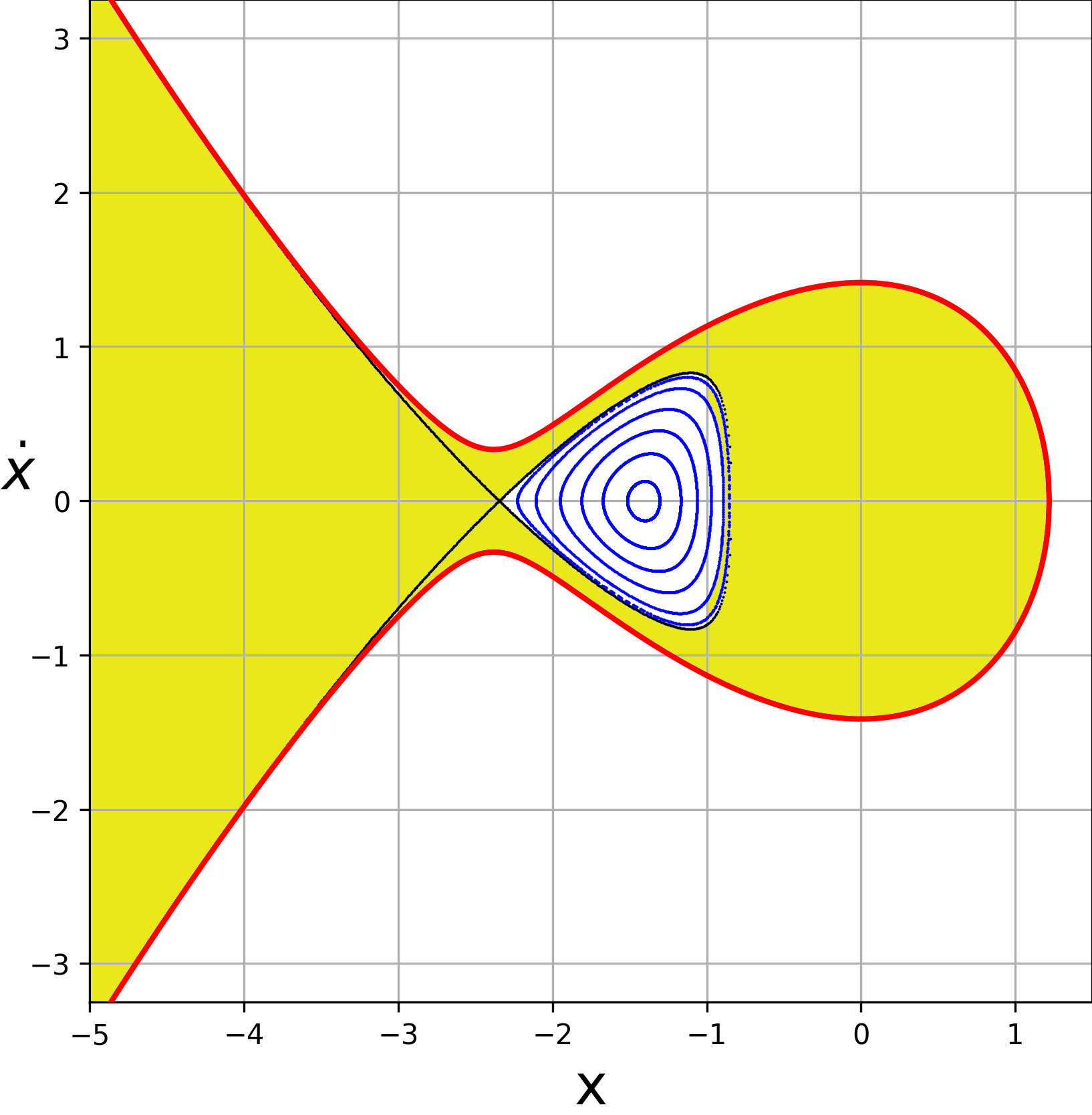}[a]
    \includegraphics[width=0.32\linewidth]{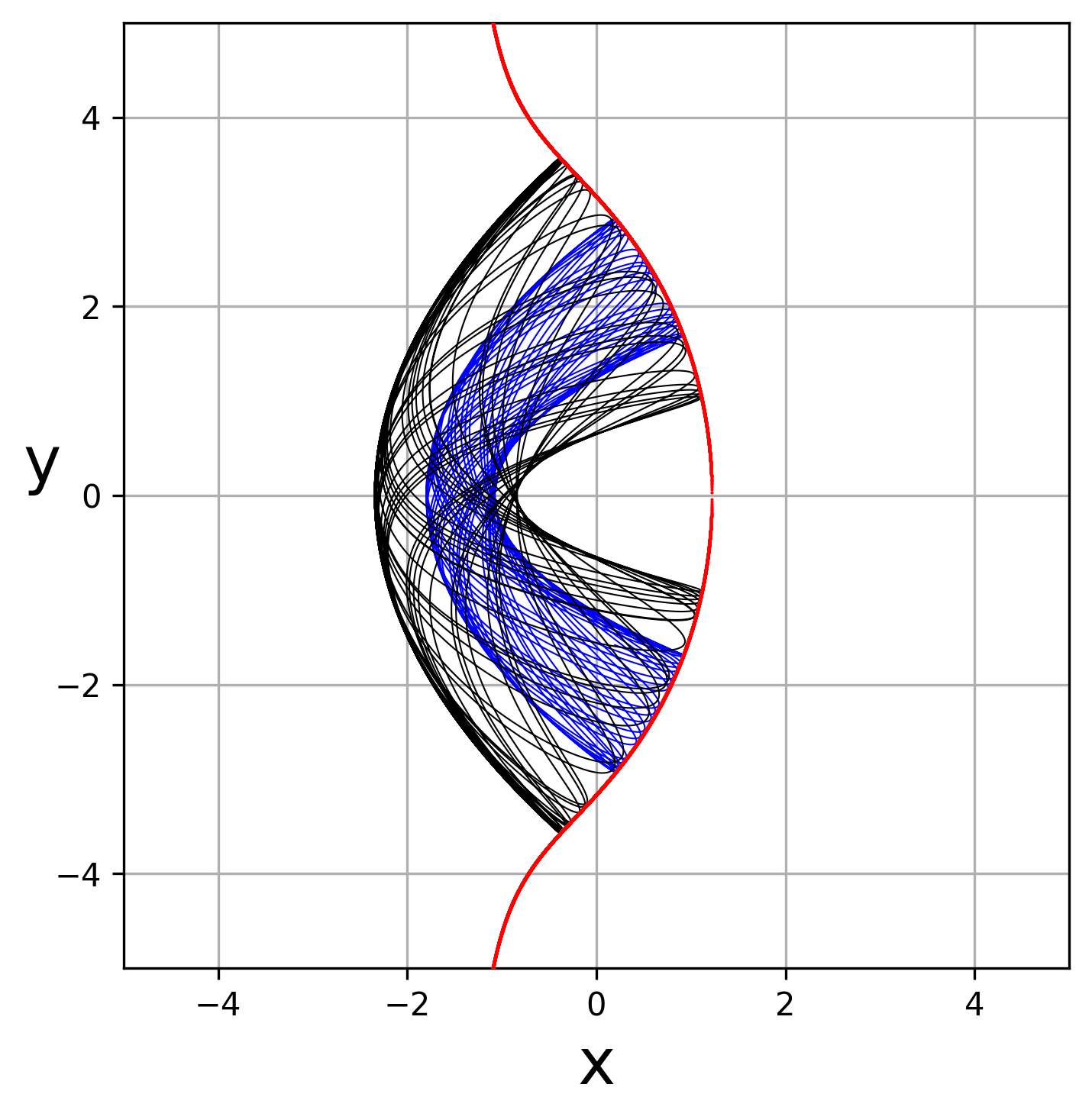}[b]
    \includegraphics[width=0.32\linewidth]{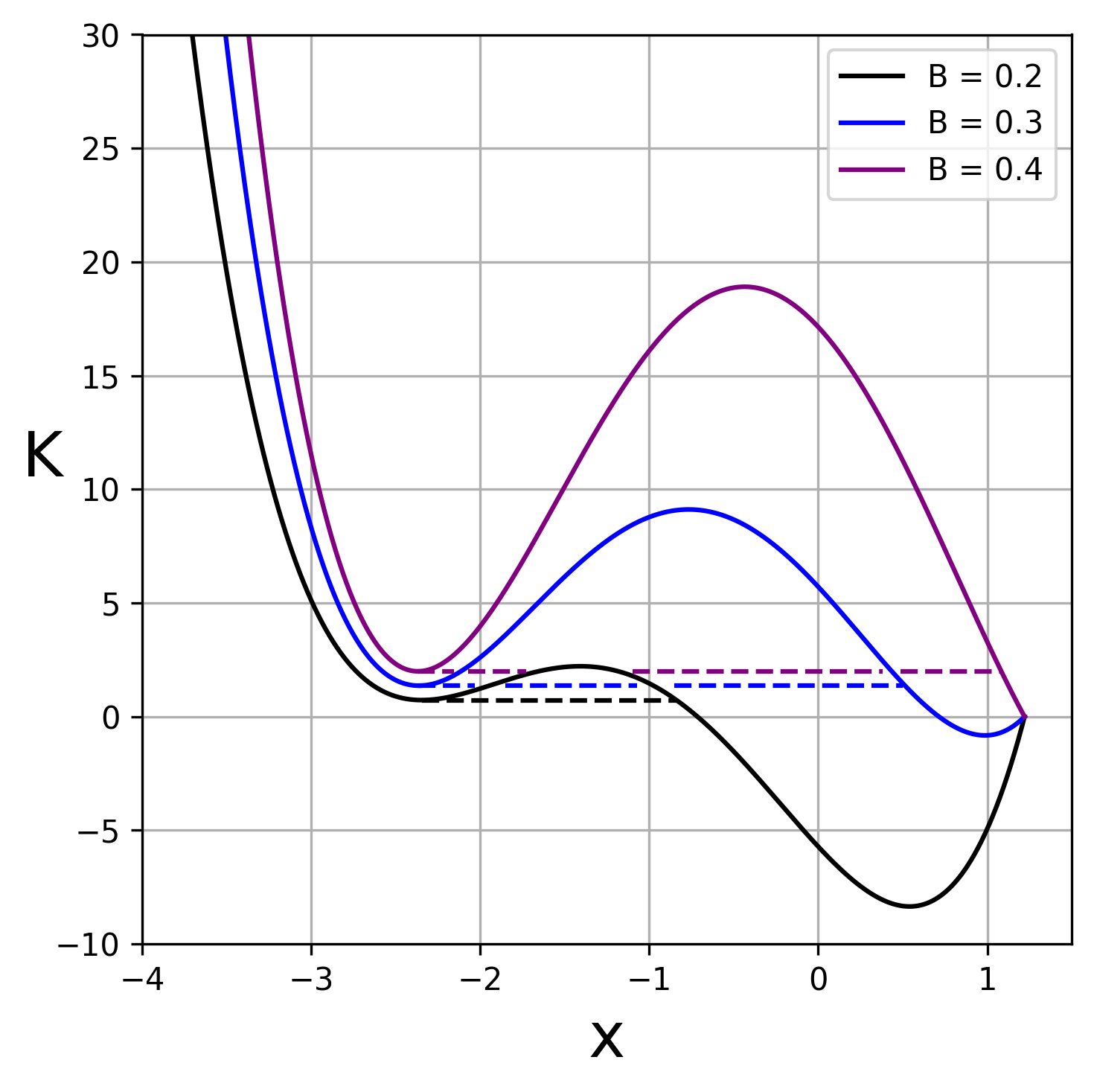}[c]
    \caption{a) Invariant curves in the case $B=0.2, \epsilon=0.07>\epsilon_\text{esc}$. The limiting curve $\dot{y}^2=2-\dot{x}^2-x^2-4\epsilon x^3=0$ is now open. The invariant curve that crosses itself forms a loop on the right that contains closed invariant curves with $K>0$. b) Lissajous orbits corresponding to invariant curves inside the loop. c) The values of $K$ as functions of $x$, for $\epsilon=0.07$ and $B=0.2$ (black), $B=0.3$ (blue) and $B=0.4$ (purple). Every curve has a  $K_\text{min}$ and on its right  a $K_\text{max}$. Beyond the value $K=K_\text{min}$ on the right the corresponding orbits are escaping for both small positive $K$ and for negative $K$ (yellow region in a)). The negative values of $K$ in the case $B=0.3$ are smaller than in the case $B=0.2$, while in the case $B=0.4$ there are no negative values.}
    \label{fig:triplet}
\end{figure}

\begin{figure}[H]
    \centering
    \includegraphics[width=0.38\linewidth]{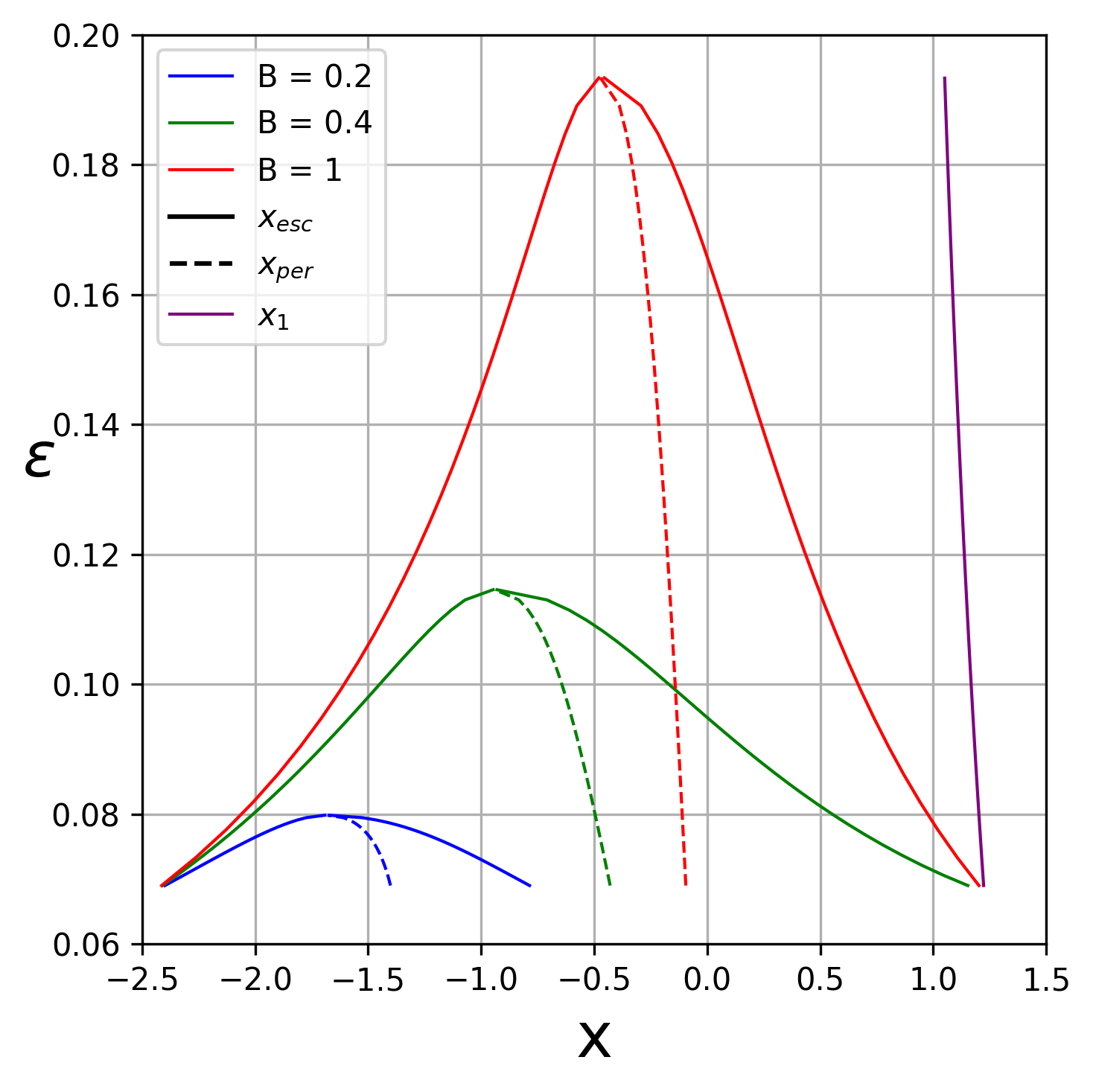}
    \caption{The values of $x=x_\text{cross}$ at the crossing point (left curves) for $B=0.2$, $B=0.4$ and $B=1$ as functions of $\epsilon$ and $\epsilon_\text{esc}$ up to a maximum $\epsilon$ beyond which there is no crossing point and no loop. On the right we give $x_\text{max}$ representing the end of the loop. The size of the loop $(x_\text{max}-x_\text{cross})$ decreases as $\epsilon$ increases and becomes zero at the top of each curve (for the corresponding $B$). The middle curves (dashed) at $x_\text{per}$ correspond to periodic orbits at the center of the loops. All these curves are on the left of $x_1$.}
    \label{fig:13}
\end{figure}

The curve $x_\text{per}$ exists not only for $\epsilon<\epsilon_\text{esc}=0.068$ but also for an interval of $\epsilon$ larger than $\epsilon_\text{esc}$ as we will see below (Fig.~\ref{fig:13}).  Moreover, all orbits starting on the left of the vertical line $K=0$ of  ($x=x_\text{trans}$) in Fig.~\ref{fig:006}a are of Lissajous  form (black in Fig.~\ref{fig:006}). The black Lissajous curves turn to the right above and below the $x$-axis and do not contain any escaping orbits, while the blue Lissajous curves turn to the left (Fig.~\ref{fig:006}b) and contain escaping orbits near their center. If $\epsilon$ is above the escape value $\epsilon_\text{esc} = 0.068$ the limiting invariant curve $\dot{y}^2 = 2-\dot{x}^2 - x^2 - 4\epsilon x^3 = 0$ and the CZV are open to the left, but we still find some Lissajous orbits. An example is the case $B=0.2$ which is shown in Figs.~\ref{fig:triplet}a,b. It is remarkable that there are Lissajous figures in this case where the CZV is very open. Furthermore,  the value of $x_{\infty}$ where the CZV goes to infinity is at $x_{\infty}=-1.43$, i.e. it crosses the Lissajous curves. However their boundaries turn to the right and reach the CZV. The largest Lissajous orbit has its upper and lower boundaries tangent to the CZV. Orbits starting at $y=0$ on the left of this orbit escape to infinity.

A remarkable connection of Figs.~\ref{fig:triplet}a,b (where $\epsilon>\epsilon_\text{esc}$) with Figs.~\ref{fig:006}a,b  (where $\epsilon<\epsilon_\text{esc}$), is in the forms of their invariant curves and some of their orbits. Namely, the invariant curves on the left of the transient value of $x=x_\text{trans}$ in Fig. \ref{fig:006}a are very similar with those of the loop formed by the self-crossing invariant curve of Fig.~\ref{fig:triplet}a. The corresponding orbits are also similar and are generalized Lissajous curves with their corners on the open CZV starting from $x_1$.

Another connection between the orbits of the case $B=0.2$ and $\epsilon=0.07$ above the escape perturbation, and the orbits of the case $B=0.2$ and $\epsilon=0.06$ below the escape perturbation, are the periodic orbits at the centers of the corresponding loops. Such orbits appear for various values of $\epsilon$ above $\epsilon_\text{esc}$ in Fig.~\ref{fig:13} and below $\epsilon_\text{esc}$ in Fig.~\ref{fig:9}. This similarity is important because these orbits are surrounded by closed invariant curves, which is in sharp contrast with the case $\alpha=16/3$ (see below) where the loops formed above the escape perturbation contain only escaping orbits.

In Fig.~\ref{fig:triplet}c we show the values of $K$ as functions of $x$ for $\epsilon=0.07$ and 3 values of $B$ ($0.2$, $0.3$, and $0.4$). The 3 curves have a $K_\text{min}$ for about the same values of $x=-2.3$, the point where the roots $x_2$ and $x_3$ join in Fig.~\ref{fig:9}. On the right of $K_\text{min}$ there is an increase of $K$ corresponding to the loop of Fig.~\ref{fig:triplet}a and larger loops for $B=0.3$ and $B=0.4$ which terminate on the right at the same $K_\text{min}$ of the corresponding $B$. Beyond that point, the values of $  K$ decrease passing through a negative $K_\text{min}$ in the cases $B=0.2$ and $B=0.3$, finally reaching the value $K=0$ at $x=x_1=1.2$. All of these parts represent escaping orbits beyond the end of the loop up to the positive root $x=x_1$. For larger values of $B$ the negative values of $K$  disappear (as in the case $B=0.4$).

In Fig.~\ref{fig:13} we plot the values of $x=x_\text{cross}$ at the crossing point (left) and $x_\text{max}$ at the end of the loop (right) and $x_\text{per}$ at a periodic orbit in the center of the loop for $B=0.2, B=0.4$ and $B=1$. For some $B<1$ and $\epsilon>\epsilon_\text{esc}$ the limiting invariant curve $\dot{y}^2=2-\dot{x}^2-x^2-4\epsilon x^3=0$ is open. But there is an invariant curve crossing itself on the $\dot{x}=0$ axis. Inside the loop we find generalized Lissajous orbits while outside of it the orbits are escaping.

Furthermore, we see that the maximum values of $\epsilon$,  for a given $B$, and the corresponding $x_\text{max}$ and $x_\text{per}$ decrease as $B$ decreases (Fig.~\ref{fig:triplet}c). All the curves of $x_\text{esc}$ start at $\epsilon = \epsilon_\text{esc} = 0.068$. But the maximum size of the loop ($x_\text{max} - x_\text{esc}$) decreases as $B$ decreases.

As $B$ decreases further the size of the loop goes to zero at about $B=0.08$. For smaller $B$ there are no loops and all the orbits for $\epsilon>\epsilon_\text{esc}$ escape.

\section{Case (iii) ($\alpha=16/3$) }
In this case the second integral of motion has the form
\begin{equation}
    Q={\dot{y}}^{4}+2\, \left( 2\,\epsilon\,x+B \right) {y}^{2}{\dot{y}}^{2}-\frac{4}{3}
\,\epsilon\,{y}^{3}\dot{x}\,\dot{y}+{B}^{2}{y}^{4}-\frac{4}{3}\,\epsilon\, \left( 
\epsilon\,x+B \right) {y}^{4}x-\frac{2}{9}\,{\epsilon}^{2}{y}^{6}.
\end{equation}
For $B=1$ the Hamiltonian  reads
\begin{equation}
    H=\frac{1}{2}\left(\dot{x}^2+16x^2+\dot{y}^2+y^2\right)+\epsilon\left(xy^2+\frac{16}{3}x^3\right)=1.\label{Hr}
\end{equation}
The invariant curves on the surface of section $y=0$ are given by the equation
\begin{equation}
    Q=\dot{y}^4=K,
\end{equation}
where
\begin{equation}
    \dot{y}^2=\left |2-\dot{x}^2-16x^2-\frac{32}{3}\epsilon x^3\right |=\sqrt{K}\geq 0.
\end{equation}

\begin{figure}[H]
    \centering
    \includegraphics[width=0.4\linewidth]{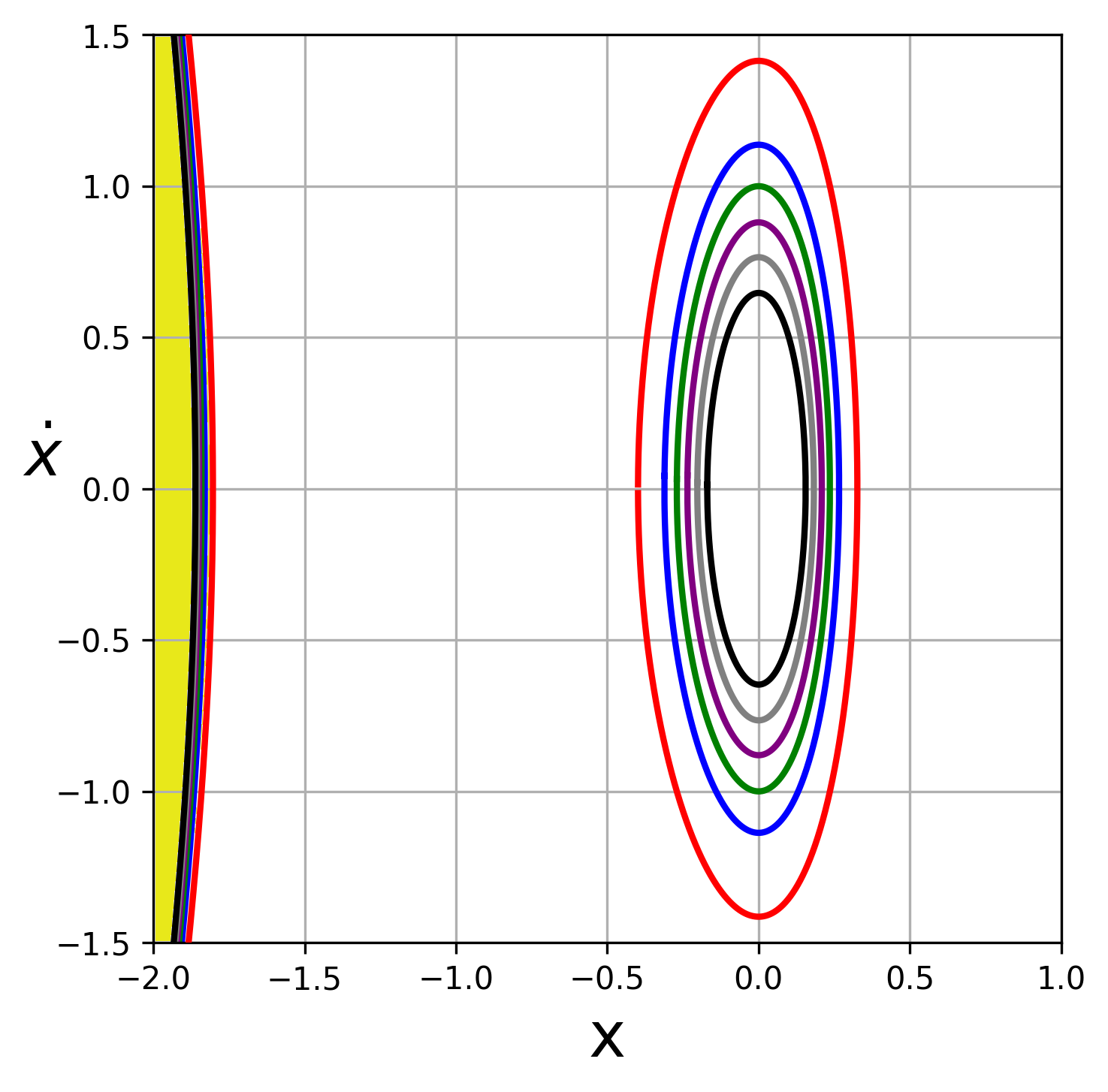}[a]
    \includegraphics[width=0.39\linewidth]{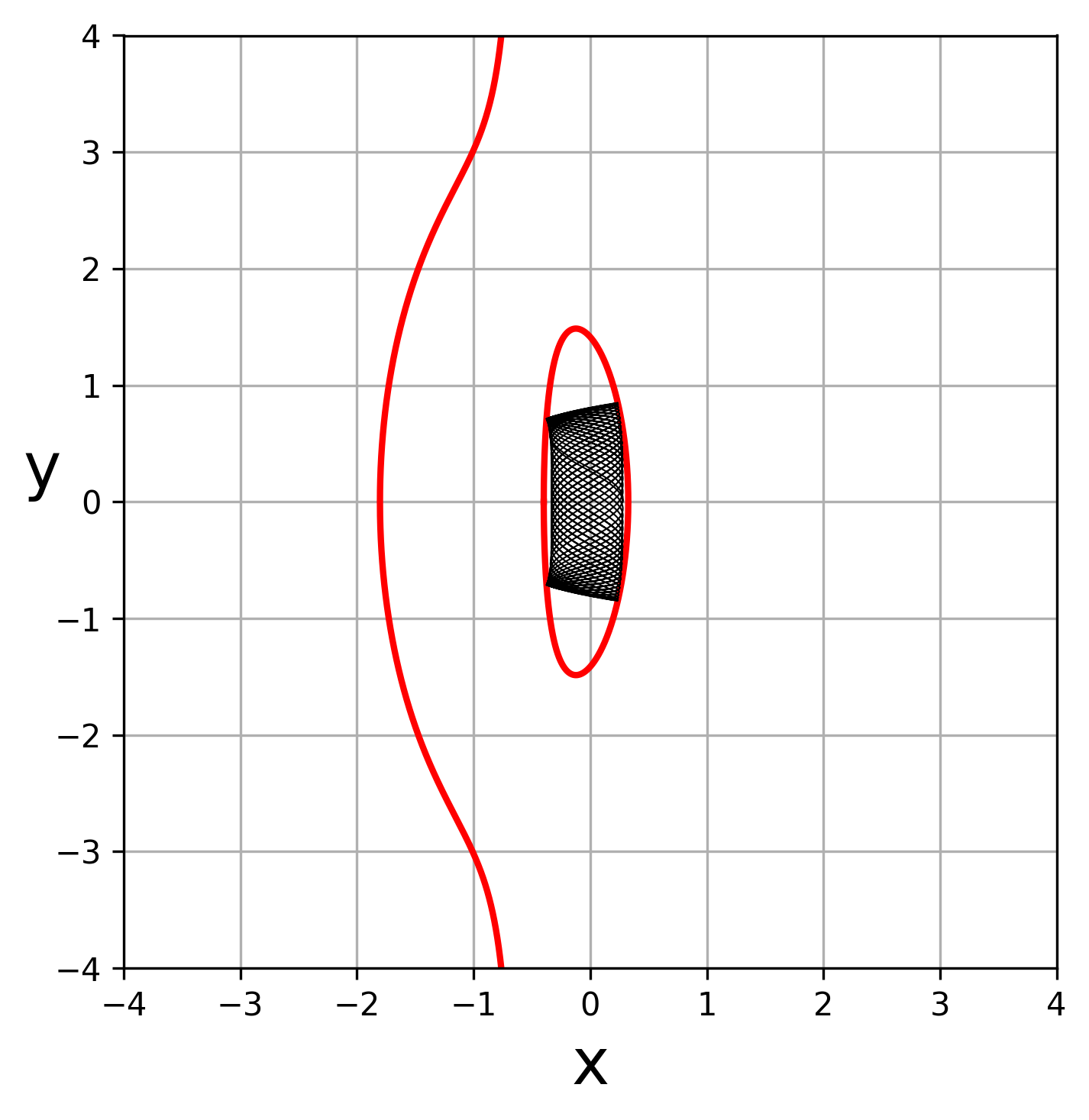}[b]
    \caption{a) Invariant curves on the $x-\dot{x}$ plane for $\epsilon=0.8$ ($A=16, B=1, \alpha=16/3$) and for various values of $K$: $K=0$ (red), $K=0.5$ (blue), $K=1$ (green), $K=1.5$ (purple), $K=2$ (gray) and $K=2.5$ (black).  The orbits on the left of $x_3$ are escaping (yellow). b) A Lissajous orbit inside the CZV in this case.}
    \label{fig:14}
\end{figure}

For $\epsilon=0$ this corresponds to an ellipse
$\dot{x}^2+16x^2=2$ elongated along the $\dot{x}$ axis, while for small $\epsilon$ the curves $\dot{y}^2=\sqrt{K}$ are deformed ellipses, plus an open curve on their left (Fig.~\ref{fig:14}a). They intersect the axis $y=0$ at the roots of the equation 
\begin{equation}
2-16x^2-3\epsilon x^3/3=0.\label{eqtr}
\end{equation}
The limiting boundary $\dot{y}^2=0$ is similar and the orbits are then Lissajous figures (Fig.~\ref{fig:14}b). The limiting curve opens if two roots are equal ($x_2=x_3$). This happens when $\dot{x}=0$ and $\frac{\partial \dot{y}^2}{\partial x}=-32 x-32\epsilon x^3=0$. Hence $x_\text{esc}=-\frac{1}{\epsilon_\text{esc}}$.

\begin{figure}[H]
    \centering
    \includegraphics[width=0.38\linewidth]{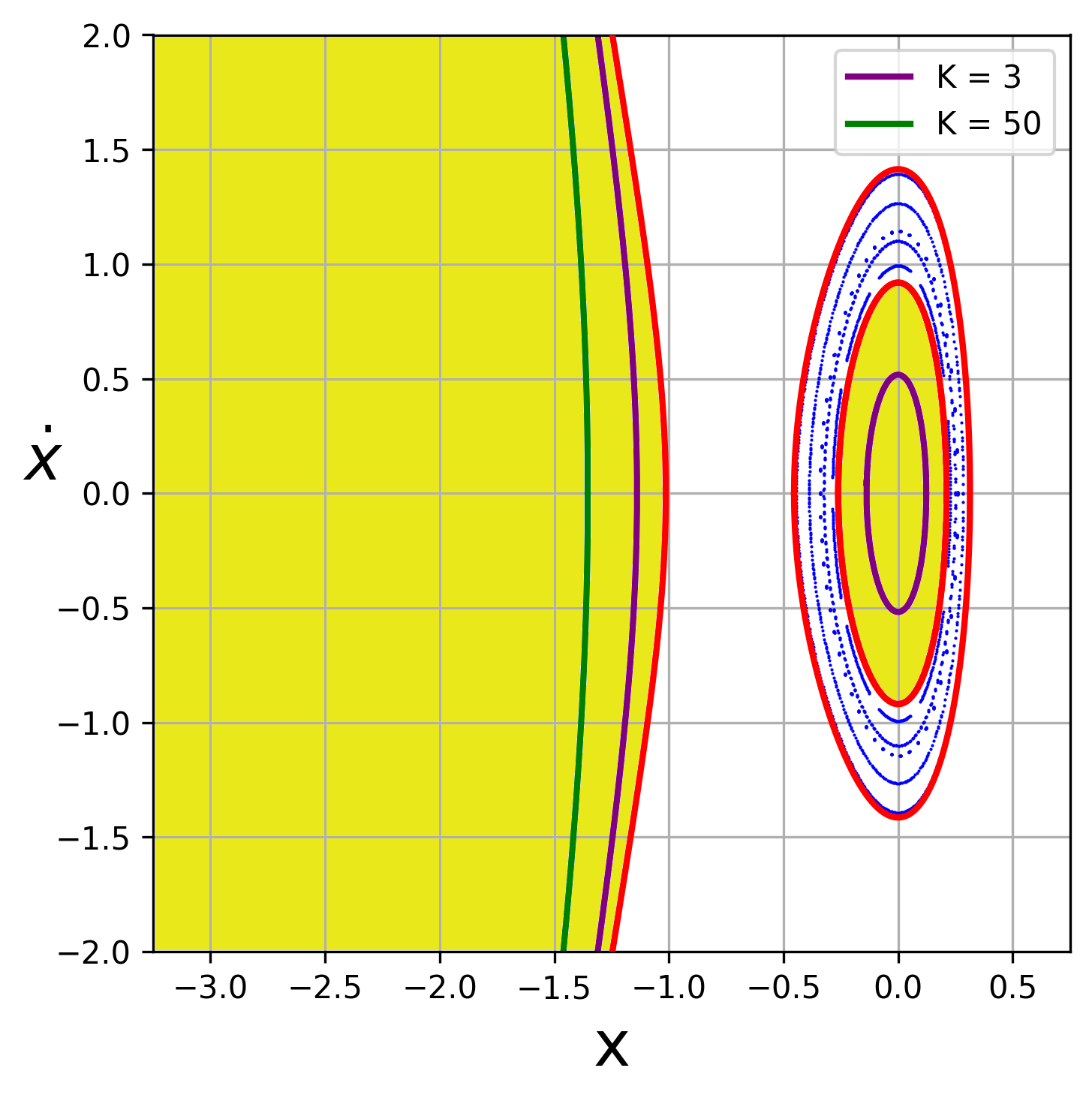}[a]
    \includegraphics[width=0.38\linewidth]{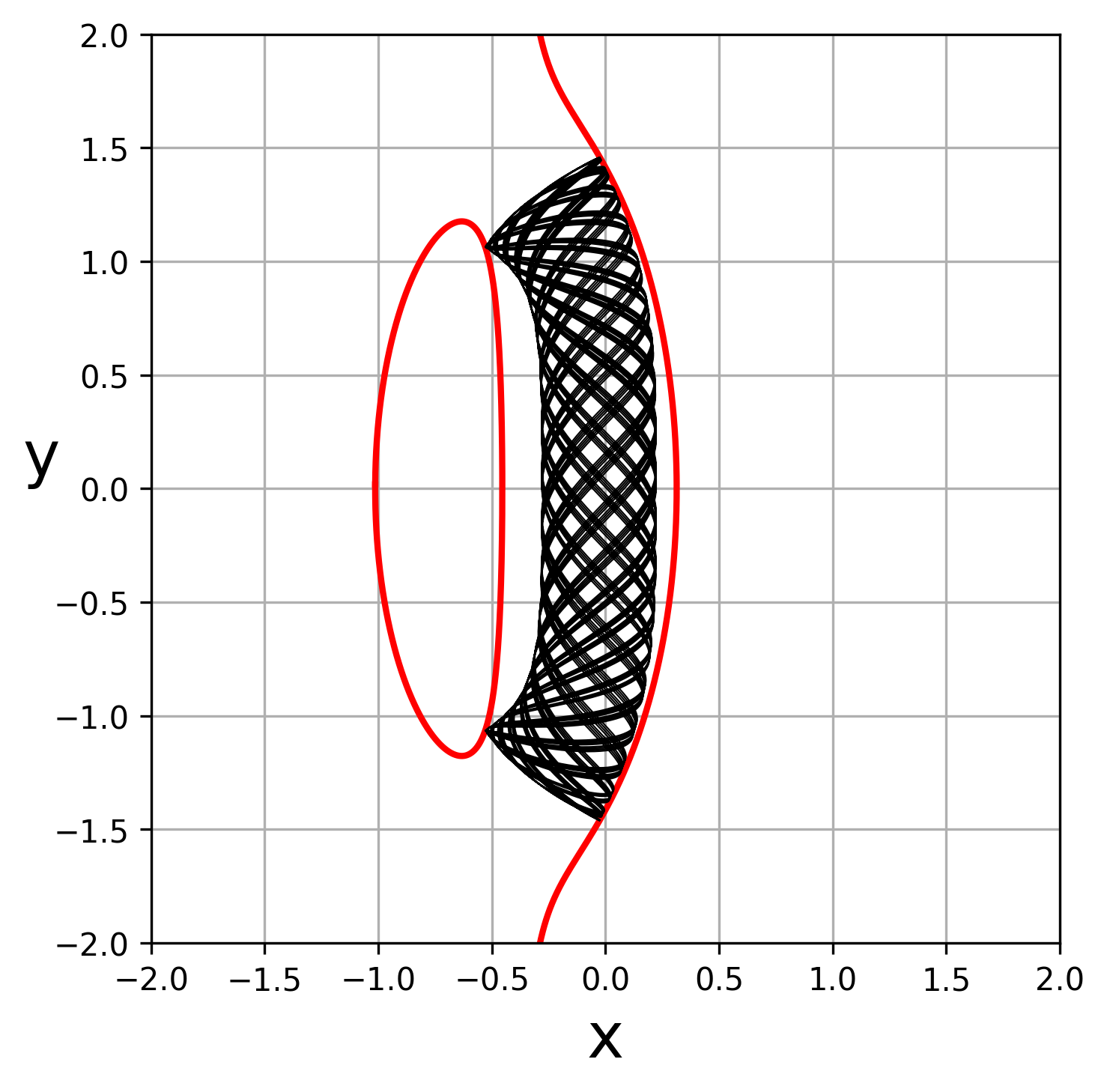}[b]
    \includegraphics[width=0.38\linewidth]{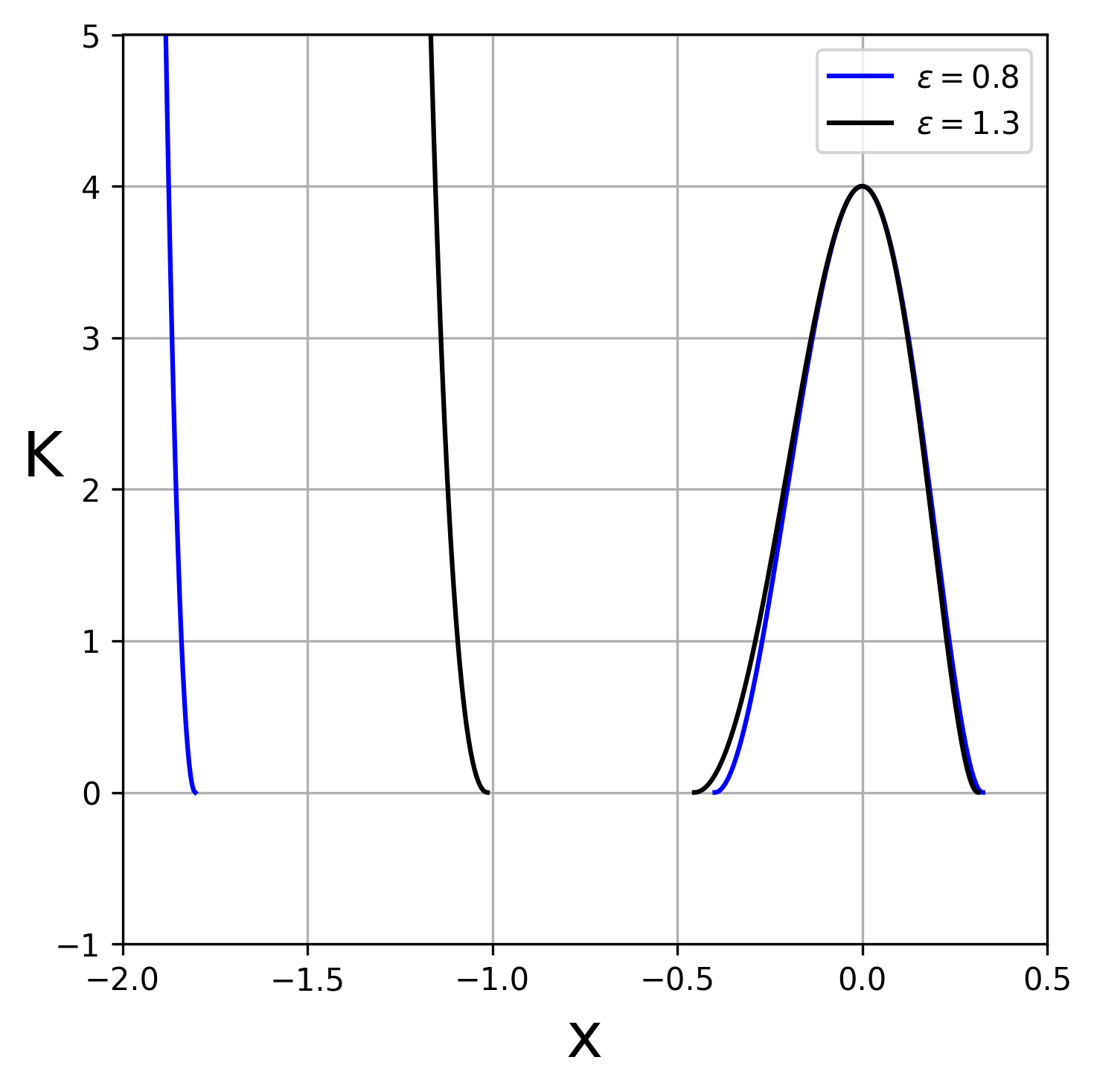}[c]
    \caption{a) Invariant curves for $\epsilon=1.3$ ($A=16, B=1, \alpha=16/3$). The inner red curve separates the Lissajous orbits (blue) from the escaping orbits (yellow). The orbits starting inside this curve as well as those on the left of the red open curve  are escaping (yellow regions)). b) A Lissajous curve in this case. c) $K$ as a function of $x$ for $\epsilon=0.8$ (blue) and $\epsilon=1.3$ (black).}
    \label{fig:16/3}
\end{figure}

Then $K=0$ and $2-\frac{16}{\epsilon_\text{esc}^2}+\frac{32}{(3\epsilon_\text{esc}^2)}=0$, 
hence $\epsilon_\text{esc}=\sqrt{\frac{8}{3}}=1.63$.
For larger $\epsilon$ the orbits can escape to $x=-\infty$. On the other hand, the CZV is given by the equation 
\begin{equation}
y^2=\frac{2-16x^2-\frac{32}{3}\epsilon x^3}{1+2\epsilon x} \label{ytet}   \end{equation} 
from Eq.~\eqref{Hr} for $\dot{x}=\dot{y}=0$. This curve goes to infinity when $x=x_{\infty}=-\frac{1}{2\epsilon}$. The numerator of Eq.~\eqref{ytet} is positive between the roots $x_2$ and $x_1$ and on the left of $x_3.$ If the value $x_{\infty}=-\frac{1}{2\epsilon}$ is on the left of $x_2$ then the CZV is a closed oval and an open curve starting at $(x=x_3, \dot{x}=0)$. All orbits starting inside the oval are Lissajous figures (Fig.~\ref{fig:14}b). But if $x_{\infty}=-\frac{1}{2\epsilon}$ is on the right of $x_2$ we may have escapes towards $y=\pm \infty$ even if the invariant curves are closed ovals. This happens if we set $x_{\infty}=-\frac{1}{2\epsilon}$ in the numerator of $y^2$ and find it positive, i.e. $2-\frac{16}{4\epsilon^2}+\frac{4}{3\epsilon^2}>0$, i.e. $\epsilon>\epsilon_\text{esc}'=\sqrt{4/3}=1.155$.

If $\epsilon>\epsilon_\text{esc}'$ and $\epsilon<\epsilon_\text{esc}$ the outermost invariant curve ($K=0$) forms an oval (Fig.~\ref{fig:16/3}a) and inside it there are Lissajous orbits (Fig.~\ref{fig:16/3}b). But there is also a region inside the oval of Fig.~\ref{fig:16/3}a with escapes to infinity. 

We note that the transition from Lissajous to escaping orbits happens in the same way as in the case $\alpha=2$. The limiting Lissajous curve has its left and right boundaries on the CZV. The corresponding invariant curve in Fig.~\ref{fig:16/3}a is a red curve inside the oval on the right of the figure. Orbits starting outside this red curve are Lissajous (Fig.~\ref{fig:16/3}b) while orbits starting inside it are escaping. The invariant curve inside the red curves correspond to larger values of $K$, but each curve consists of the initial conditions of different escaping orbits. In (Fig.~\ref{fig:16/3}c) we give $K(x)$ for $\epsilon=0.8$ and $\epsilon=1.3$.

\begin{figure}[H]
    \centering
    \includegraphics[width=0.38\linewidth]{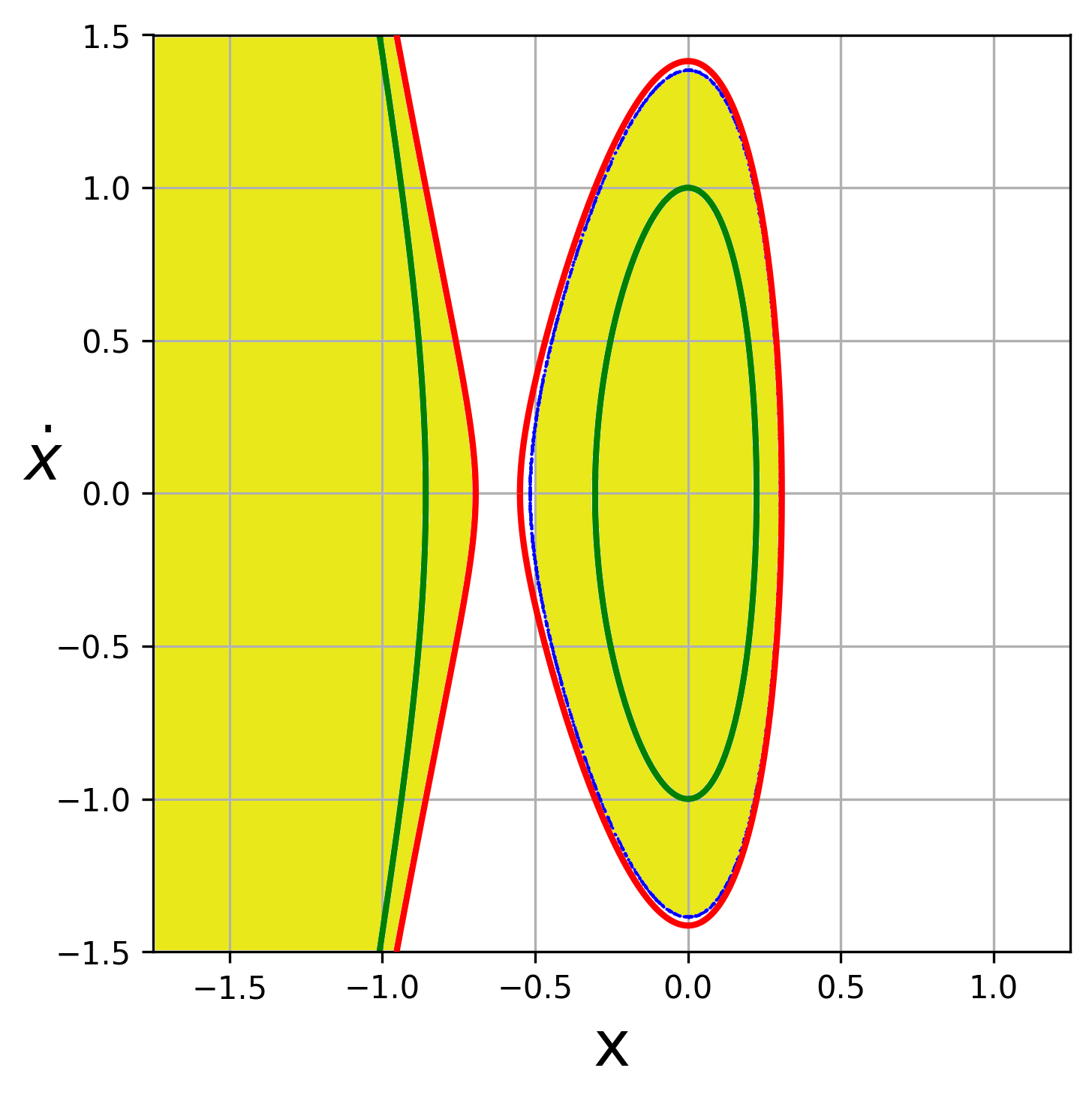}[a]
    \includegraphics[width=0.38\linewidth]{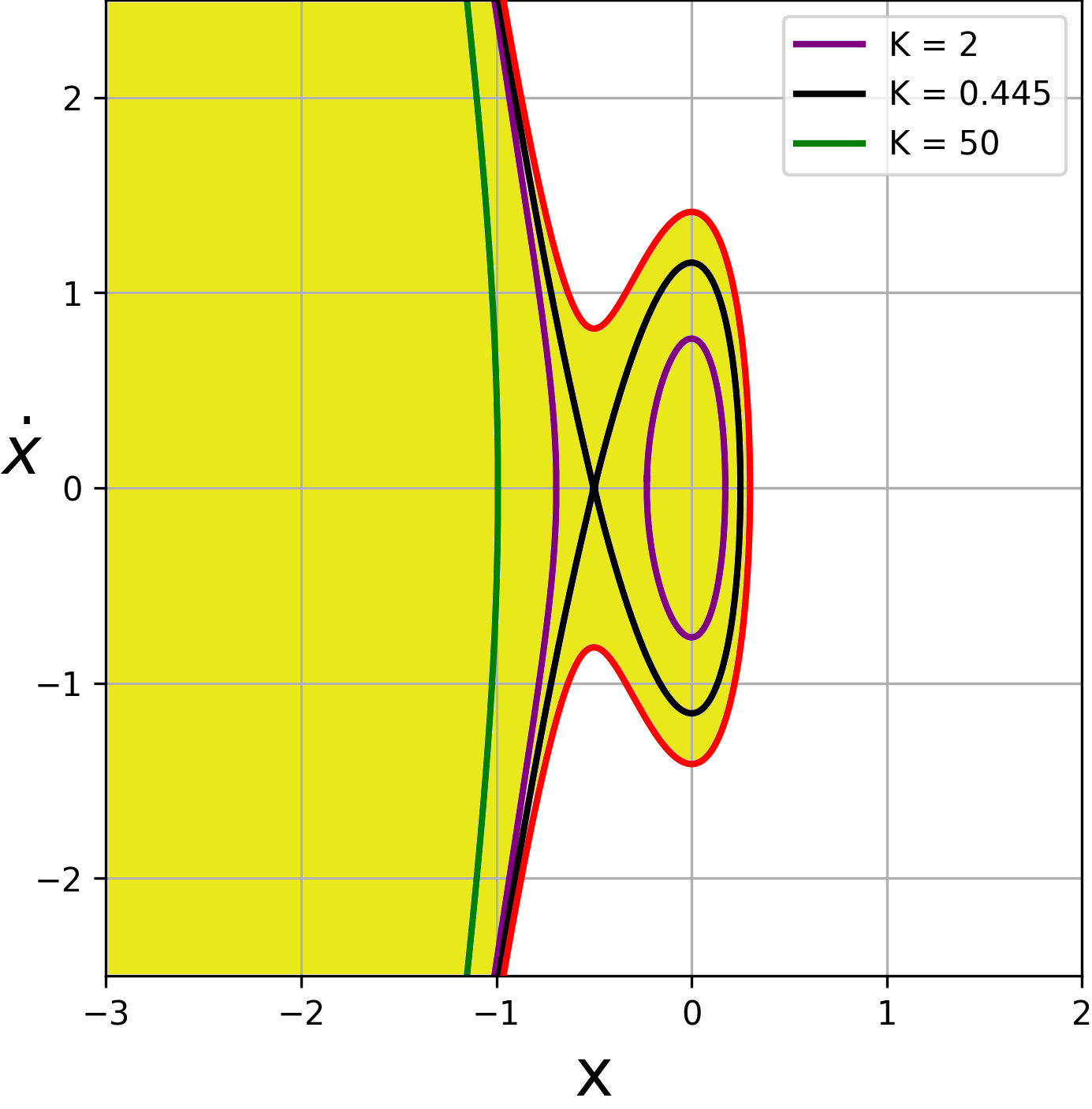}[b]
    \includegraphics[width=0.38\linewidth]{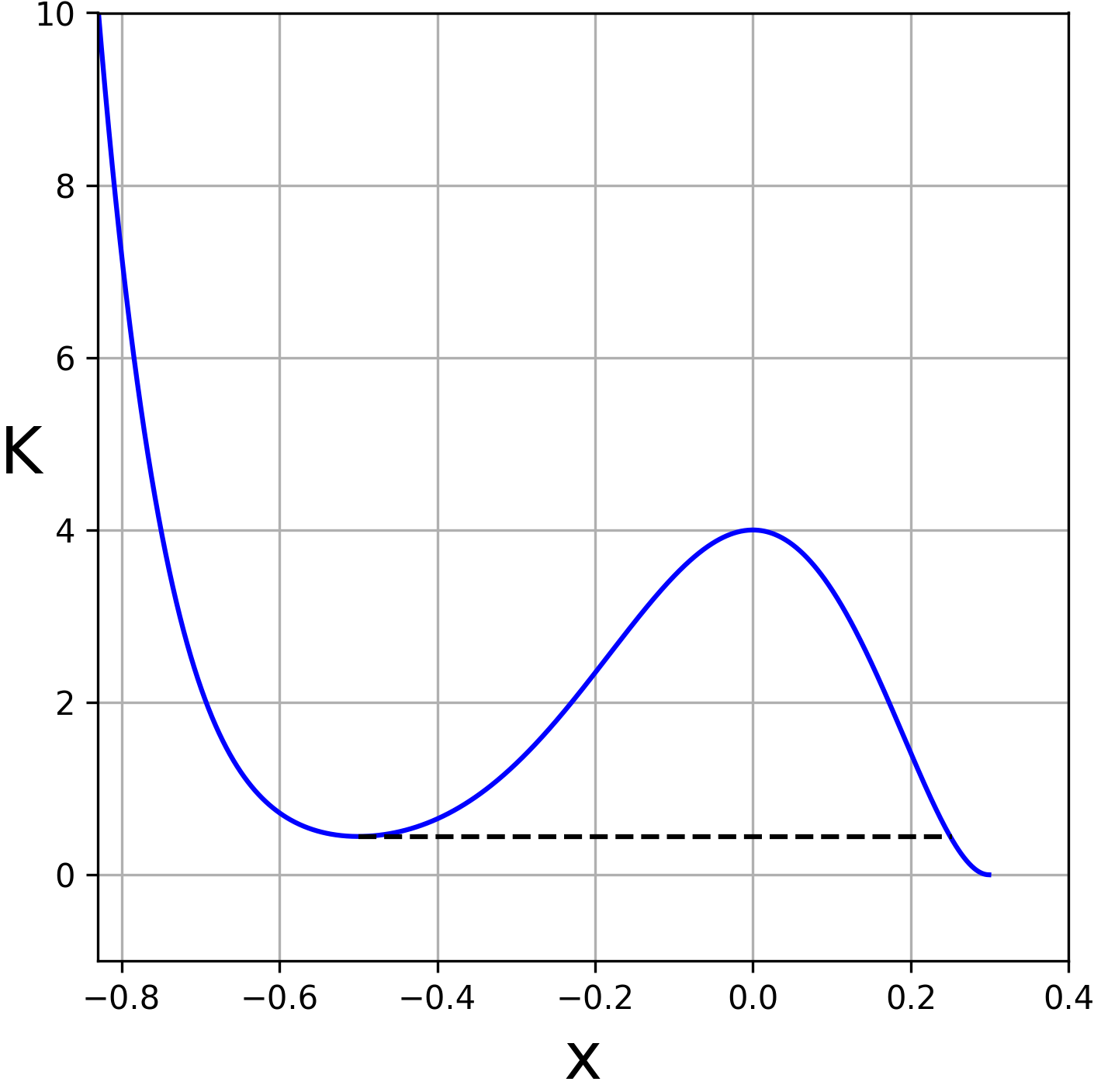}[c]
    \caption{a) Invariant curves for $\epsilon=1.6$ ($A=16, B=1, \alpha=16/3$) close to $\epsilon_\text{esc}$. Between the limiting curve $\dot{y}^2=2-\dot{x}^2-16x^2-\frac{32}{3}x^3=0$ on the right (red) and the yellow area of the oval (which contains escaping orbits)  there is a very thin layer containing Lissajous orbits (blue).  On the left of the open curve there are only escaping orbits. b) Invariant curve for $\epsilon=2>\epsilon_\text{esc}$. Although there is a crossing point and a loop on its right all the orbits inside the open CZV are escaping in this case. c) The values of $K$ as a function of $x$ for $\epsilon=2$. The values above/below $K_\text{min}$ are inside/outside the loop. But all the orbits are escaping.}
    \label{fig:fig17}
\end{figure}

If $\epsilon$ approaches the escape value, most orbits inside the oval (Fig.~\ref{fig:fig17}a) are escaping to ${y}^2=\infty$ (yellow region) and only a few Lissajous orbits (blue) are located in a small layer near the boundary of the oval.

For $\epsilon>\epsilon_\text{esc}$, the outermost invariant curve is  a single curve corresponding to $\dot{y}^2=\sqrt{K}=0$ ($y=0$) and it is given by the equation
\begin{equation}
    \dot{y}^2=2-\dot{x}^2-16x^2-\frac{32}{3}\epsilon x^3=0.
\end{equation}

Then for any particular value of $\epsilon$ there is a corresponding value of $K$ that represents an invariant curve that crosses itself and forms a loop on the right (Fig.~\ref{fig:fig17}b). All the orbits starting both inside and outside the loop escape to infinity. The same happens for all larger values of $\epsilon$. The corresponding values of $K$ as a function of $x$ are given in Fig.\ref{fig:fig17}c. $K_\text{min}$ corresponds to the invariant curve crossing itself on the $\dot{x}=0$ axis. The maximum $x$ of this curve $x$ is at $x=0.25$ close to the point $x_1$ and orbits starting beyond this point in $x$ escape to $x=-\infty$. But the orbits that have $K>K_\text{min}$ above the dashed line in Fig.~\ref{fig:fig17}c also escape towards to $y$ direction. 

Thus we conclude that the case with $A=16B$ and $\alpha=16/3$ is different from the cases $\alpha=2$ as regards the orbits with $\epsilon$ beyond the escape perturbation. While in both cases we find an invariant curve crossing itself and forming a loop on the right, in the case ($\alpha=2$) the orbits are Lissajous inside the loop, while in the case $\alpha=16/3$ they are escaping (towards $y^2\to \infty$).

\section{A summary of the case (i) ($\alpha=1/3$)}
In our previous work \cite{contopoulos2024} we studied the integrable case (i) with $A=B=E=1, \alpha=1/3$. We found both Lissajous and escaping orbits and the transition between them.  In this case  the second integral of motion reads
\begin{equation}
    Q=\dot{x}\dot{y}+Axy+\epsilon(xy^2+y^3/3).
\end{equation}
 and we have similar equations concerning the invariant curves to those in cases ii) and (iii). Namely, for $y=0$  we have 
\begin{equation}
    \dot{x}^2\dot{y}^2=K^2\label{gin}
\end{equation}
with 
\begin{equation}
    \dot{y}^2=2-\dot{x}^2-x^2-\frac{2}{3}\epsilon x^3.\label{gin2}
\end{equation}
On the other hand, the CZV  is given by the equation
\begin{equation}
    y^2=\frac{2-x-\frac{2}{3}\epsilon x^2}{1+2\epsilon x}.\label{eqf}
\end{equation}
However, the forms of the Lissajous orbits are different. In fact, they are parallel to the diagonals $x\pm y=0$, since the boundaries of the orbits are given by the  equation \cite{contopoulos2024}.
\begin{equation}
    2-(x\pm y)^2-\frac{2\epsilon}{3}(x+y)^4=2K,
\end{equation}
This gives particular solutions for $(x\pm y)$ depending on $\epsilon$ and $K$.

\begin{figure}[H]
    \centering
    \includegraphics[width=0.4\linewidth]{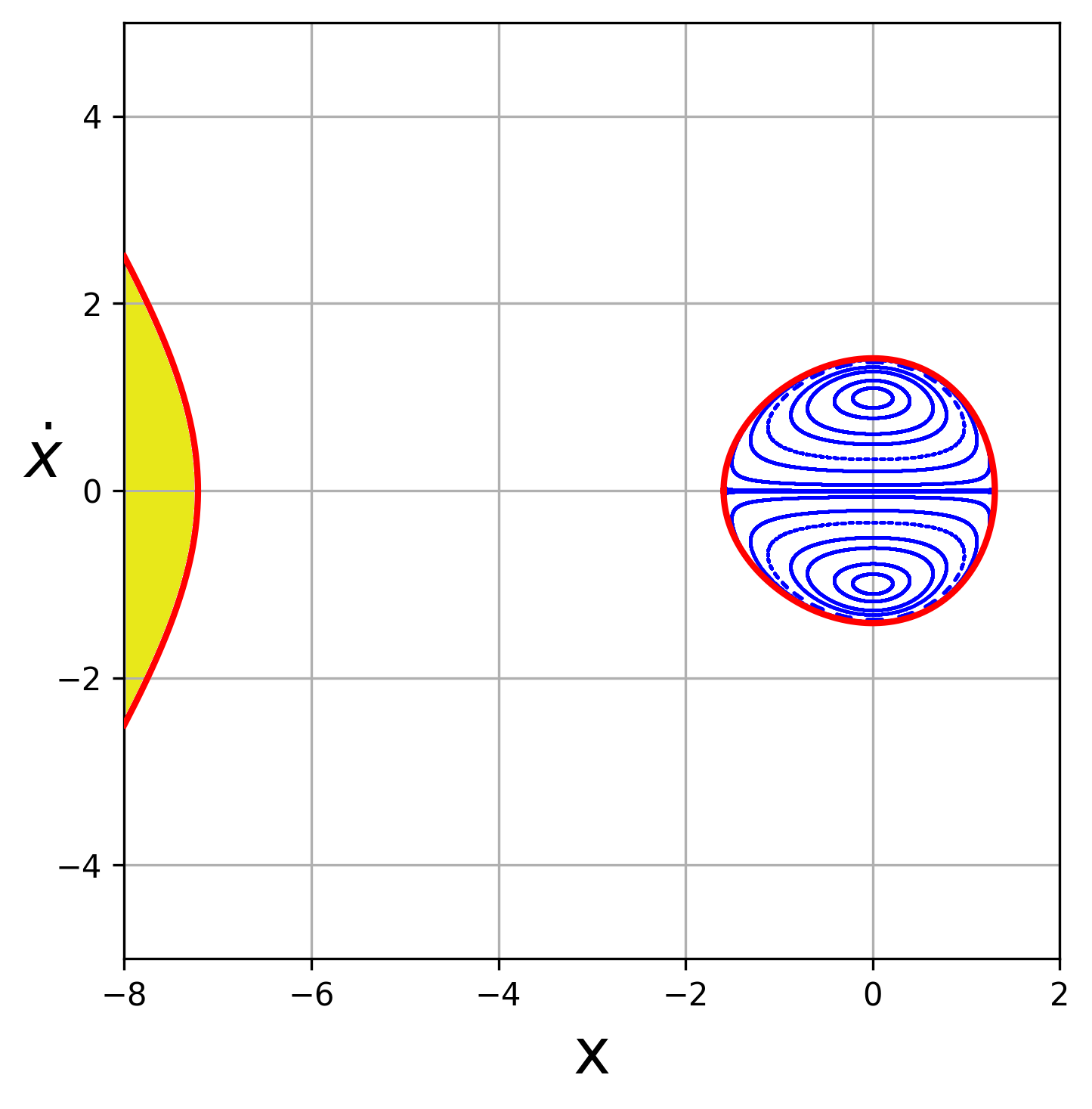}[a]
    \includegraphics[width=0.4\linewidth]{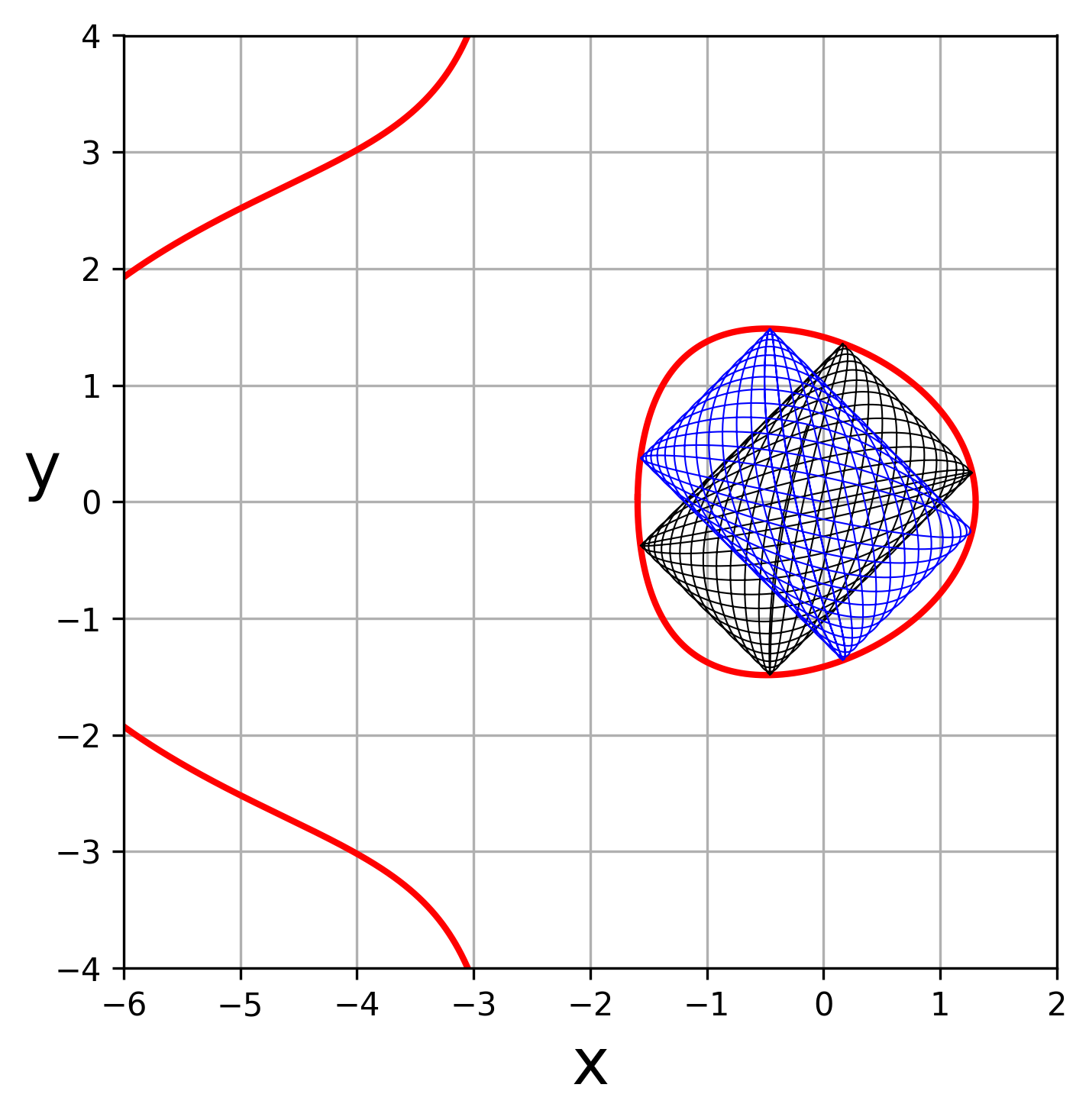}[b]
    \caption{Case $A=B=1, \alpha=1/3$ a) Invariant curves around two points $(x=0,\dot{x}=\pm1)$ for $\epsilon=0.2$. b) Two corresponding Lissajous orbits with boundaries parallel to $x=\pm y$. All the orbits inside the oval are Lissajous. The orbits on the left of the open curve  are escaping (yellow).}
    \label{fig:figpala}
\end{figure}

For small $\epsilon$  the invariant curves on the $(x-\dot{x})$ plane are inside a limiting curve $\dot{y}^2=0$, between the roots $x_2$ and $x_1$ of Eq.~\eqref{gin2} for $\dot{x}^2=\dot{y}^2=0$, and form closed curves around two points ($x=0, \dot{x}=\pm 1$) for  $K=0$  up to $K_\text{max}=1$  (Fig.~\ref{fig:figpala}a). There are also invariant curves starting for $\dot{x}=0$ on the left of the third root $x_3$. The corresponding orbits are Lissajous figures inside the CZV which  is composed of an oval between the roots $x_2$ and $x_1$ (Fig.~\ref{fig:figpala}b) and has also an open curve starting at the third root $x_3$ on the left. The orbits on the left of the open curve are escaping.

\begin{figure}[H]
    \centering
    \includegraphics[width=0.4\linewidth]{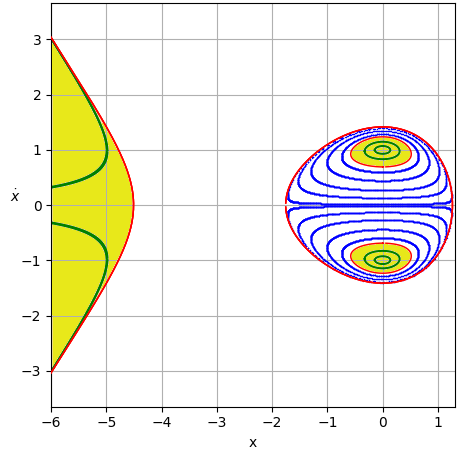}[a]
    \includegraphics[width=0.4\linewidth]{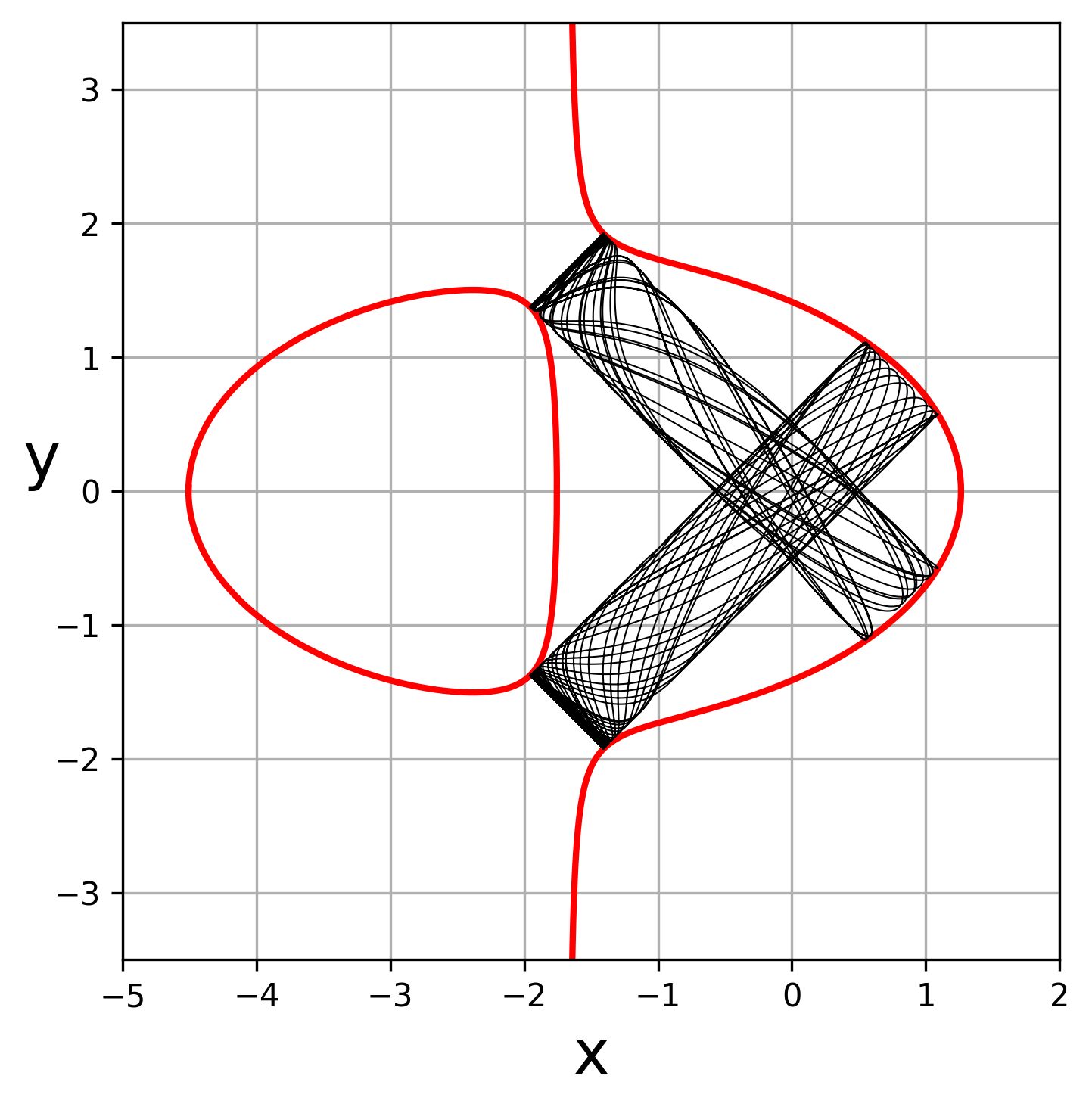}[b]
    \caption{Case $A=B=1, \alpha=1/3$: a) Invariant curves for $\epsilon=0.3$.  Around the points $(x=0, \dot{x}=\pm 1)$ there are two red curves containing escaping orbits (yellow regions).  The orbits on the left of the open red curve are escaping. b) Two  limiting Lissajous orbits in this case with their boundaries tangent to the two parts of the CZV. }
    \label{fig:figpalc}
\end{figure}

In this case there is an escape perturbation $\epsilon=\epsilon_{esc}=1/\sqrt{6}=0.408$ for which the roots $x_2$ and $x_3$ join, and for $\epsilon>\epsilon_{esc}$ there is only an open limiting invariant curve $(\dot{y}^2=0)$  in Eq.~\eqref{gin2} to the left of $x_1$. There is also another escape perturbation $\epsilon_\text{esc}'=\frac{1}{2\sqrt{3}}=0.288$, when the CZV reaches $y^2=\infty$.  For $\epsilon$ between $\epsilon_\text{esc}'$ and $\epsilon_\text{esc}$ there are both Lissajous orbits and escaping orbits. Then there is an escape value of $K_\text{esc}(\epsilon)$. For larger $K$, the orbits escape to infinity while for smaller $K$ the orbits are Lissajous figures like those shown in Fig.~\ref{fig:figpalc}b where $\epsilon=0.3$. The escape value of $K$ is $K_\text{esc}(0.3)=0.852$. The transition from Lissajous to escaping orbits happens when the upper and lower boundaries of the Lissajous orbits become tangent to the CZV and escapes are from the upper or the lower opening of the CZV. The corresponding invariant curves are shown in Fig.~\ref{fig:figpalc}a. The transition value of $K$ corresponds to two red curves on the surface of section $y=0$, one around ($x=0,\dot{x}=1$) and the other around ($x=0,\dot{x}=-1$).

\begin{figure}[H]
    \centering
    \includegraphics[width=0.39\linewidth]{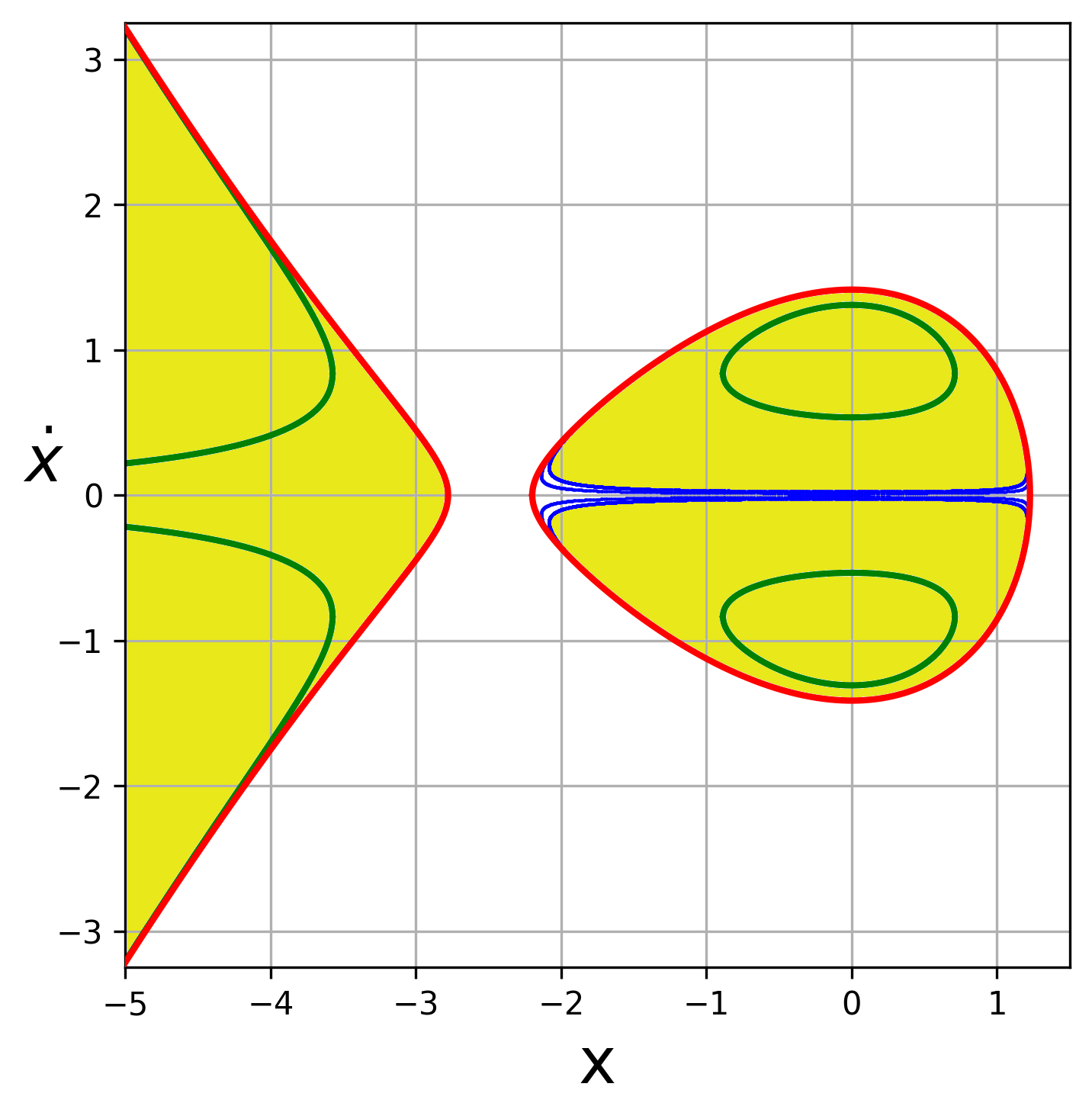}[a]
    \includegraphics[width=0.4\linewidth]{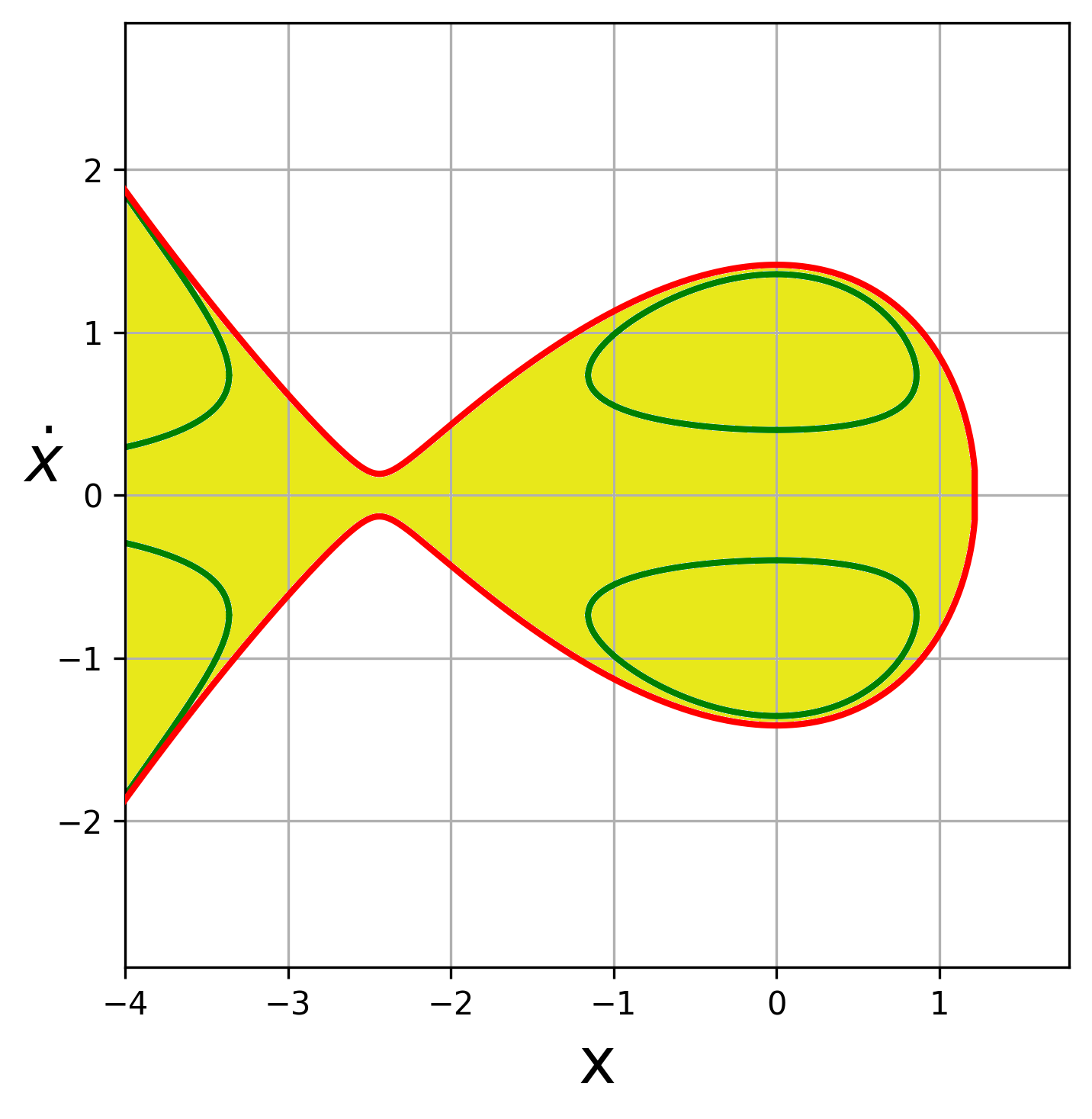}[b]
    \includegraphics[width=0.4\linewidth]{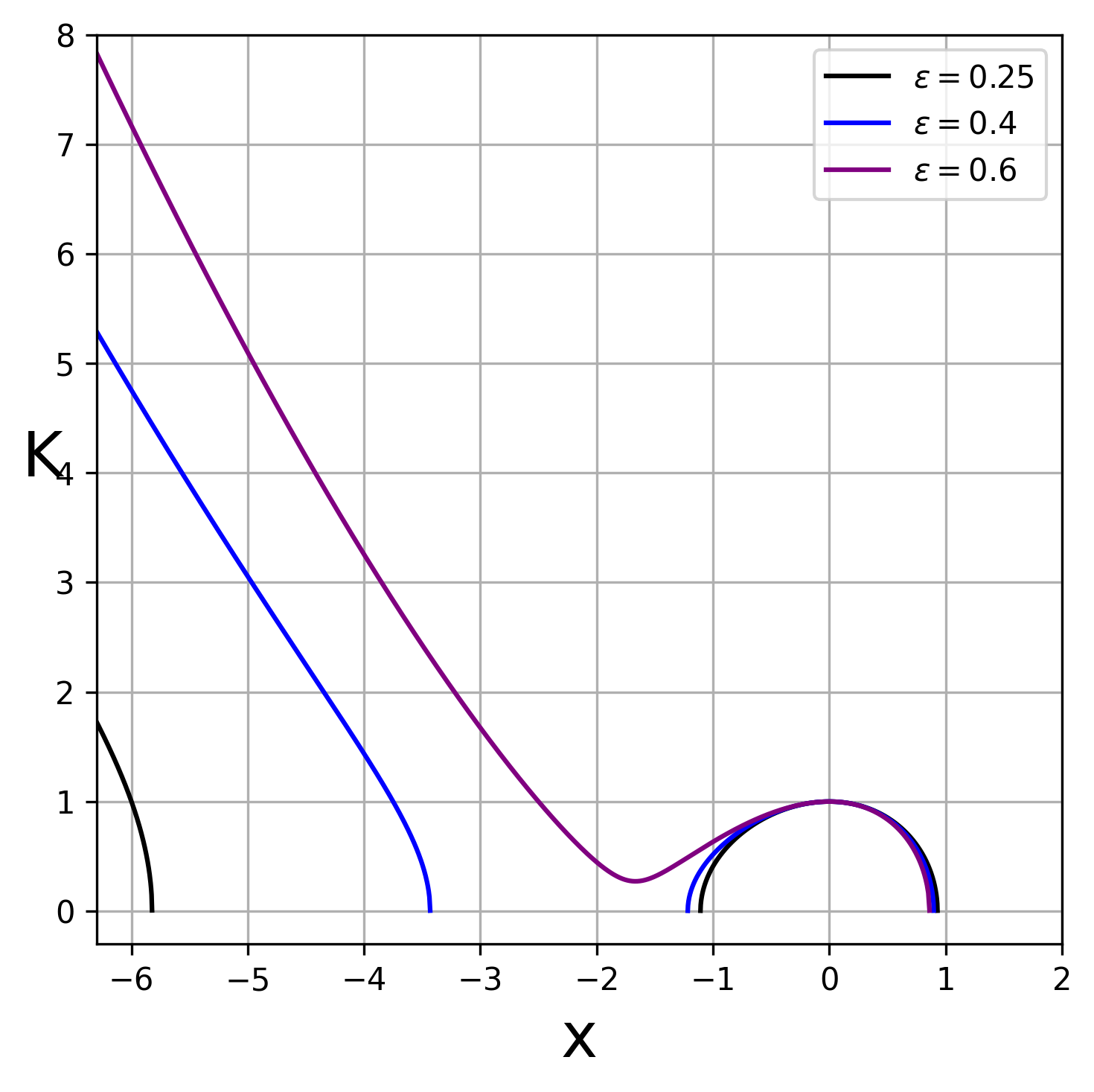}[c]
    \caption{Case $A=B=1, \alpha=1/3$. a) Invariant curves in the case $\epsilon=0.4$ (close to $\epsilon_\text{esc}=0.408$). Only a few invariant curves exist close to the $x$-axis inside the oval of the CZV (around the upper or the lower yellow region of escapes). All the orbits on the left of the open CZV are escaping. b) In the case $\epsilon=0.41$ above the escape value $\epsilon_\text{esc}=0.408$ all the orbits are escaping c)  $K(x)$ for $\epsilon=0.25, 0.4$ and $\epsilon=0.6$.}
    \label{fig:figc2}
\end{figure}

As $\epsilon$ increases the area of escaping orbits increases (Fig.~\ref{fig:figc2}a for $\epsilon=0.4$) and only a few Lissajous orbits for $K$ close to zero remain. Then for $\epsilon$ beyond the escape value $\epsilon_\text{esc}=0.408$ all the orbits escape to infinity (Fig.~\ref{fig:figc2}b).  Finally, the values of $K$ as functions of $x$ in this case are given in Fig.~\ref{fig:figc2}c. 

We conclude that the behavior of the case (i) as regards the escaping orbits is similar to the behavior of the case (iii) and not to that of the case (ii).

\section{Summary and conclusions}

The orbits in all the integrable cases of 2-d H\'{e}non-Heiles systems are either of generalized Lissajous forms, or they escape to infinity. The generalized Lissajous orbits fill a curvilinear parallelogram whose corners are on the CZV. The main purpose of the present paper is to classify the various cases when we change the parameters of the system.

The Hamiltonian is of the form $H=\frac{1}{2}\left(\dot{x}^2+Ax^2+\dot{y}^2+By^2\right)+\epsilon(xy^2+\alpha x^3)=E$ and the 3 integrable cases are i) with $A=B$ and $\alpha=1/3$, ii) with $A,B$, arbitrary and $\alpha=2$ and finally (iii) with $A=16B$ and $\alpha=16/3$. We worked with $E=1$ in all cases and $A=1$ in the first two cases and with $B=1$ in the third case.

We calculated the invariant curves on the surface of section $(x, \dot{x})$ at $y=0$, and the orbits  on the plane $(x,y)$ inside the CZV. The three cases have many similarities that we summarize here.
The invariant curves on the surface of section $y=0$ are given by an equation of the form

\begin{equation}
    F\dot{y}^2=K,
\end{equation}
where 
\begin{equation}
    \dot{y}^2=-2-\dot{x}^2-Ax^2-2\epsilon \alpha x^3=0
\end{equation}
and $K$ is the value of the second integral. The above equation for $\dot{x}=\dot{y}=0$ and small $\epsilon$ has 3 roots $x_1>0>x_2>x_3$. Thus for $\dot{x}^2=0, \dot{y}^2$ is positive for  $x_2x<x_1$ and for $x<x_3$. The coefficient $F$ is equal to $\dot{x}^2$ in (i), equal to $4B-A-4\epsilon x$ in (ii) and  equal to  $\dot{y}^2$ in case (iii).

The CZV $\dot{x}=\dot{y}=0$ is found from the Hamiltonian, which for $\dot{x}=\dot{y}=0$ gives
\begin{equation}
    y^2=\frac{2E-Ax^2-\epsilon \alpha x^3}{B+2\epsilon x}\label{cuzv}
\end{equation}
(where $E=1$ in all our examples).

For small values of $\epsilon$ the equation $y^2=0$ has the same  3 roots $x_1>0>x_2>x_3$. Then the CZV consists of an oval between $x_1$ and $x_2$ and an open curve beyond $x_3$ on the left, symmetric with respect to the $x$-axis.

As $\epsilon$ increases and goes beyond a critical value $\epsilon_\text{esc}$ the roots $x_2$ and $x_3$ join and then become complex, thus the only real root is $x_1$. Then many orbits (or all the orbits) escape to minus infinity.

However, we may also have escapes to $y=\pm\infty$ when the denominator of Eq.~\eqref{cuzv} becomes zero for $x_{\infty}=-\frac{B}{2\epsilon}$. If this value of $x$ is between the roots of $x_1$ and $x_2$, the escapes along $y$ may occur even inside the closed oval formed by the CZV. This happens beyond a certain value of $\epsilon_\text{esc}'$ when $\epsilon_\text{esc}' <\epsilon_\text{esc}$.

The separation between the Lissajous and the escaping orbits depends on the value of the second integral $K$.
In particular, the intersections of orbits by the surface of section $y=0$ give an invariant curve on the plane ($x, \dot{x}$) for every $K$. The limiting invariant curves are found by setting $\dot{y}^2=0$. For initial conditions beyond that curve we have $\dot{y}^2<0$, i.e. such an orbit does not exist.

The limiting invariant curve on the surface of section also consists of two parts for small $\epsilon$: a  closed oval between $x_2$ and $x_1$ and an open curve from $x_3$ and beyond on the left. If $\epsilon>\epsilon_\text{esc}$ the two parts of the limiting curve join and form an open curve from $x_1$ open to the left.

The escapes to $y\to\pm \infty$ occur when the value of $x_{\infty}=-\frac{B}{2\epsilon}$ is on the right of $x_2$. In such a case the CZV consists of two parts. One starting at $x_1$ symmetric with respect to the $x$-axis and going to $y=\pm \infty$ at $x=x_{\infty}=-\frac{B}{2\epsilon}$ and another one forming an oval between $x_2$ and $x_3$ (inside which $\dot{x}^2+\dot{y}^2<0$) and no orbits exist. Between the two parts of the CZV there are two openings, above and below the $x$-axis, from which some orbits may escape to $y=\pm \infty$.

Whether or not an orbit will escape depends on the form of its  boundary. The boundary is found by eliminating $\dot{x}$ and $\dot{y}$ from the equations $H=E, Q=K$  and $   \frac{\partial Q}{\partial \dot{x}}\dot{y}-\frac{\partial Q}{\partial \dot{y}}\dot{x}=0$. If its 4 corners  reach the CZV then the orbit is a Lissajous figure. But if the boundary escapes to infinity along the $y=\pm \infty$ direction  then the orbit escapes. The separation of the Lissajous orbits from the escaping orbits occurs at a value of $K$ for which the boundaries of the orbit become tangent to the CZV. Examples of such a transition are given  for various values of $A,B$ and $E=1$ in the previous examples.

Now we point out the similarities and the differences between the cases (i), (ii) and (iii).

The boundaries of the Lissajous figures are along curves perpendicular to the $x$-axis in cases (ii) and (iii), while they are exactly straight lines parallel to the axes $x=\pm y$ in (i) because in that case the equation $\frac{\partial Q}{\partial \dot{x}}\dot{y}-\frac{\partial Q}{\partial \dot{y}}\dot{x}=0$ takes a particularly simple form $\dot{x}^2=\dot{y}^2$. This does not happen in the (ii) and (iii) cases.

If the value of $x_{\infty}$ is to the left of the root $x_2$ then all the orbits inside the oval between $x_2$ and $x_1$ are Lissajous. This happens for all the values of $\epsilon$ up to the escape value $\epsilon_\text{esc}$ where $x_2$ and $x_3$ join, and only one root, $x_1$, remains. But even for a range of values of $\epsilon$ above $\epsilon_\text{esc}$ in the case (ii) there are some orbits  in some ranges of the value of $B$ forming a loop in the surface of section for $y=0$ that do not escape but form Lissajous curves. The orbits starting inside it are Lissajous figures, while the orbits starting outside the loop are escaping. As $\epsilon$ increases the size of the loop decreases and for a particular maximum $\epsilon$ the loop shrinks to a point. Then for even larger $\epsilon$ all the orbits are escaping. However in the cases (i) and (iii) all the orbits for $\epsilon$ beyond $\epsilon_\text{esc}$ escape. This is due to the fact that a proportion of orbits escape towards $y=\pm \infty$ even for $\epsilon<\epsilon_\text{esc}$ and their proportion tends to $100\%$ when $\epsilon\to\epsilon_\text{esc}$.

The present paper conducted a thorough investigation of the integrable cases of the generalized Hénon-Heiles system. We studied the structure and properties of the invariant curves and the form of the corresponding orbits, bounded and escaping. In particular,  we studied the influence of parameter variations on the forms and the positions of the invariant curves and of the CZVs and  on the transition from bounded to escaping orbits.  This approach reinforces the significance of the integrability conditions in shaping phase-space structures and lays the groundwork for future studies exploring the boundaries of integrable behavior in more generalized  Hamiltonians.

\section*{Acknowledgements}
This research was conducted in the framework of the programme of the Research
Committee of the Academy of Athens, “Study of order and chaos in quantum dynamical systems”
(No. 200/1026).

\bibliographystyle{iopart-num}
\bibliography{bibliography}

\providecommand{\newblock}{}
\begin{thebibliography}{10}
\expandafter\ifx\csname url\endcsname\relax
  \def\url#1{{\tt #1}}\fi
\expandafter\ifx\csname urlprefix\endcsname\relax\def\urlprefix{URL }\fi
\providecommand{\eprint}[2][]{\url{#2}}

\bibitem{stackel1890}
St\"{a}ckel P 1890 {\em Math. Ann.\/} {\bf 35} 91

\bibitem{stackel1893}
St\"{a}ckel P 1893 {\em Math. Ann.\/} {\bf 42} 537

\bibitem{lynden1962stellar}
Lynden-Bell D 1962 {\em Month. Not. R. Astron. Soc.\/} {\bf 124} 95

\bibitem{contopoulos2002order}
Contopoulos G 2002 {\em Order and chaos in dynamical astronomy\/} vol~21
  (Springer)

\bibitem{giorgilli2022notes}
Giorgilli A 2022 {\em Notes on Hamiltonian dynamical systems\/} vol 102
  (Cambridge Univ. Press)

\bibitem{kolmogorov1954conservation}
Kolmogorov A~N 1954 {\em Dokl. akad. nauk Sssr\/} {\bf 98} 527

\bibitem{arnold1961}
Arnold V 1961 {\em Sov. Math. Dokl.\/} {\bf 2} 247

\bibitem{Möser:430015}
Möser J 1962 {\em Nachr. Akad. Wiss. Göttingen, II\/}  1

\bibitem{Moser1956}
Moser J 1956 {\em Comm. Pure Appl. Math.\/} {\bf 9} 673

\bibitem{Moser1958}
Moser J 1958 {\em Comm. Pure Appl. Math.\/} {\bf 11} 81

\bibitem{Giorgilli2001}
Giorgilli A 2001 {\em Discrete and Continuous Dyn. Syst.\/} {\bf 7} 855

\bibitem{Henon1964}
H\'{e}non M and Heiles C 1964 {\em Astron. J.\/} {\bf 69} 73

\bibitem{conte2005explicit}
Conte R, Musette M and Verhoeven C 2005 {\em J. Math. Phys.\/} {\bf 12} 212

\bibitem{zhao2007threshold}
Zhao H and Du M 2007 {\em Phys. Rev. E\/} {\bf 76} 027201

\bibitem{blesa2012escape}
Blesa F, Seoane J~M, Barrio R and Sanjuan M~A 2012 {\em Int. J. Bif. Chaos\/}
  {\bf 22} 1230010

\bibitem{Waite1981}
Waite B and Miller W 1981 {\em J. Quant. Chem.\/} {\bf 74} 3910

\bibitem{sengupta1996quantum}
Sengupta S and Chattaraj P 1996 {\em Phys. Lett. A\/} {\bf 215} 119

\bibitem{contopoulos2024}
Contopoulos G and Tzemos A~C 2024 {\em Particles\/} {\bf 7} 1062

\bibitem{ablowitz1980connection}
Ablowitz M~J, Ramani A and Segur H 1980 {\em J. Math. Phys.\/} {\bf 21} 1006

\bibitem{ramani1989painleve}
Ramani A, Grammaticos B and Bountis T 1989 {\em Phys. Rep.\/} {\bf 180} 159

\bibitem{lakshmanan1993painleve}
Lakshmanan M and Sahadevan R 1993 {\em Phys. Rep.\/} {\bf 224} 1

\bibitem{wang2016integrability}
Wang D~S and Wei X 2016 {\em Appl. Math. Lett.\/} {\bf 51} 60

\bibitem{chang1982analytic}
Chang Y, Tabor M and Weiss J 1982 {\em J. Math. Phys.\/} {\bf 23} 531

\bibitem{grammaticos1982painleve}
Grammaticos B, Dorizzi B and Padjen R 1982 {\em Phys. Lett. A\/} {\bf 89} 111

\bibitem{bountis1982integrable}
Bountis T, Segur H and Vivaldi F 1982 {\em Phys. Rev. A\/} {\bf 25} 1257

\bibitem{fordy1991henon}
Fordy A~P 1991 {\em Phys. D\/} {\bf 52} 204

\bibitem{sarlet1991new}
Sarlet W 1991 {\em J. Phys. A\/} {\bf 24} 5245

\bibitem{hietarinta1987direct}
Hietarinta J 1987 {\em Phys. Rep.\/} {\bf 147} 87

\bibitem{smirnov1998integrability}
Smirnov R 1998 {\em Appl. Math. Lett.\/} {\bf 11} 71

\bibitem{contarx}
Contopoulos G, Tzemos A~C and Zanias F 2025 {\em Phys. Scr.\/} {\bf 100} 045225

\bibitem{Contopoulos1960}
Contopoulos G 1960 {\em Z. Astrophys.\/} {\bf 49} 273

\bibitem{Rosenbluth1966}
Rosenbluth M~N, Sagdeev R~A, Taylor J~B and Zaslavsky G~M 1966 {\em Nucl.
  Fusion\/} {\bf 6} 297

\bibitem{Contopoulos1966c}
Contopoulos G 1966 Resonance phenomena and the non-applicability of the "third"
  integral {\em Les Nouvelles M\'{e}thodes de la Dynamique Stellaire\/} vol~2
  ed H\'{e}non M and Nahon F p 223 Bull. Astron. (3)(1967)

\bibitem{ContopMouts1965}
Contopoulos G and Moutsoulas M 1965 {\em Astron. J.\/} {\bf 70} 817

\end{thebibliography}

\end{document}